\RequirePackage[T1]{fontenc}
\documentclass[11pt, urlcolor=blue, linkcolor=blue]{article} 
\usepackage{cite}
\usepackage{float}
\usepackage[utf8]{inputenc}

\usepackage{multicol}
\usepackage{multirow}
\usepackage{tablefootnote}
\usepackage{amsmath, amsthm, amssymb,slashed,mathtools,tabu}
\usepackage{ifpdf}
\ifpdf
  \usepackage[pdftex]{graphicx}
  \usepackage{epstopdf}
\else
  \usepackage[dvips]{graphicx}
\fi
\parskip=\baselineskip

\usepackage[usenames, dvipsnames]{color}
\usepackage[svgnames]{xcolor}
\usepackage[colorlinks,citecolor=RoyalBlue, urlcolor=RoyalBlue, linkcolor=RoyalBlue ]{hyperref} 

\allowdisplaybreaks[1]

\numberwithin{equation}{section}

\usepackage{booktabs}

\theoremstyle{definition}

\newcommand{\cblue}[1]{\textcolor{blue}{#1}}

\newcommand{\ccblue}[1]{\textcolor{blue}{#1}}
\newcommand{\ccred}[1]{\textcolor{red}{#1}}
\newcommand{\ccgreen}[1]{\textcolor{dgrn}{#1}}
\newcommand{\ccpurple}[1]{\textcolor{purple}{#1}}

\definecolor{mygray}{gray}{0.6}

\usepackage{upgreek}
\usepackage{bbm}

\newenvironment{myfont}[2][]{\csname#2\endcsname[#1]}{}

\usepackage{slashed}
\usepackage[makeroom]{cancel}
\usepackage[normalem]{ulem}
\usepackage{soul}
\newcommand{\stkout}[1]{\ifmmode\text{\sout{\ensuremath{#1}}}\else\sout{#1}\fi}

\usepackage{sseq}
\usepackage[all,cmtip]{xy}
\usepackage{tikz-cd}
\usepackage{tikz}
\usetikzlibrary{matrix}

\newcommand{\bea}{\begin{eqnarray}}
\newcommand{\eea}{\end{eqnarray}}
\def\be{\begin{equation}}
\def\ee{\end{equation}}

\newcommand{\e}{\hspace{1pt}\mathrm{e}}

\newcommand{\ii}{\hspace{1pt}\mathrm{i}\hspace{1pt}}

\def\CP{{\mathbb{CP}}}

\newcommand{\nn}{\nonumber}

\definecolor{red}{rgb}{1,0,0}
\definecolor{blue}{rgb}{0,0,1}
\definecolor{dblue}{rgb}{0,0,0.4}
\definecolor{green}{rgb}{0,1,0}
\definecolor{black}{rgb}{0,0,0}
\definecolor{white}{rgb}{1,1,1}

\definecolor{brn}{rgb}{.8,.4,.0}
\definecolor{redo}{rgb}{1,.5,.0}
\definecolor{ddgrn}{rgb}{0,0.4,0}
\definecolor{dgrn}{rgb}{0,0.55,0}
\definecolor{dbl}{rgb}{0,0,0.5}

\usepackage[bbgreekl]{mathbbol}
\usepackage{amscd}

\newcommand{\one}{\mathbf{1}}

\newcommand{\Z}{\mathbb{Z}}
\newcommand{\C}{\mathbb{C}}
\newcommand{\R}{\mathbb{R}}

\newcommand{\dd}{\hspace{1pt}\mathrm{d}}

\newcommand{\Refe}[1]{Ref.~\cite{#1}}
\newcommand{\Eq}[1]{(\ref{#1})} 
\newcommand{\eq}[1]{(\ref{#1})} 
 
\newcommand{\Eqn}[1]{Eqn.~(\ref{#1})} 

\newcommand{\Tr}{{\rm Tr}} 
 
\renewcommand{\Im}{{\rm Im}} 
\renewcommand{\Re}{{\rm Re}}

\newcommand{\prt}{\partial}

\newcommand{\bpm}{\begin{pmatrix}}
\newcommand{\epm}{\end{pmatrix}}
\newcommand{\bmm}{\begin{matrix}}
\newcommand{\emm}{\end{matrix}}

\newcommand{\cA}{ {\cal A} } 
\newcommand{\cB}{ {\cal B} }

\def\CA{{\cal A}}
\def\CB{{\cal B}}

\def\Z{{\mathbb{Z}}}

\def\R{{\mathbb{R}}}
\def\C{{\mathbb{C}}}

\def\Tr{{\mathrm{Tr}}}

\def \Hom{\operatorname{Hom}}

\def \Im{\operatorname{Im}}
\def \H{\operatorname{H}}

\def \Z{\mathbb{Z}}
\def \Pin{\mathrm{Pin}}

\def \CP{\mathbb{CP}}

\newcommand{\Sec}[1]{Sec.~\ref{#1}}

\usetikzlibrary{decorations.markings}
\usetikzlibrary{positioning}
\usetikzlibrary{shadings}

\newcommand{\SO}{{\rm SO}}
\newcommand{\Spin}{{\rm Spin}}
\newcommand{\U}{{\rm U}}
\newcommand{\SU}{{\rm SU}}
\newcommand{\Sp}{{\rm Sp}}
\renewcommand{\O}{{\rm O}}
\newcommand{\USp}{{\rm USp}}

\def\Sq{\mathrm{Sq}}

\def\TP{\mathrm{TP}}

\usepackage{datetime}
\usepackage{enumitem} 
\usepackage{moreenum}

\newcommand{\Wfootnote}[1]{%
\let\oldthefootnote=\thefootnote%
\stepcounter{mpfootnote}%
\addtocounter{footnote}{-1}%
\renewcommand{\thefootnote}{{W}} 
\footnote{#1}%
\let\thefootnote=\oldthefootnote%
}

\newcommand{\naturalfootnote}[1]{%
\let\oldthefootnote=\thefootnote%
\stepcounter{mpfootnote}%
\addtocounter{footnote}{-1}%
\renewcommand{\thefootnote}{{W$^-\natural$}}
\footnote{#1}%
\let\thefootnote=\oldthefootnote%
}

\newcommand{\flatfootnote}[1]{%
\let\oldthefootnote=\thefootnote%
\stepcounter{mpfootnote}%
\addtocounter{footnote}{-1}%
\renewcommand{\thefootnote}{{W$^-\flat$}}
\footnote{#1}%
\let\thefootnote=\oldthefootnote%
}

\DeclareRobustCommand\sWang
{\cblue{\includegraphics[height=3.2ex]{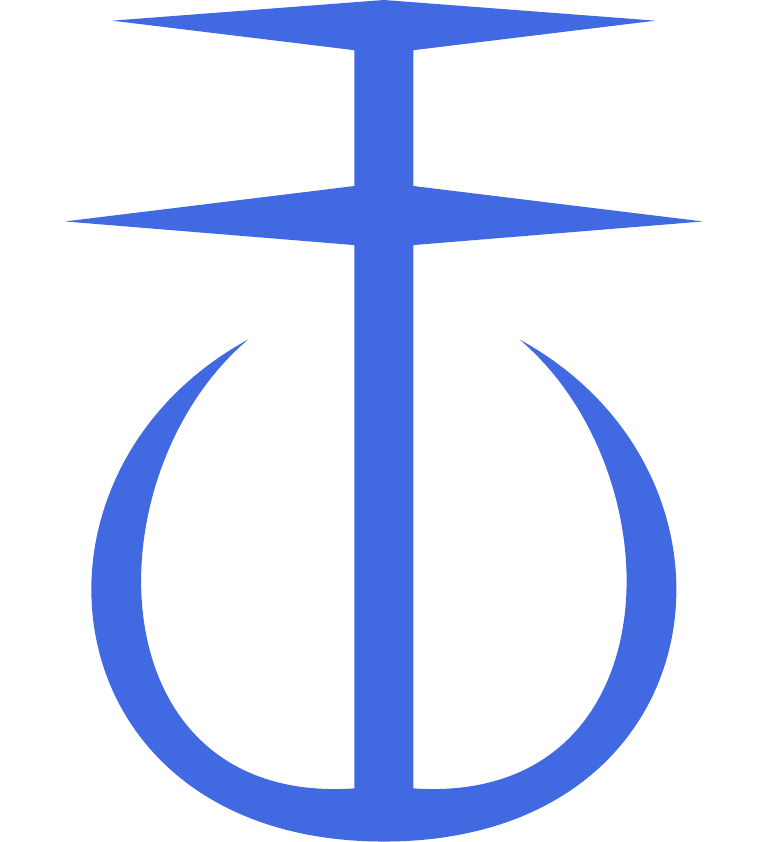}}}

\DeclareRobustCommand\sYou
{\cblue{\includegraphics[height=3.3ex]{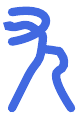}}}

\newcommand{\Wangfootnote}[1]{%
\let\oldthefootnote=\thefootnote%
\stepcounter{mpfootnote}%
\addtocounter{footnote}{-1}%
\renewcommand{\thefootnote}{\sWang}
\footnote{#1}%
\let\thefootnote=\oldthefootnote%
}

\newcommand{\Youfootnote}[1]{%
\let\oldthefootnote=\thefootnote%
\stepcounter{mpfootnote}%
\addtocounter{footnote}{-1}%
\renewcommand{\thefootnote}{\sYou}
\footnote{#1}%
\let\thefootnote=\oldthefootnote%
}

\usepackage{comment}
\def\bZ{{\mathbf{Z}}}

\usepackage{upgreek}
\newcommand{\Fig}[1]{Fig.~\ref{#1}}

\newcommand{\SM}{{\rm SM}} 
\newcommand{\PD}{{\rm PD}} 
 
\newcommand{\GUT}{{\rm GUT}}

\def \DSpin{\mathrm{DSpin}}
\def \DPin{\mathrm{DPin}}

\newcommand{\PS}{\mathrm{PS}}
\newcommand{\LR}{\mathrm{LR}}
\newcommand{\GG}{\mathrm{GG}}
\newcommand{\rL}{\mathrm{L}}
\newcommand{\rR}{\mathrm{R}}
\newcommand{\rA}{\mathrm{A}}

\newcommand{\rC}{\mathrm{C}}

\def\ra{\mathrm{a}}

\def\EM{\mathrm{EM}}

\usepackage{mathrsfs}
\usepackage{esint} 

\newcommand{\Table}[1]{Table \ref{#1}}

\newcommand\ointint{\begingroup \displaystyle \unitlength 1pt
\int\mkern-7mu\begin{picture}(0,3)\put(0,3){\oval(10,8)}\end{picture}
\mkern-8mu\int\endgroup}

\usepackage{colortbl}

\begin{document}

\begin{titlepage}

\vspace*{0cm}
\begin{center}


{\bf\LARGE{ 
Gauge Enhanced Quantum Criticality 
\\[8mm]
Beyond the Standard Model
}}

\vskip0.5cm 
\Large{\quad Juven Wang$^{1}$\Wangfootnote{
{\tt jw@cmsa.fas.harvard.edu} \quad\quad\quad\quad\; 
 \includegraphics[width=2.2in]{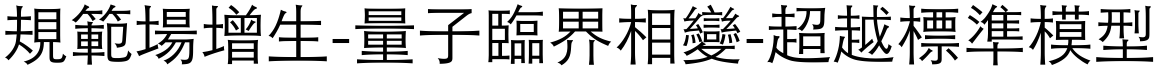}
} 
}
 \Large{\quad\quad Yi-Zhuang You$^{2}$\Youfootnote{
 {\tt  yzyou@ucsd.edu}
 \hfill Dedicate to Subir Sachdev (60) and Xiao-Gang Wen (60),\\[2mm]
  \strut \hfill Edward Witten (70) and Shing-Tung Yau (72),\\[2mm]
 \strut \hfill \quad \quad \quad 
and anniversaries of various researchers mentioned in: \href{https://www.youtube.com/results?search_query=quantum+crticiality+standard+model+Juven+Wang+Yizhuang+You}{Related presentation 
videos available online}
 }
 } 
\\[2.75mm]  
\vskip.5cm
{ {\small{\textit{$^{1}${Center of Mathematical Sciences and Applications, Harvard University,  Cambridge, MA 02138, USA}}}}
}\\
{ {\small{\textit{$^{2}${Department of Physics, University of California, San Diego, CA 92093, USA}}}}
}

\end{center}

\vskip 0.5cm
\baselineskip 16pt
\begin{abstract}

Standard lore ritualizes our quantum vacuum in the 4-dimensional spacetime (4d) governed by one of the candidate Standard Models (SMs), while lifting towards some Grand Unification-like structure (GUT)  
at higher energy scales. 
In contrast, in our work, we introduce an alternative view that the SM is a low energy quantum vacuum arising from various neighbor vacua competition in an immense quantum phase diagram. 
In general, we can regard the SM arising near the gapless quantum criticality 
(either critical points or critical regions) between the competing neighbor vacua. In particular detail, we demonstrate how the $su(3)\times su(2)\times u(1)$ SM with 16n Weyl fermions arises near the quantum criticality between the GUT competition of Georgi-Glashow (GG) $su(5)$ and Pati-Salam (PS) $su(4) \times su(2) \times su(2)$. 
We propose two enveloping toy models.
Model I is the conventional $so(10)$ GUT with a Spin(10) gauge group plus GUT-Higgs potential inducing various Higgs transitions.
Model II modifies Model I by adding a new 4d discrete torsion class of Wess-Zumino-Witten-like term built from GUT-Higgs field
(that matches a nonperturbative global mixed gauge-gravity anomaly captured by a 5d invertible topological field theory $w_2w_3(TM)=w_2w_3(V_{\rm SO(10)})$),
which manifests a  
Beyond-Landau-Ginzburg quantum criticality between GG and PS models, with extra Beyond-the-Standard-Model (BSM) excitations emerging near a quantum critical region. 
If the internal symmetries were treated as global symmetries (or weakly coupled to probe background fields), 
we show an analogous gapless 4d deconfined quantum criticality with new BSM 
fractionalized fragmentary excitations of Color-Flavor separation, 
and gauge enhancement including a Dark Gauge force sector, 
altogether requiring a double fermionic Spin structure named DSpin. 
If the internal symmetries are dynamically gauged (as they are in our quantum vacuum), we show the 4d criticality as a boundary criticality such that only appropriately gauge enhanced dynamical $so(10)$ GUT gauge fields can propagate into an extra-dimensional 5d bulk. The phenomena may be regarded as SM  deformation or
 ``morphogenesis.'' 


\flushright

\end{abstract}

\end{titlepage}

\pagenumbering{arabic}
\setcounter{page}{2}
    
\tableofcontents

\newpage


\hfill  \includegraphics[width=1.9in]{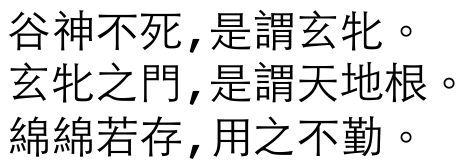}\\
\hspace*{\fill}  ``The Valley Spirit (Void Spirit) never dies;\\
\hspace*{\fill} It is named the Mysterious Female.\\
\hspace*{\fill} And the gateway of the Mysterious Female;\\
\hspace*{\fill} It is called the root of Heaven and Earth.\\
\hspace*{\fill} Dimly visible, it is there within us all the while;\\
\hspace*{\fill} Draw upon it as you will, yet use will never drain it.''\\

\hspace*{\fill} \includegraphics[width=.9in]{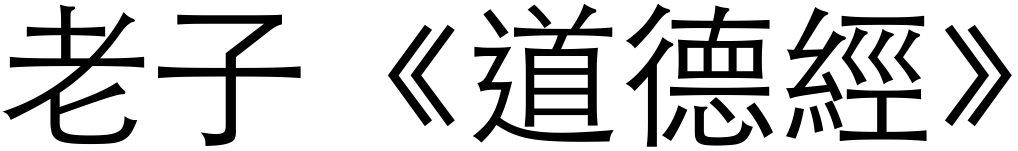}\\
\hspace*{\fill}   Laozi (B.C. 600) - Dao De Jing - an excerpt \\[14mm]
 \begin{center}
  \;\includegraphics[width=7.in]{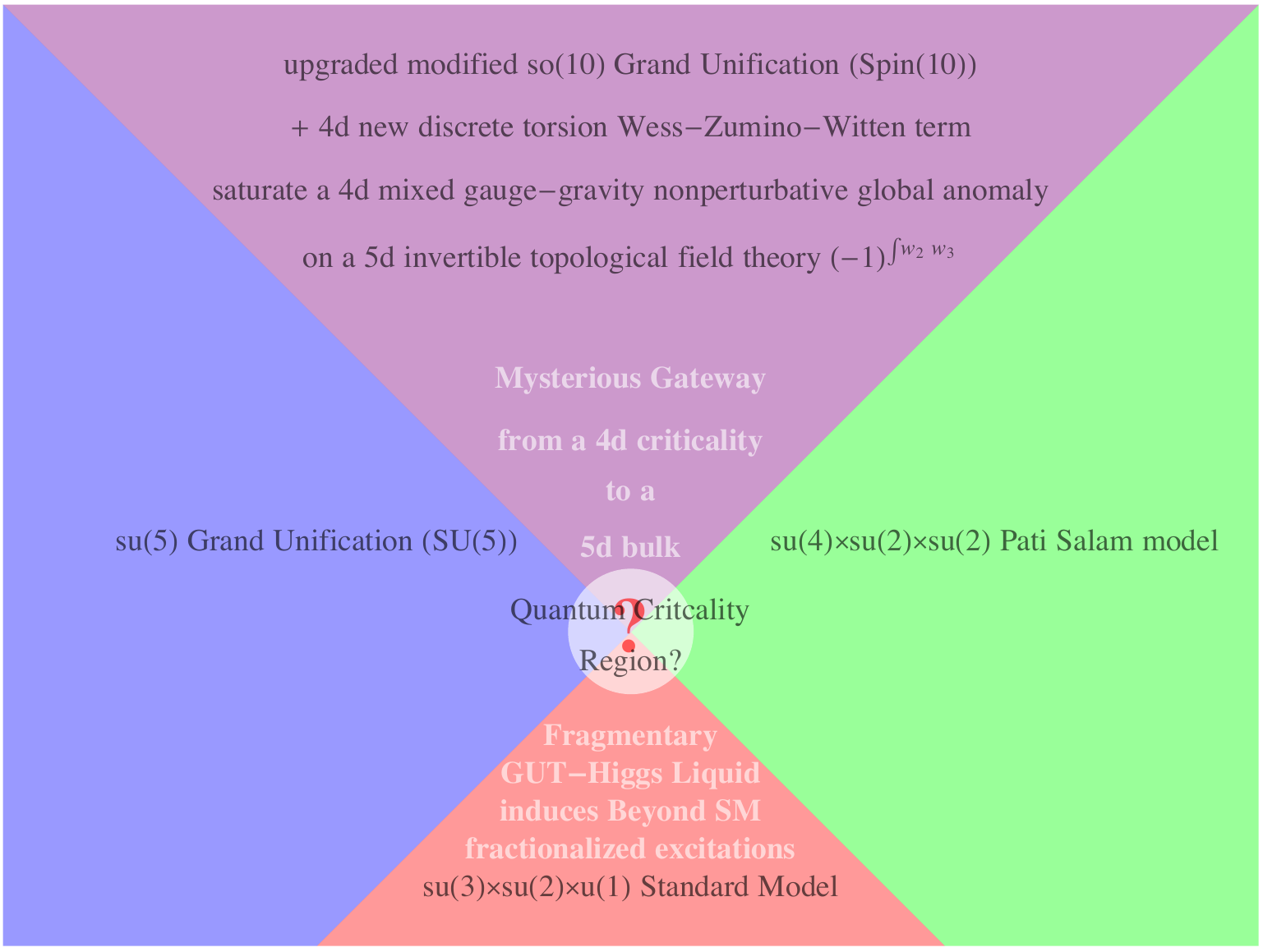}\\
  \includegraphics[width=5.in]{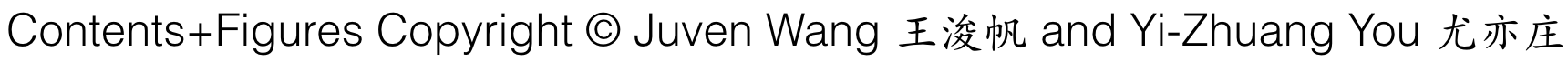}
 \end{center}
\newpage

\section{Introduction, Motivation, and Summary}
\label{sec:IntroductionandSummary}

\begin{figure}[!ht] 
\centering
(a) \includegraphics[height=0.25\textwidth]{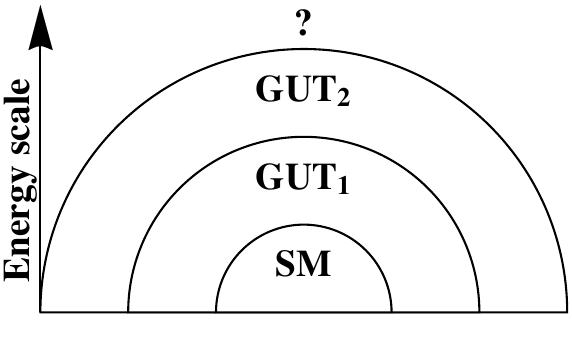}
(b) \includegraphics[height=0.25\textwidth]{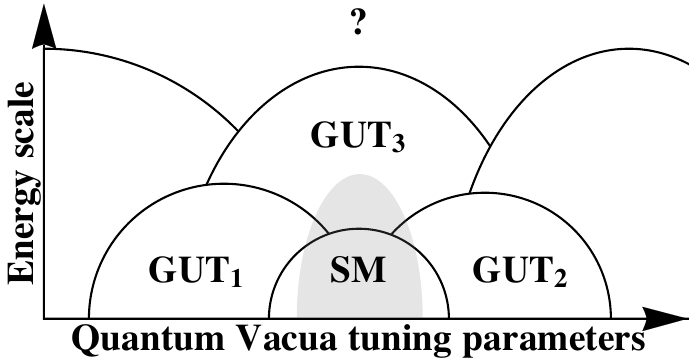}
\quad
\caption{(a) Standard lore seeks for 
a single unified dynamically gauged internal symmetry at high energy. 
One probes the shorter distance and higher energy scales to look for the GUT, SUSY, or String Theory evidence.
The vertical axis shows an energy scale, while the horizontal axis plays no physical role.
(b) We propose an alternative view: SM is just one of many possible low energy phases of the quantum vacua of our universe.
By introducing a horizontal axis that represent many possible quantum vacua tuning parameters,
we can show that SM phase can tune to other GUT phases,
even at a fix energy scale (without the necessity to go to higher energy) and at zero temperature.
SM arises near the gapless quantum 
critical region (shown as the gray area).
}
\label{fig:view-1}
\end{figure}

It is a common ritual practice in high-energy physics (HEP) to 
regards our quantum vacuum in the 4-dimensional spacetime (denoted as 4d or 3+1d) governed by one of the candidate $su(3)\times su(2)\times u(1)$ Standard Models (SMs) \cite{Glashow1961trPartialSymmetriesofWeakInteractions, Salam1964ryElectromagneticWeakInteractions, Salam1968, Weinberg1967tqSMAModelofLeptons}
as a quantum field theory (QFT) and an effective field theory (EFT) suitable below a certain energy scale,
while lifting towards one of some Grand Unification-like structure (GUT) 
\cite{Georgi1974syUnityofAllElementaryParticleForces, Pati1974yyPatiSalamLeptonNumberastheFourthColor,
Fritzsch1974nnMinkowskiUnifiedInteractionsofLeptonsandHadrons} or String Theory at higher energy scales,\footnote{{Throughout our article,
we denote $n$d for $n$-dimensional spacetime, or $n'+1$d as an $n'$-dimensional space and 1-dimensional time. 
We also denote the Lie algebra in the lower case such as $so(10)$, and denote the Lie group in the capital case such as Spin(10).
For example, we follow the convention to call the model \cite{Fritzsch1974nnMinkowskiUnifiedInteractionsofLeptonsandHadrons} as the $so(10)$ GUT, but it requires the Spin(10) gauge group.}} see \Fig{fig:view-1} (a).
 {Although many non-supersymmetric GUT models had been ruled out by experiments due to no evidence yet on the predicted proton decay (proton lifetime $> 10^{34}$ years) \cite{ParticleDataGroup2020ssz}, many physicists still speculate that GUT plays a certain crucial role in a higher energy unification \cite{Zee2003book}.
How can we remedy the conventional GUTs other than seeking for their supersymmetry (SUSY) variants or String Theory modifications at higher energy?}

\begin{figure}[!ht] 
\centering
\includegraphics[height=0.5\textwidth]{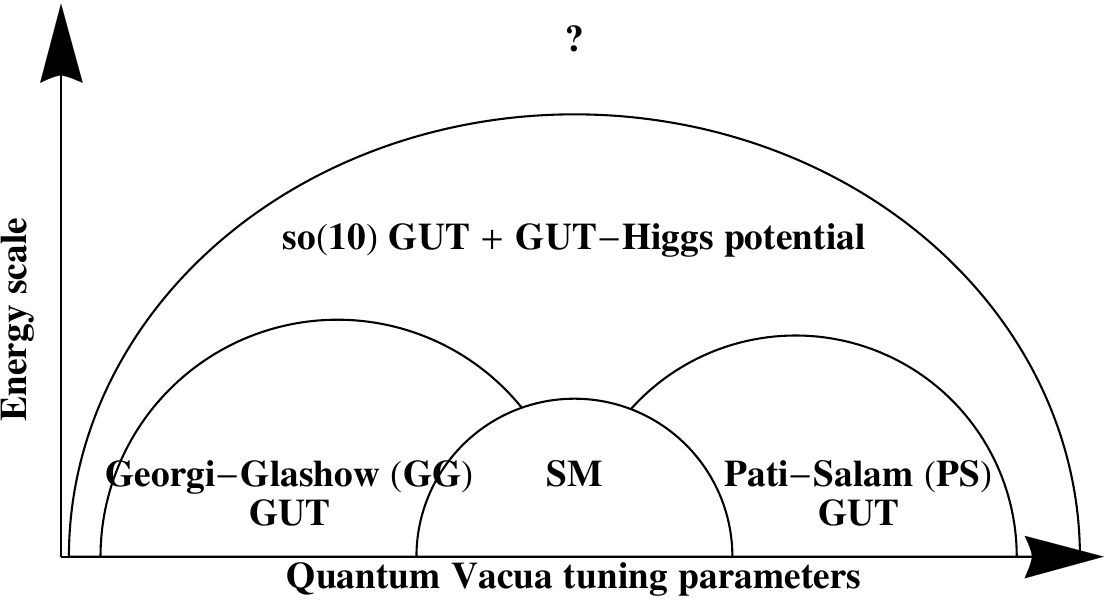}
\caption{Schematic phases for the Toy Model I: The parent EFT is the conventional $so(10)$
GUT with a Spin(10) gauge group plus GUT-Higgs potential inducing various Higgs transitions to GG, PS, or SM.
}
\label{fig:view-2}
\end{figure}

\begin{figure}[!ht] 
\centering
\includegraphics[height=0.52\textwidth]{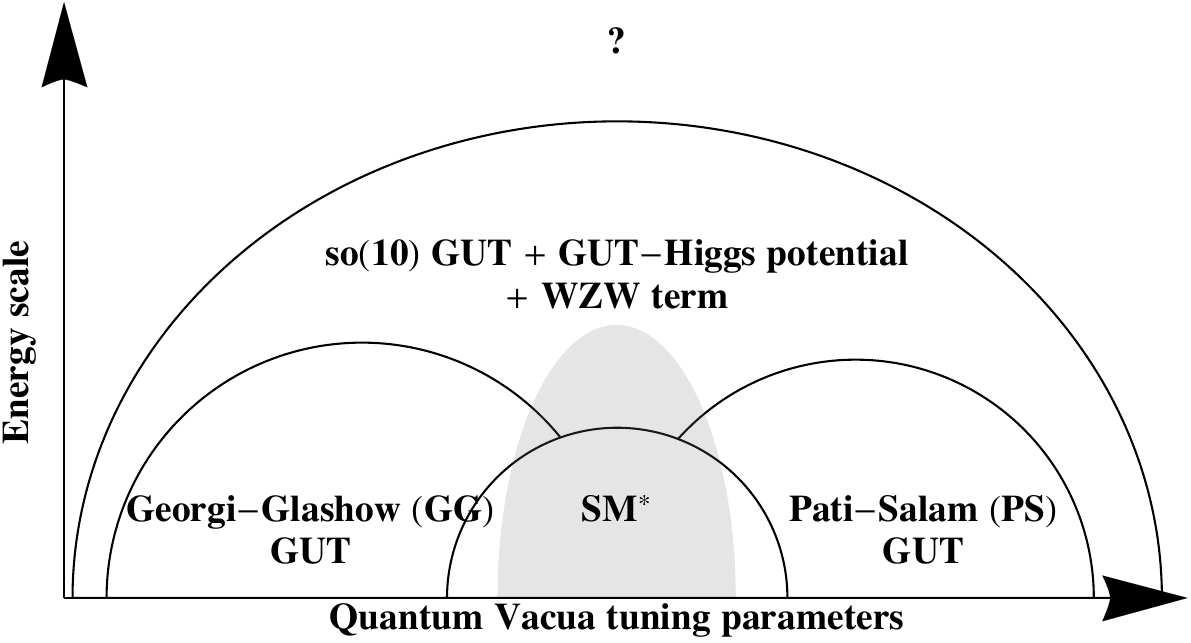}
\caption{Schematic phases for the Toy Model II:
The parent EFT is
a modified $so(10)$ GUT with a Spin(10) gauge group, 
plus not only a GUT-Higgs potential but also a new 4d discrete torsion class of Wess-Zumino-Witten-like (WZW) term built from GUT-Higgs fields
that saturates a nonperturbative global mixed gauge-gravity anomaly captured by a 5d invertible topological field theory $w_2w_3(TM)=w_2w_3(V_{\rm SO(10)})$,
which manifests a  
Beyond-Landau-Ginzburg quantum critical region 
(shown in a gray area) between GG and PS models, with extra Beyond-the-Standard-Model (BSM) excitations emerging near the quantum criticality. 
The SM + BSM physics is denoted as SM$^*$.
}
\label{fig:view-3}
\end{figure}

\begin{figure}[h!] 
  \centering
  \hspace{-1.9cm}
  \;\includegraphics[width=5in]{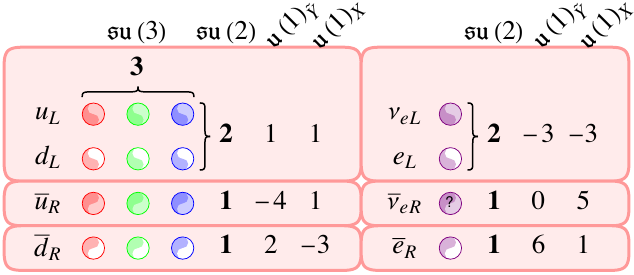}
  \caption{{Standard Model (SM).}
  The 15n Weyl fermions of SM contain the representation
$(\overline{\bf 3},{\bf 1})_{2,L} \oplus ({\bf 1},{\bf 2})_{-3,L}$  
$\oplus$
 $({\bf 3},{\bf 2})_{1,L} \oplus (\overline{\bf 3},{\bf 1})_{-4,L} \oplus ({\bf 1},{\bf 1})_{6,L}$.
 The 16n Weyl fermions of SM add an extra ${({\bf 1},{\bf 1})_{0,L}}$.
}
  \label{fig:SM-GUT-fig}
\end{figure}

\begin{figure}[h!] 
  \centering
  \hspace{-1.9cm}
  \;\includegraphics[width=5.in]{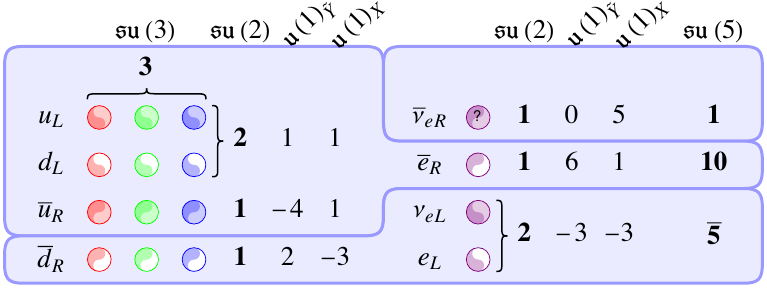}
  \caption{{Georgi-Glashow SU(5) model and the $su(5)$ GUT.
  The 15 Weyl fermions of SM are $ \overline{\bf 5} \oplus {\bf 10}$  of SU(5); namely,
$(\overline{\bf 3},{\bf 1})_{2,L} \oplus ({\bf 1},{\bf 2})_{-3,L} \sim \overline{\bf 5}$
and $({\bf 3},{\bf 2})_{1,L} \oplus (\overline{\bf 3},{\bf 1})_{-4,L} \oplus ({\bf 1},{\bf 1})_{6,L}  \sim {\bf 10}$  of SU(5).
Also $({\bf 1},{\bf 1})_{0,L}  \sim {\bf 1}$ of SU(5),
so the 16 Weyl fermions of SM are $ \overline{\bf 5} \oplus {\bf 10} \oplus {\bf 1}$  of SU(5).
}}
  \label{fig:SU5-GUT-fig}
\end{figure}

\begin{figure}[h!] 
  \centering
  \hspace{-1.9cm}
  \;\includegraphics[width=5.in]{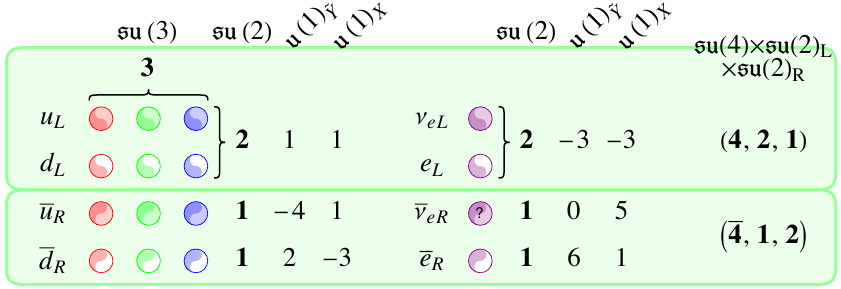}
  \caption{{Pati-Salam (PS) model:
$G_{\PS_{q'}}\equiv \frac{\SU(4)\times \SU(2)_{\rm L} \times \SU(2)_{\rm R}}{\Z_{q'}}
\equiv \frac{\Spin(6)\times \Spin(4)}{\Z_{q'}}$ with $q'=1,2$.
The 16 Weyl fermions of SM are $ ({\bf 4}, {\bf 2}, {\bf 1}) \oplus (\overline{\bf 4}, {\bf 1}, {\bf 2})$ of $su(4) \times su(2)_{\rL} \times su(2)_{\rR}$,
and the ${\bf 16}$ of $so(10)$ (or Spin(10)).
These L and R are \emph{internal} symmetry group indices. They are different from (but correlated with) the \emph{spacetime} symmetry $L$ and $R$. 
So $({\bf 3},{\bf 2})_{1,L} \oplus ({\bf 1},{\bf 2})_{-3,L}\sim ({\bf 4}, {\bf 2}, {\bf 1})_L$, 
and
$(\overline{\bf 3},{\bf 1})_{2,L} 
\oplus (\overline{\bf 3},{\bf 1})_{-4,L} \oplus ({\bf 1},{\bf 1})_{6,L} \oplus ({\bf 1},{\bf 1})_{0,L} 
\sim (\overline{\bf 4}, {\bf 1}, {\bf 2})_L$
of PS model.
}}
  \label{fig:PS-GUT-fig}
\end{figure}

{To address the above question, we propose to seek for a new viewpoint. 
In our present work, instead of viewing GUT only as some higher-energy theory of SM,
we suggest that various GUTs may be neighbor quantum vacua next to SM
in an immense quantum phase diagram\footnote{Here quantum phases mean 
that we focus on the zero temperature physics where the quantum effect is dominant, see for example an overview \cite{subirsachdev2011book}.
The quantum phase diagram at zero temperature behaves more quantum than the thermal phase diagram at finite temperature.
}
shown schematically 
in \Fig{fig:view-1} (b), 
with an underlying larger quantum vacua tuning parameter space 
(i.e., the horizontal axis in \Fig{fig:view-1} (b), \ref{fig:view-2} and  \ref{fig:view-3}). 
We provide two explicit Toy Models in \Fig{fig:view-2} and  \Fig{fig:view-3}:
 SM arises near the gapless quantum 
critical point (for \Fig{fig:view-2}) or critical region (gray area for \Fig{fig:view-3}) between the competing neighbor GUT vacua.
Readers may be puzzled:
What precisely can be the quantum vacua tuning parameters? 
What can we gain from this viewpoint? What are the motivations? Let us address these issues one by one.
}
\begin{itemize}
\item   \emph{Quantum vacua tuning parameters} can be as familiarly simple as the tuning of the
GUT-Higgs potential $(\Big(r_{{\mathbf{R}}} (\Phi_{\mathbf{R}})^2 +\lambda_{{\mathbf{R}}} (\Phi_{{\mathbf{R}}})^4\Big))$
of some GUT-Higgs field $\Phi_{\mathbf{R}}$
that can induce a Higgs condensation\footnote{Throughout our work,
whenever we mention Higgs field or Higgs transition, we normally mean the GUT-Higgs instead of the electroweak Higgs. 
Namely, we always focus on the SM gauge group $su(3)\times su(2)\times u(1)$ 
as above the electroweak scale instead of $su(3)\times u(1)_{\rm EM}$
below the electroweak scale. \quad \quad\quad }
  phase transition 
 via tuning from $r_{{\mathbf{R}}} >0$ to $r_{{\mathbf{R}}}<0$. 
The {quantum vacua tuning parameters} can be those triggering a scalar condensation  $\langle \Phi_{\mathbf{R}} \rangle \neq 0$ in the $r_{{\mathbf{R}}}<0$ region.
{The possibility to access the GUT vacua from the SM vacuum by tuning certain model parameters has been largely overlooked in the existing literature, 
because some of these tuning parameters appear to be \emph{perturbatively irrelevant} at the SM fixed point. 
A key proposal of this work is to investigate the \emph{non-perturbative} effect of these tuning parameters 
in driving quantum phase transitions from the SM phase to adjacent GUT phases.}
\item \emph{Deformation class of QFT}:
Given the importance of 
symmetry and its associated 't Hooft anomaly of QFT,
Seiberg \cite{NSeiberg-Strings-2019-talk} and others\footnote{In fact the related concept has been used in arguing that the fermion doubling problem 
(occurred in regularizing chiral fermions nonperturbatively on the lattice with a chiral $G$ symmetry) 
can be resolved by gapping the mirror chiral fermion if and only if the chiral fermion
is anomaly free in $G$ 
(tautologically, the mirror fermion is also anomaly free in $G$), see \cite{WangWen2018cai1809.11171, RazamatTong2009.05037} and reference therein.
The argument follows directly from the fact that the gapless anomaly-free $G$-symmetric chiral fermion theory
is in the \emph{same deformation class} of the gapped anomaly-free $G$-symmetric theory.}
conjectured that 
seemly different $d$d QFTs within the same symmetry $G$ and same 't Hooft anomaly $\bZ_{d+1}$ of symmetry $G$  \cite{tHooft1979ratanomaly}
can indeed be deformed to each other via adding degrees of freedom at short distances
that preserve the same symmetry and that maintain the same overall anomaly.
Namely, the whole system allows all symmetric interactions between the original QFT and any new symmetric QFTs
brought down from high energy. 
This organization principle that connects a large class of QFTs together 
within the same data $(G, \bZ_{d+1})$
via any symmetric deformation (possibly with discontinuous or continuous quantum phase transitions \cite{subirsachdev2011book} between different phases)
is called the \emph{deformation class of QFTs} in $d$d \cite{NSeiberg-Strings-2019-talk},
which is indeed controlled by the \emph{cobordism or deformation class of invertible topological quantum field theory} $\bZ_{d+1}$ in $d+1$d \cite{2016arXiv160406527F}.
One can further define the deformation class for 4d SM \cite{WangWanYou2112.14765}.

As we will see, our viewpoint in \Fig{fig:view-1} (b) (also in \Fig{fig:view-2} and  \Fig{fig:view-3}) is not only compatible with this 
\emph{symmetric deformation class of QFT} \cite{NSeiberg-Strings-2019-talk}, but also that we allow \emph{symmetry-breaking deformations}, 
along the quantum vacua tuning parameter space. We may refer to all these deformations of the SM to other neighbor vacua
as ``morphogenesis'' of the SM.

\item   \emph{Proton decay}: The aforementioned issue of GUT proton decay may be resolved in our framework by two ways.
First, the change of viewpoint --- instead of looking for GUT proton decay in our vacuum (or in a higher energy GUT along the \emph{vertical axis}, as in \Fig{fig:view-1}),
we may look for GUT proton decay by first moving to the appropriate quantum vacuum along the \emph{horizontal axis} in \Fig{fig:view-1} (b)
that already lives this specific GUT.\footnote{Take Georgi-Glashow $su(5)$ GUT \cite{Georgi1974syUnityofAllElementaryParticleForces}
as an example.
The conventional viewpoint may be problematic because this specific GUT
may \emph{not} be the correct higher energy theory of our vacuum along the \emph{vertical axis}, in \Fig{fig:view-2} and  \Fig{fig:view-3}.
If we want to detect any proton decay in $su(5)$ GUT, hypothetically we may imagine to create a small bubble within the domain wall
such that inside the bubble resides any possible deformation of the SM (e.g., any models along the horizontal axis in \Fig{fig:view-2} and  \Fig{fig:view-3}).
Although changing the large-scale quantum vacuum structure of our SM universe is likely energetically impossible,
changing the quantum vacuum inside a small-scale bubble is possibly feasible experimentally.} 
Second, a modified parent EFT that controls all possible deformation of SM in the phase diagram may give rise to a different proton decay rate.\footnote{For example, two different toy-model parent EFTs in \Fig{fig:view-2} and  \Fig{fig:view-3} respectively can give different proton decay rates.
We do not attempt to compute the explicit proton decay rate in this work, because so far we only have two Toy Models
that control a ${\rm p}= \{0,1\} \in \Z_2$ deformation class labeled by a $\Z_2$ nonperturbative global anomaly in 4d.
The two Toy Models describe only a partial deformation class of the SM. 
There is also a $\Z_{16}$ deformation class for SM \cite{WangWanYou2112.14765}, etc.
To compute a experimentally sensible proton decay rate for our vacuum, it will be the best that we (1) locate the specific point on the phase diagram that precisely labels our vacuum,
and (2) compute from the general enveloping parent EFT that includes all physically relevant deformations.} 
{The experimental bound on proton decay rate only rules out the possibility to access non-supersymetic GUT phases from the SM phases by \emph{thermal phase transitions} (i.e.~by raising the energy or temperature scales), but it does not say anything about accessing these GUT phases by \emph{quantum phase transitions} (by tuning parameters 
near ground states at low-energy). This work exactly focuses on the later possibility of quantum phase transitions among the SM and GUTs.}

\end{itemize}
The above three arguments summarize the motivation and philosophy behind our viewpoint.
Namely, in our present work, we initiate and introduce an alternative complementary perspective --- we propose that the SM vacuum can be 
a low energy quantum vacuum arising from the quantum competition of various neighbor GUT vacua in a quantum phase diagram.
SM is just one possible phase allowed by the deformation class of SM \cite{WangWanYou2112.14765}. Let us list down some key results of our work:
\begin{itemize}
\item
In general, we propose that the SM may arise {as one adjacent phase} from the vicinity of 
gapless quantum criticality (either a critical point for Model I in \Fig{fig:view-2}, or a critical region for Model II in \Fig{fig:view-3}) between the competing 
neighbor GUT vacua.
\item In particular, we demonstrate how
the $su(3)\times su(2)\times u(1)$ SM  
\cite{Glashow1961trPartialSymmetriesofWeakInteractions, Salam1964ryElectromagneticWeakInteractions, Salam1968, Weinberg1967tqSMAModelofLeptons}
with 16n Weyl fermions (\Fig{fig:SM-GUT-fig}) could emerge near the quantum criticality 
between two neighbor vacua of
Georgi-Glashow $su(5)$ model (GG) \cite{Georgi1974syUnityofAllElementaryParticleForces} (\Fig{fig:SU5-GUT-fig})
and Pati-Salam $su(4) \times su(2) \times su(2)$ model (PS) \cite{Pati1974yyPatiSalamLeptonNumberastheFourthColor} (\Fig{fig:PS-GUT-fig}), which represents two distinct Higgs phases of the further unified $so(10)$ GUT (with a Spin(10) gauge group).

\item We propose two explicit Toy Models.
The two models are differed by whether they can carry a
4d nonperturbative global anomaly of mixed gauge-gravitational (i.e., gauge-diffeomorphism) probes,
captured by a 5d invertible topological quantum field theory (TQFT):\footnote{The 
$w_j$ is the $j$-th Stiefel-Whitney (SW) characteristic class.
The $w_j(TM)$ is the SW class of spacetime tangent bundle $TM$ of manifold $M$. 
The $w_j(V_{G})$ is the SW class of the principal $G$ bundle.
This mod 2 class $w_2w_3$ global anomaly has been checked to be absent in the $so(10)$ GUT by \Refe{WangWen2018cai1809.11171, WangWenWitten2018qoy1810.00844}.
This mixed gauge-gravitational anomaly is tightly related to \emph{the new SU(2) anomaly} \cite{WangWenWitten2018qoy1810.00844} due to the bundle constraint
$w_2w_3(TM) = w_2w_3(V_{G})$ with $G$ can be substituted by $\SO(3) \subset \SO(10)$ related to the embedding $\SU(2) = \Spin(3) \subset \Spin(10)$.
However, as we will see, it is natural to introduce a new 4d WZW term (appending to the $so(10)$ GUT)
with this $w_2w_3$ global anomaly in order to realize the SM vacuum as the quantum criticality phenomenon between the neighbor SU(5) GUT and Pati-Salam vacua.\\[2mm]
The $w_2w_3$ global anomaly also occurs on 
a certain $\Z_2$ gauge theory with fermionic strings \cite{Thorngren1404.4385} 
and all-fermion U(1) electrodynamics \cite{WangPotterSenthil1306.3238, KravecMcGreevySwingle1409.8339} 
which is a pure U(1) gauge theory 
whose electric, magnetic, and dyonic objects are all fermions. For these $\Z_2$ and U(1) gauge theories, they do have the spacetime tangent bundle constraints on $TM$, but do
\emph{not} have the analogous gauge bundle constraints on $V_G$. So this $w_2w_3=w_2w_3(TM)$ anomaly
becomes a pure gravitational anomaly for these $\Z_2$ and U(1) gauge theories.\\[2mm]
We recommend the following references 
\cite{GarciaEtxebarriaMontero2018ajm1808.00009, WanWang2018bns1812.11967, 2019arXiv191011277D, WW2019fxh1910.14668} 
or this seminar video \cite{CMSAJWUltraUnification}
for readers who wish to overview some modern perspectives 
about the anomalies of SM and GUT relevant gauge theories. In particular, we follow closely \Refe{WW2019fxh1910.14668, CMSAJWUltraUnification}. 
In summary, we may address anomalies with different adjectives to characterize their properties:
\begin{itemize}
\item  invertible vs noninvertible: We only focus on the invertible anomalies,
which follow the standard definition of anomalies (also in high-energy physics) 
captured by one higher-dimensional invertible TQFT as the low energy theory of invertible topological phases.
The $d$d invertible anomalies (also the $(d+1)$d invertible TQFTs) are classified by the cobordism group data
$\Omega^{d}_{G} \equiv \TP_d(G)$ defined in Freed-Hopkins 
\cite{Freed2016}.
The partition function ${\bf Z}$ of a $(d+1)$d invertible TQFT satisfies ${\bf Z}(M^{d+1})=1$ on a closed $M^{d+1}$-manifold.\\[1mm]
In contrast, the noninvertible anomalies are non-standard (usually not named as anomalies in high-energy physics), 
characterized by non-invertible topological phases with intrinsic topological orders.
\item perturbative local vs nonperturbative global anomalies: Whether the anomalies are local (or global), is determined by whether the gauge or diffeomorphism transformations 
are infinitesimal (or large) transformations, continuously deformable (or not deformable) to the identity element. The classifications of local vs global anomalies are
the integer $\Z$ vs the finite torsion $\Z_n$ classes respectively. 
\item gauge anomaly vs mixed gauge-gravity anomaly vs gravitational anomaly: 
The adjective, gauge or gravity, refers to the types of couplings or probes that we require to detect them -- whether the probes depends on the internal gauge bundle/connection
or the spacetime geometry. 
\item background fields or dynamical fields: 
Anomalies of global symmetries probed by non-dynamical background fields are known as \emph{'t Hooft anomalies}.
Anomalies coupled to dynamical fields must lead to \emph{anomaly cancellations} to zero for consistency.
\end{itemize}
\label{footnote:w2w3anomaly}
}
\bea
 (-1)^{\int {\rm p} \; w_2w_3(TM)}=(-1)^{\int {\rm p} \; w_2w_3(V_{\SO(10)})} \text{  with ${\rm p} \in \{0,1\} =\mathbb{Z}_2$. }
\eea

{\bf Toy Model I as the ${\rm p}=0$ class without $w_2w_3$ anomaly:} Its parent EFT is the conventional $so(10)$
GUT 
with a Spin(10) gauge group \cite{Fritzsch1974nnMinkowskiUnifiedInteractionsofLeptonsandHadrons}
plus a GUT-Higgs potential inducing various Higgs transitions to GG, PS, or SM,
schematically shown in \Fig{fig:view-2}. The first model has no $w_2w_3$ or any other anomaly within the Spin(10).

{\bf Toy Model II as the ${\rm p}=1$ class with $w_2w_3$ anomaly and WZW term:}
To introduce non-trivial competitions between GG and PS phases, we consider a new parent EFT of
a modified $so(10)$ GUT with a Spin(10) gauge group, 
which includes not only the familiar $so(10)$ GUT 
plus a GUT-Higgs potential, but also
a new extra 4d discrete torsion class of Wess-Zumino-Witten-like (WZW) term that saturates a mod-2 class $w_2w_3$ anomaly within the Spin(10).

The WZW term introduces nonperturbative interaction effects between different GUT-Higgs fields, which cause a substantial change of the deformation class of QFT vacuum that cannot be smoothly connected to the conventional $so(10)$ GUT vacuum. There are distinct ${\rm p} \in \{0,1\} =\mathbb{Z}_2$ deformation classes of QFT.

We propose a schematic quantum phase diagram, shown in \Fig{fig:phase-schematic},  
interpolating between different quantum vacua:
the modified $so(10)$ GUT + WZW term, the $su(5)$ GG GUT, 
the $su(4) \times su(2)_{\rL} \times su(2)_{\rR}$ PS model, and 
the $su(3)\times su(2)\times u(1)$ SM. 
{In fact, 
this $w_2w_3$ global anomaly (hereafter $w_2w_3$ as a shorthand for the precise bundle constraint $w_2w_3(TM)=w_2w_3(V_{\SO(10)})$) 
does  \emph{not} occur when the internal symmetry is within $su(5)$ (for the GG $su(5)$ GUT), 
\emph{nor} occur within $su(4) \times su(2) \times su(2)$  (for the PS model),
\emph{nor} occur within $su(3)\times su(2)\times u(1)$ (for the SM).
Alternatively, we can also regard this $w_2w_3$ anomaly is matched in the GG, PS, and SM via the symmetry breaking.
This $w_2w_3$ global anomaly 
\emph{only} occurs when the internal symmetry is Spin(10) (for the modified $so(10)$ GUT + WZW term), but this anomaly still 
constrains the full quantum phase diagram (\Fig{fig:phase-schematic}).
}

For {\bf Toy Model I without WZW term and without $w_2 w_3$ anomaly}, we should remove the whiten quantum critical region in \Fig{fig:phase-schematic}, but we are left with a quantum critical point at the origin.

For {\bf Toy Model II with WZW term and with $w_2 w_3$ anomaly}, 
we encounter the whiten quantum critical region near the origin in \Fig{fig:phase-schematic}.

\begin{enumerate}[leftmargin=.mm, label=\textcolor{blue}{Case (\arabic*)}., ref={Case (\arabic*)}]
\item \label{Case1}
If the internal symmetries were pretended to be global symmetries (or weakly gauged by probe background fields), 
then we are dealing with the quantum criticality between Landau-Ginzburg global symmetry breaking phases in 4d.
Conventionally, the global symmetry breaking pattern can be triggered by the GUT-Higgs fields.
Surprisingly, for Model II (\Fig{fig:view-3}), we discover a gapless quantum phase with fractional excitations and deconfined emergent gauge structure
in analogy to 4d 
\emph{deconfined quantum criticality}\footnote{The concept 
of deconfined quantum criticality was first developed in the condensed matter community\cite{SenthildQCP0311326}, to describe a class of direct continuous transition between two distinct symmetry breaking phases with fractionalized excitations and gauge structures emerging in the low-energy spectrum at and only at the transition. It occurs when a quantum system with global symmetry $G$ has the tendency to spontaneously break the symmetry to its distinct subgroups $G_{\text{sub},1}$ and $G_{\text{sub},2}$, while the low-energy effective field theory has $G$-anomaly but not $G_{\text{sub},1}$- or $G_{\text{sub},2}$-anomalies, in terms of 't Hooft anomalies. 
Then the two symmetry breaking phases cannot share a trivial $G$-symmetric intermediate phase, paving ways for gapless phase transition and fractionalized excitations to emerge.
\\[2mm]
Several recent works explore the possible deconfined quantum criticality in 4d spacetime (see 
\cite{BiSenthil1808.07465, Wan2018djlW2.1812.11955,BiLakeSenthil1910.12856,WangYouZheng1910.14664} and References therein).
A hint toward our construction of 4d deconfined quantum criticality between symmetry breaking phase is the fact that the Spin(10) (treated as global symmetry) can have a 't Hooft anomaly of gauge-gravity anomaly type (due to the aforementioned $w_2w_3$ anomaly);
while the smaller subgroups with Lie algebras $su(5)$  of GG, $su(4) \times su(2) \times su(2)$ of PS, or $su(3)\times su(2)\times u(1)$ of SM, 
have no such $w_2w_3$ anomaly. So the \emph{anomalous} spacetime-internal Spin(10) symmetry hints a possible fractionalization of the GUT-Higgs field as a deconfined quantum criticality.\\[2mm]
A crucial idea of deconfined quantum criticality construction is that {``the 
$G_{\PS}$-\emph{symmetry-breaking} topological defect of the GG GUT-Higgs model traps the fractionalized quantum number of 
\emph{unbroken} GG internal symmetry group;
while vice versa, the $G_{\GG}$-\emph{symmetry-breaking} topological defect of the 
PS GUT-Higgs model traps the quantum number of \emph{unbroken} PS internal symmetry group.''
Here $G_{\PS}$-symmetry-breaking and $G_{\GG}$-symmetry-breaking respectively refer to the internal symmetry groups $G$ (i.e., gauge group) 
of PS and GG models are \emph{partly} broken.}
\\[2mm]
The terminology \emph{gauge enhanced quantum criticality} is introduced in \cite{WangYouZheng1910.14664}.
}
beyond the Landau-Ginzburg-Wilson-Fisher critical phenomena. 
Specifically, we propose a 4d mother effective field theory,
where the GUT-Higgs \emph{bosonic} fields
can be fractionalized to new fragmentary \emph{fermionic} excitations, with extra \emph{gauge enhancement}. 
An example of such gauge enhancement introduces a new U(1) gauge sector called {$[\U(1)']^{\text{emergent}}_{\text{gauge}}$}, 
different from the SM electrodynamics $\U(1)_{\EM}$.
We name such a new theory as
a {\bf Fragmentary GUT-Higgs Liquid model} with emergent new fermions and new gauge fields, \emph{emergent only near the quantum criticality}. 

\item \label{Case2}
If the internal symmetries are dynamically gauged (as they are not global symmetries but indeed are gauged in our quantum vacuum), we show the gauge-enhanced 4d criticality
not merely has the emergent {$[\U(1)']^{\text{emergent}}_{\text{gauge}}$}, but also has the enhanced Spin(10) gauge group. 
The Spin(10) gauge group and {$[\U(1)']^{\text{emergent}}_{\text{gauge}}$} forms 
a gauge enhancement of the smaller gauge groups of the SM, GG or PS models, 
only near the quantum criticality, see \Fig{fig:phase-schematic}. 

Because the 5d invertible TQFT has the bundle constraint $w_2w_3(TM)=w_2w_3(V_{\SO(10)})$,
once the internal symmetries (such as the Spin(10)) are dynamically gauged, 
the 5d bulk is \emph{no longer} an invertible TQFT. 
The Spin(10) gauge fields have also to be dynamically gauged in the 5d bulk. 
The Spin(10) gauge fields contribute \emph{deconfined gapless modes} in 5d\footnote{The reason 
that the non-abelian gauge theory can become gapless in 5d can be understood
simply by analyzing
the renormalization group (RG) fixed point at the 5d Yang-Mills term,
the dimensional analysis says  
$[|F|^2] \sim [F][F] \sim [\dd A][\dd A] + [\dd A][A]^2 +[A]^4$.
 The kinetic term $[\dd A][\dd A]$ has the canonical scaling dimension 5 in 5d (i.e., $E^5$ in energy $E$).
 The $[\dd]$ has a dimension 1 and the $[A]$ has a dimension $3/2$. 
The $[\dd A][A]^2$ has a dimension $11/2$, while the $[A]^4$
has a dimension 6, which means that the $[\dd A][A]^2$ and $[A]^4$ become irrelevant at low energy.
Thus, the 5d non-abelian Yang-Mills term $|F|^2$ behaves like the \emph{gapless} 5d abelian Maxwell term $|\dd A|^2$.
} 
 (in contrast to the \emph{confined} non-abelian gauge fields being \emph{gapped} in 4d).
Remarkably, the Spin(10) gauge fields in 5d turns the previous TQFT $w_2w_3(TM)=w_2w_3(V_{\SO(10)})$ 
into a 5d gapless bulk criticality!


In summary, when the internal symmetries are dynamically gauged (as in our gauged quantum vacuum),
\begin{itemize}
\item {\bf  \emph{4d gauge fields}}:
The gauge fields of SM, GG, and PS GUT ($su(3)\times su(2)\times u(1)$, $su(5)$, and $su(4) \times su(2)_{\rL} \times su(2)_{\rR}$)
are still restricted in 4d in their respective regions of quantum phase diagram (\Fig{fig:phase-schematic}).
There is still some emergent {$[\U(1)']^{\text{emergent}}_{\text{gauge}}$} gauge field, also restricted in 4d, as 
a 4d boundary deconfined quantum criticality (the same as the previous \ref{Case1} when internal symmetry is not gauged).
\item {\bf  \emph{5d gauge fields}}:
However, when and only when the GUT gauge fields 
are appropriately gauge enhanced (to the Spin(10) gauge fields in our \Fig{fig:phase-schematic}), 
then they can propagate into the extra-dimensional 5d bulk, and they can induce 
a 5d bulk criticality.
\end{itemize}

Indeed our proposal manifests additional Beyond-the-Standard-Model (BSM) excitations.
After all, what are these BSM excitations near the quantum criticality in our theory?
\begin{itemize}
\item 
{\bf \emph{Dark Gauge} force sector}: the emergent {$[\U(1)']^{\text{emergent}}_{\text{gauge}}$} gauge fields correspond to analogous Dark Photon.
However, our {$[\U(1)']^{\text{emergent}}_{\text{gauge}} 
\equiv [\U(1)']^{\text{dark}}_{\text{gauge}}$} does not directly interact with the SM gauge forces, nor interact with the SM quarks and leptons.
This Dark Photon sector can be a  {\bf \emph{light Dark Matter}} candidate.
The {$[\U(1)']^{\text{dark}}_{\text{gauge}}$} only interacts with the 
fractionalized new fragmentary \emph{fermionic} excitations that we name \emph{colorons} and \emph{flavorons}.
\item {\bf Fragmentary fermionic \emph{colorons} and \emph{flavorons}}:
These are  fractionalized excitations as the \emph{fermionic patrons}. We implement the parton construction, 
where two (or multiple) of {patrons} ($\xi_a, \xi_b, \dots$) can combine with emergent gauge fields to form the GUT-Higgs $\Phi$:
\bea
{\Phi_{ab} \sim \xi_a^\dagger \xi_b, \text{ or more precisely } 
\Phi_{ab}(x) \sim \xi_a^\dagger (x) \exp(\ii \int_{x}^{x} a_{\mu,\text{gauge}}^{\text{dark}} \dd x^{\mu} ) \xi_b (x) }.
\eea
The GUT-Higgs $\Phi$ is also the basic degrees of freedom for the 4d WZW term that saturates the $w_2w_3$ anomaly.
To rephrase what we had said, the GUT-Higgs $\Phi$ is split into the 
fractionalized fragmentary \emph{colorons} and \emph{flavorons}.
Just as the GUT-Higgs $\Phi$ can interact with the SM particles and SM gauge forces,
the fragmentary \emph{colorons} and \emph{flavorons} can also  interact with the SM particles and SM gauge forces.
The \emph{colorons} carries the SM's $\SU(3)_c$ strong gauge charge,
while the  \emph{flavorons} carries the SM's $\SU(2)_{\rm L}$ weak gauge charge.
Just like the GUT-Higgs are made to be very heavy,
these \emph{colorons} and \emph{flavorons} are also heavy and
can also be the  {\bf   \emph{heavy Dark Matter}} candidates.
{This fractionalization accompanies the emergent dark gauge field $a_{\mu,\text{gauge}}^{\text{dark}}$.}
\end{itemize}

\end{enumerate}


\item \emph{The number of generations/families $N_{f}$}:
{So far we have not yet specified the role of the number of generations $N_{f}$ of quarks and leptons in our theory. 
If each generation of 16 SM Weyl fermions associates with its own GUT-Higgs field and its WZW term,
then the generation number $N_{f}$ times of 16 SM Weyl fermions with $N_{f}$ GUT-Higgs field requires a constraint 
$N_{f}  = 1 \mod 2$ to match the $w_2 w_3$ anomaly, where $N_{f} = 3$ generation indeed works. 
However, regardless the $N_{f}$ of SM,
in general, we can just introduce a single (or any odd number) of GUT-Higgs field and WZW sector
to match the $1 \mod 2$ class of $w_2 w_3$ anomaly.}
In any case,
it is inspiring to confirm
our proposal on the gauge enhanced quantum criticality can really happen between our $N_{f}  =3$ SM quantum vacuum 
and the neighbor GUT vacua.  
In this article, we focus on $N_{f}  =1$ for simplicity, but we can also triplicate $N_{f}  =1$ to $N_{f}  =3$.

\end{itemize}


%

\newpage

\begin{figure}[h!] 
  \centering
  \hspace{-1.9cm}
  \;\includegraphics[width=5.in]{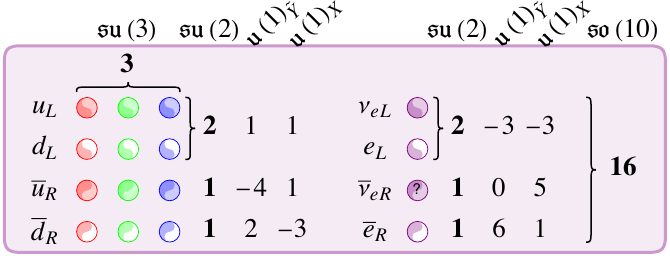}
  \caption{{The $so(10)$ GUT model:
The 16 Weyl fermions of Spin(10),
form the ${\bf 16}$-dimensional representation of Spin(10).
}}
  \label{fig:so10-GUT-fig}
\end{figure}

\begin{figure}[h!] 
\centering
\includegraphics[width=0.7\textwidth]{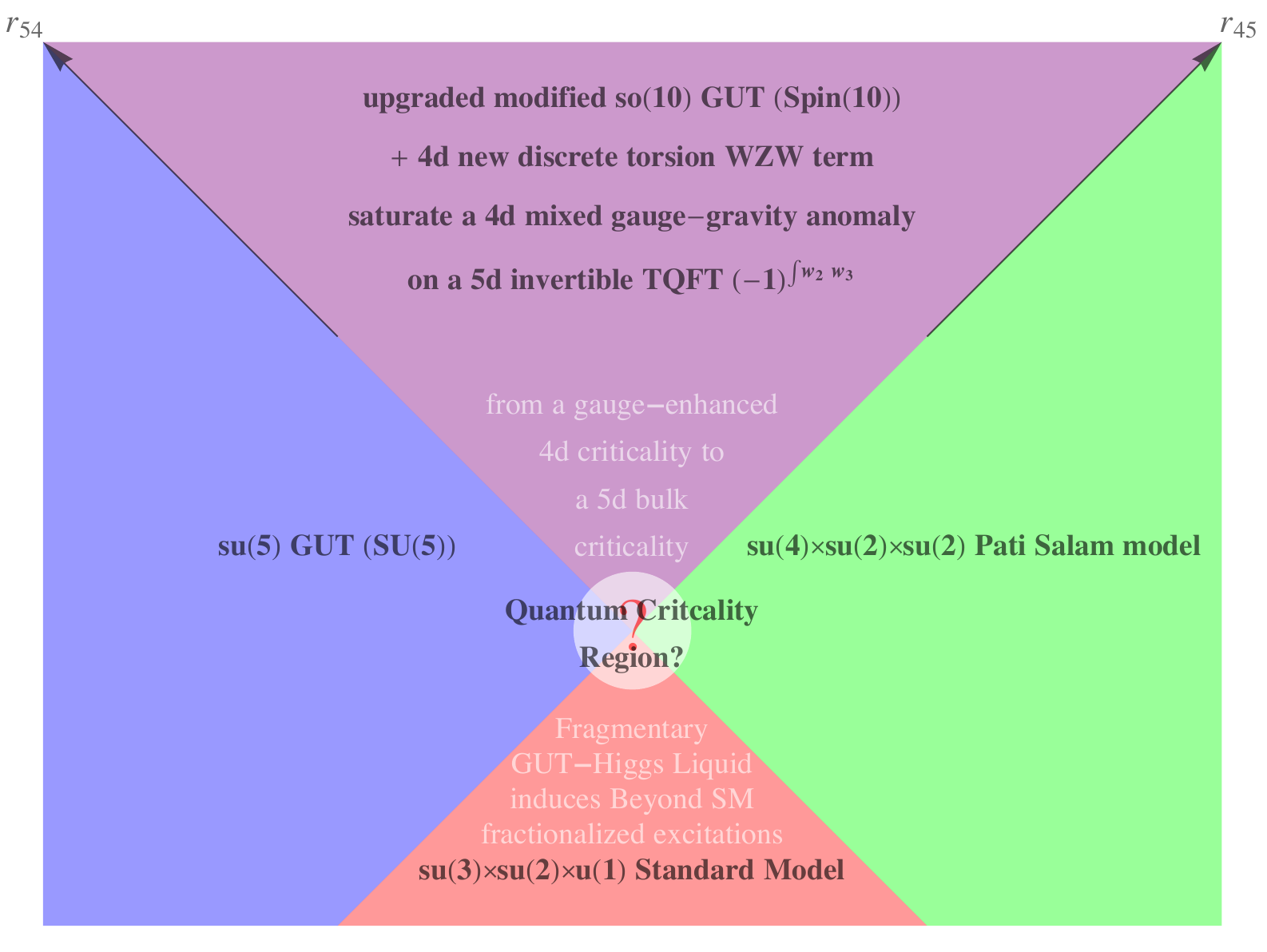}
\caption{One of our research motif is proposing and investigating
this schematic quantum phase diagram. 
The phase diagram interpolates between different quantum EFT vacua:
the $so(10)$ GUT (Spin(10) group), the $su(5)$ GUT (SU(5) group), 
the $su(4) \times su(2)_{\rL} \times su(2)_{\rR}$ Pati-Salam model (PS), and 
the $su(3)\times su(2)\times u(1)$ Standard Model (SM). 
{We will explore the nature of phase transitions later in \Sec{sec:EFTHiggs}.
We propose the whiten region as a possible quantum critical region, which we explored in \Sec{sec:EFTHiggs} and \Sec{sec:EFTBSMGEQC}.}
Here $r_{\mathbf{R}}$ denotes the coefficient of the effective quadratic potential of $\Phi_{\mathbf{R}}$ field in the representation ${\mathbf{R}}$. 
The corresponding GUT-Higgs $\Phi_{\mathbf{R}}$ field will condense in the representation-${\mathbf{R}}$ if $r_{\mathbf{R}}<0$.
Relatively speaking, the infrared (IR) low energy is drawn with the \ccred{red} color (for SM), 
the intermediate neighbor phases are drawn with the \ccgreen{green} or \ccblue{blue} color (for PS or SU(5) models),
while the ultraviolet (UV) higher energy is drawn with the \ccpurple{violet purple} color (for Spin(10));
although the readers should keep in mind that we really explore the near-ground-state, zero-energy and zero-temperature quantum phase diagram.
These colors are also designed to match the colors of partitions of representations in \Fig{fig:SM-GUT-fig} to \Fig{fig:so10-GUT-fig}.
For Toy Model I without WZW term and without $w_2 w_3$ anomaly, we should remove the whiten quantum critical region, but we are left with a quantum critical point at the origin.
For Toy Model II with WZW term and with $w_2 w_3$ anomaly, 
we encounter the whiten quantum critical region near the origin.
The quantum critical region can have dynamical consequences such as emergent deconfined dark gauge force {$[\U(1)']^{\text{emergent}}_{\text{gauge}}$},
see \Sec{sec:QED4Parton}.
}
\label{fig:phase-schematic}
\end{figure}

In the remaining part of Section \ref{sec:IntroductionandSummary}, we start from an overview on the basic required ingredients of SM and GUT in \Sec{sec:SM-GUT}.
The outline of this article is given in the table of Contents.

\newpage

\subsection{Various Standard Models and Grand Unifications as Effective Field Theories}
\label{sec:SM-GUT}
Unification, as a central theme in the modern fundamental physics, 
is a theoretical framework aiming to embody the ``elementary'' excitations and forces
into a common origin.
Assuming 
without any significant dynamical gravity effect at the subatomic scale
(i.e., we are only limited to probe the underlying quantum theory
by placing the quantum systems on any curved spacetime geometry, but without significant gravity back-reactions), 
the quantum field theory (QFT) provides a suitable framework for such a unification.
Furthermore, assuming that we look at the QFT description valid below a certain energy scale (thus we are ignorant above that energy scale),
we shall also implement the effective field theory (EFT) perspective.

In fact, from the EFT perspective, we should remind ourselves the ``elementary'' excitations are only 
``elementary'' respect to a given EFT quantum vacuum. Moving away from the EFT vacuum (by tuning appropriate physical parameters) to a new quantum vacuum,
we shall see that the ``elementary'' excitations of the new vacuum may be drastically different from the original ``elementary''  excitations of the previous EFT.
So the ``elementary'' excitations reveal the \emph{limitations} of our EFT descriptions of quantum vacua.\footnote{Prominent 
examples occur in various systems with the duality descriptions and the order/disorder operators, 
such as in the Ising model and Majorana fermion system in 1+1d.}
Several examples of such 3+1d QFT and EFT paradigms for high energy physics (HEP) 
include Standard Model ({SM}) and
Grand Unification (Grand Unified Theory or GUT)
 \cite{Glashow1961trPartialSymmetriesofWeakInteractions, Salam1964ryElectromagneticWeakInteractions, Salam1968, Weinberg1967tqSMAModelofLeptons,
 Georgi1974syUnityofAllElementaryParticleForces, Pati1974yyPatiSalamLeptonNumberastheFourthColor,
Fritzsch1974nnMinkowskiUnifiedInteractionsofLeptonsandHadrons}:
\begin{enumerate}
\item \emph{Standard Model} ({SM}) :
Glashow-Salam-Weinberg (GSW) \cite{Glashow1961trPartialSymmetriesofWeakInteractions, Salam1964ryElectromagneticWeakInteractions, Salam1968, Weinberg1967tqSMAModelofLeptons} 
proposed the electroweak theory of the unified electromagnetic and weak forces 
between elementary particles.
The GSW theory together with the strong force \cite{Gross1973id, Politzer1973fx} becomes the Standard Model (SM), 
which is essential to describe the subatomic particle physics. 
The SM gauge group can be 
$$G_{\SM_q} \equiv \frac{{\SU(3)}_c \times {\SU(2)}_{\rm L} \times \U(1)_{\tilde{Y}}}{\Z_q}$$
with the mod $q=1,2,3,6$ so far undetermined by the current experiments (see an overview \cite{Tong2017oea1705.01853, Wan2019sooWWZHAHSII1912.13504}
on this global structure of SM Lie group issue).
The subscript $c$ is for color, the L is for the internal SU(2) (L for internal symmetry and its spinor)
locked with the left-handed Weyl fermion ($L$ for spacetime symmetry and its spinor) in the standard HEP convention, and ${\tilde{Y}}$ for electroweak hypercharge. 
The ``elementary'' particle excitations of this SM EFT, with 15n or 16n Weyl fermions, are constrained by the representation of 
$su(3)\times su(2)\times u(1)$ as (see \Fig{fig:SM-GUT-fig}):\footnote{Here we use the integer quantized ${\U(1)_{\tilde{Y}}}$.
If we use the phenomenology hypercharge $\U(1)_{{Y}}$ which is 1/6 of $\U(1)_{\tilde{Y}}$, namely 
$q_{\U(1)_{{Y}}}=\frac{1}{6} q_{\U(1)_{\tilde{Y}}}$, to write \eq{eq:SMrep}, then we have instead:
$${(\overline{\bf 3},{\bf 1})_{\frac{1}{3},L} \oplus ({\bf 1},{\bf 2})_{-\frac{1}{2},L}  
\oplus
 ({\bf 3},{\bf 2})_{\frac{1}{6},L} \oplus (\overline{\bf 3},{\bf 1})_{-\frac{2}{3},L} \oplus ({\bf 1},{\bf 1})_{1,L} \oplus {({\bf 1},{\bf 1})_{0,L}}}.$$} 
\bea \label{eq:SMrep}
{(\overline{\bf 3},{\bf 1})_{2,L} \oplus ({\bf 1},{\bf 2})_{-3,L}  
\oplus
 ({\bf 3},{\bf 2})_{1,L} \oplus (\overline{\bf 3},{\bf 1})_{-4,L} \oplus ({\bf 1},{\bf 1})_{6,L} \oplus {({\bf 1},{\bf 1})_{0,L}}}
.
\eea
The 16th Weyl fermion ${({\bf 1},{\bf 1})_{0,L}}$ is an extra sterile neutrino, sterile to the SM gauge force, also called the 
right-handed neutrino.
We will focus on the 16n Weyl fermion model in this present work.\footnote{In our present work, we shall focus on the SM or GUT with 16n Weyl fermions.\\
In contrast, 
Ref.~\cite{JW2006.16996, JW2008.06499, JW2012.15860} considers the SM or GUT 
with 15n Weyl fermions and with a discrete variant of baryon minus lepton number ${\bf B}-{\bf L}$ symmetry preserved. 
Ref.~\cite{JW2006.16996, JW2008.06499, JW2012.15860} then suggests that the missing 16th Weyl fermions can be substituted 
by additional 4d or 5d gapped topological quantum field theories (TQFTs), or by 4d gapless interacting conformal field theories (CFTs) to saturate a certain $\Z_{16}$ global anomaly.
On the other hand, our present work does \emph{not} introduce these $\Z_{16}$-class anomalous sectors, 
because we already have implemented 
the 16n Weyl fermion models that already make the $\Z_{16}$ global anomaly fully cancelled.
\label{footnote:Z16-anomaly}
} In our convention, we write Weyl fermions in the left-handed ($L$) basis which means that each is a 2-component ${\bf 2}_L$ spinor 
of the spacetime symmetry group Spin(1,3).

\item \emph{The $su(5)$ Grand Unification} ({$su(5)$ GUT}):
Georgi-Glashow (GG)
\cite{Georgi1974syUnityofAllElementaryParticleForces}
hypothesized that at a higher energy, 
the three SM gauge interactions 
merged into a single electronuclear force
under a simple Lie algebra $su(5)$, or precisely a Lie group 
$$
G_{\GG}\equiv  \SU(5)
$$ 
gauge theory. 
The {su(5) GUT} works for 15n Weyl fermions, also for 16n Weyl fermions (i.e., 15 or 16 Weyl fermions per generation).
The ``elementary'' particle excitations of this SU(5) EFT, with 15n or 16n Weyl fermions, are constrained by the representation of SU(5) as (see \Fig{fig:SU5-GUT-fig}):
\bea
\overline{\bf 5} \oplus {\bf 10} \oplus {\bf 1},
\eea
again written all in the left-handed ($L$) Weyl basis.
The 16th Weyl fermion is an extra sterile neutrino, sterile to the SU(5) gauge force, also called the right-handed neutrino. 
\item \emph{The Pati-Salam model} ({PS model}):
Pati-Salam (PS) \cite{Pati1974yyPatiSalamLeptonNumberastheFourthColor}
hypothesized that the lepton forms the fourth color, extending SU(3) to SU(4). 
The PS also puts the left $\SU(2)_{\rL}$ and a hypothetical right $\SU(2)_{\rR}$ on equal footing.
The PS gauge Lie algebra is
$su(4) \times su(2)_{\rL} \times su(2)_{\rR}$, 
and the PS gauge Lie group is 
$$G_{\PS_{q'}}\equiv\frac{\SU(4)_c\times(\SU(2)_\rL\times \SU(2)_\rR)}{\mathbb{Z}_{q'}}
= \frac{\Spin(6) \times \Spin(4)}{\mathbb{Z}_{q'}}$$ with the mod $q'=1,2$ depending on the global structure of Lie group.
The ``elementary'' particle excitations of this PS EFT, with 16n Weyl fermions, are constrained by the representation of 
$G_{\PS_{q'}}$as (see \Fig{fig:PS-GUT-fig}):
\bea
({\bf 4}, {\bf 2}, {\bf 1}) \oplus (\overline{\bf 4}, {\bf 1}, {\bf 2}),
\eea
written all in the left-handed ($L$) Weyl basis.\footnote{To be clear, 
we have the Weyl spacetime spinor ${\bf 2}_L$ of Spin(1,3) for $({\bf 4}, {\bf 2}, {\bf 1}) \oplus (\overline{\bf 4}, {\bf 1}, {\bf 2})$ of $su(4) \times su(2)_{\rL} \times su(2)_{\rR}$.
In contrast, we can also write the:
$$
\text{{${\bf 2}_L$ of Spin(1,3) for $({\bf 4}, {\bf 2}, {\bf 1})$ of $su(4) \times su(2)_{\rL} \times su(2)_{\rR}$,}\quad\quad 
{${\bf 2}_R$ of Spin(1,3) for $({\bf 4}, {\bf 1}, {\bf 2})$ of $su(4) \times su(2)_{\rL} \times su(2)_{\rR}$,}}
$$
then the representations of spacetime spinor $L$ (or $R$) would lock exactly with the internal spinor L (or R).\\
Here we use the $L$ and $R$ to specify the left/right-handed spacetime spinor of Spin(1,3).
We use the L and R to specify the left or right internal spinor representation of $su(2)_{\rL} \times su(2)_{\rR}$.
\label{footnote:LR}
}

\item \emph{The $so(10)$ Grand Unification} ({$so(10)$ GUT}):
Georgi and Fritzsch-Minkowski
\cite{Fritzsch1974nnMinkowskiUnifiedInteractionsofLeptonsandHadrons}
hypothesized that quarks and leptons become the 16-dimensional spinor representation 
\bea
\text{${\bf 16}^+$ of $G_{so(10)} \equiv \Spin(10)$ gauge group} 
\eea (with a local Lie algebra $so(10)$). 
Thus, the 16n Weyl fermions can interact via the Spin(10) gauge fields at a higher energy.
In this case, the 16th Weyl fermion, previously a sterile neutrino to the SU(5),
is \emph{no longer sterile} to the Spin(10) gauge fields; it also carries a charge 1, thus not sterile, under the gauged center subgroup $Z(\Spin(10))=\Z_4$.
\end{enumerate}

We relegate several tables of data relevant for SMs and GUTs into Appendix \ref{sec:SM-GUT-table}, for readers' convenience to check 
the quantum numbers of various elementary particles or field quanta of SMs and GUTs.

\newpage
\section{Standard Models from the competing phases of Grand Unifications}
\label{sec:SM-GUT-more}

In \Sec{sec:SM-GUT-more}, we start by enlisting and explaining some group embedding structures from some of relevant GUTs to SM in \Sec{sec:SM-GUT-group-branch}.

\subsection{Spacetime-Internal Symmetry Group embedding of SMs and GUTs, and the $w_2w_3$ anomaly}
\label{sec:SM-GUT-group-branch}

Here we use the \emph{inclusion} notation $G_{\text{large}} \hookleftarrow G_{\text{small}}$ to imply that: 
\begin{itemize}
\item $G_{\text{large}} \supset G_{\text{small}}$, namely
the $G_{\text{large}}$ contains $G_{\text{small}}$ as a subgroup, or equivalently $G_{\text{small}}$ can be embedded in $G_{\text{large}}$.
\item $G_{\text{large}}$ can be broken to $G_{\text{small}}$ via \emph{symmetry breaking} of Higgs condensation (which we will explore).
\end{itemize}
The \emph{internal symmetry} group embedding structure has been explored, for example summarized in \cite{BaezHuerta0904.1556}:
\bea 
\xymatrix{
G_{\SM_6} \equiv \frac{{\SU(3)}_c \times {\SU(2)}_{\rm L} \times \U(1)_Y}{\Z_6} \ar@{^{(}->}[r] \ar@{^{(}->}[d] & 
{G}_{\GG} \equiv  \SU(5) \ar@{^{(}->}[d] \\
\displaystyle
G_{\PS_2} \equiv\frac{\Spin(6) \times \Spin(4)}{\Z_2}  \ar@{^{(}->}[r] & \Spin(10) \\
}
\eea
We further include both the complete \emph{spacetime-internal symmetry} group embedding structure as follows:
\bea
\bar{G} \equiv {{G_{\text{spacetime} }} \times_{{N_{\text{shared}}}}  {{G}_{\text{internal}} } }
\equiv  ({\frac{{G_{\text{spacetime} }} \times  {{G}_{\text{internal}} } }{{N_{\text{shared}}}}}).
\eea
\bea \label{eq:spacetime-internal-embedding-1}
\xymatrix{
{\bar{G}_{\SM_6} \equiv \Spin \times_{\Z_2^F} \Z_{4,X} \times \frac{{\SU(3)}_c \times {\SU(2)}_{\rm L} \times \U(1)_Y}{\Z_6} }\ar@{^{(}->}[r] \ar@{^{(}->}[d] &  
\bar{G}_{\GG} \equiv \Spin \times_{\Z_2^F} \Z_{4,X} \times\SU(5) \ar@{^{(}->}[d] \\
\displaystyle
{ \bar{G}_{\PS_2} \equiv\Spin \times_{\Z_2^F} \frac{\Spin(6) \times \Spin(4)}{\Z_2}}  \ar@{^{(}->}[r] & 
\bar{G}_{so(10)} \equiv  \Spin \times_{\Z_2^F}\Spin(10) \\
}.
\eea
Some comments about \eq{eq:spacetime-internal-embedding-1} follow:\\[-10mm]
\begin{enumerate}[leftmargin=.mm]
\item
The $\Spin$ means the spacetime rotational symmetry group
$\Spin \equiv \Spin(1,3)$ for 4d Lorentz signature (or $\Spin \equiv\Spin(4)$ for 4d Euclidean signature).
The $\Spin$  contains the fermionic parity $\Z_2^F$ at the center subgroup thus 
${\Spin}/{\Z_2^F}=\SO$ where the $\SO$ is the bosonic spacetime (special orthogonal) rotational symmetry group
(similarly, $\SO \equiv \SO(1,3)$ for 4d Lorentz signature, or $\SO \equiv\SO(4)$ for 4d Euclidean signature).
The notation $G_1 \times_{N_{\text{shared}}} G_2 \equiv \frac{G_1 \times G_2}{{N_{\text{shared}}}}$ means modding out their common normal subgroup 
${N_{\text{shared}}}$. So $ \Spin \times_{\Z_2^F} G \equiv \frac{\Spin \times G}{{\Z_2^F}}$ means modding out their common normal subgroup ${\Z_2^F}$.

\item The $\Z_{4,X}$ has the $X$-symmetry generator such that its square $(X)^2=(-1)^F$ is the fermion parity operator, so 
$\Z_{4,X} \supset \Z_2^F$. 
Wilczek-Zee \cite{Wilczek1979hcZee} 
firstly noticed that the $X \equiv 5({ \mathbf{B}-  \mathbf{L}})-4Y$, with the baryon minus lepton number ${ \mathbf{B}-  \mathbf{L}}$ and the electroweak hypercharge $Y$,
is a good global symmetry respected by SM and the $su(5)$ GUT. 
All known quarks and leptons carry a charge 1 of $\Z_{4,X}$, in the left-handed Weyl spinor basis.
The center of Spin(10) can be chosen exactly as $Z(\Spin(10))=\Z_{4,X}$.
We summarize how $\Z_{4,X}$ can be obtained in 
\Table{table:SMfermion} and \Table{table:PSfermionAll}. 
See more discussions on $\Z_{4,X}$ in
\cite{GarciaEtxebarriaMontero2018ajm1808.00009, WW2019fxh1910.14668,JW2006.16996, JW2008.06499, JW2012.15860}.
\item  The $(X)^2=(-1)^F$ relation is obeyed in the non-supersymmetric SM and GUT models, so it is natural to introduce the $\Spin \times_{\Z_2^F} \Z_{4,X}$ structure in \eq{eq:spacetime-internal-embedding-1}. However, it is possible to have new fermions, such as in supersymmetric SMs or GUTs, 
which does not necessarily obey $(X)^2=(-1)^F$ relation.
In that case, we can introduce just $\Spin \times_{} \Z_{4,X}$ structure.
See a footnote for the alternative symmetry embedding with the $\Spin \times_{} \Z_{4,X}$ structure.\footnote{Another version of the \emph{spacetime-internal symmetry} group embedding (that is more suitable for supersymmetric SMs or GUTs) is
\bea \label{eq:spacetime-internal-embedding-2}
\xymatrix{
{\bar{G}_{\SM_6} \equiv \Spin \times_{} \Z_{4,X} \times \frac{{\SU(3)}_c \times {\SU(2)}_{\rm L} \times \U(1)_Y}{\Z_6} }\ar@{^{(}->}[r] \ar@{^{(}->}[d] &  
\bar{G}_{\GG} \equiv \Spin \times_{} \Z_{4,X} \times\SU(5) \ar@{^{(}->}[d] \\
\displaystyle
{ \bar{G}_{\PS_2} \equiv\Spin \times_{} \frac{\Spin(6) \times \Spin(4)}{\Z_2}}  \ar@{^{(}->}[r] & 
\bar{G}_{so(10)} \equiv  \Spin \times_{}\Spin(10) \\
}.
\eea
}
\item
In this \eq{eq:spacetime-internal-embedding-1}, we keep a structure of $\Spin \times_{\Z_2^F} \Z_{4,X}$ which is essential to produce a mixed gauge-gravity nonperturbative global anomaly constraint of a $\Z_{16}$ class. 
As already mentioned in footnote \ref{footnote:Z16-anomaly}, in this article,
we keep the 16n Weyl fermions in all our SM and GUT models, thus the $\Z_{16}$ global anomaly is already cancelled by 16n chiral fermions.

\item In this \eq{eq:spacetime-internal-embedding-1}, we also keep a structure of $ \Spin \times_{\Z_2^F}\Spin(10)$
---
the cobordism group
$\Omega^{d}_{G} \equiv \TP_d(G)$  shows \cite{WangWen2018cai1809.11171, WanWang2018bns1812.11967}
\bea \label{eq:Spin10-anomaly}
\text{$\TP_5(\Spin \times_{\Z_2^F}\Spin(10))=\Z_2$, \quad
but $\TP_5(\Spin \times_{}\Spin(10))=0$.}
\eea
This implies only the $\Spin \times_{\Z_2^F}\Spin(10)$ structure
offers a possible $\Z_2$ class global anomaly in 4d that is captured by a 5d invertible TQFT 
with a partition function on a 5d manifold $M^5$:\footnote{The invertible TQFT means that the TQFT path integral or partition function
${\bf Z}(M)$ on any closed manifold $M$ has its absolute value $|{\bf Z}(M)|=1$. Thus the dimension of its Hilbert space is always 1 also any closed spatial manifold, there is no topological ground state degeneracy. Here ${\bf Z}(M^5)=(-1)^{\int w_2w_3}=\pm 1$ on any closed $M^5$ thus it is an invertible TQFT, such that when $M^5$ is a Dold manifold $\CP^2 \rtimes S^1$ or
a Wu manifold $\SU(3)/\SO(3)$ generating a ${\bf Z}(M^5)=-1$ \cite{WangWenWitten2018qoy1810.00844, WanWang2018bns1812.11967}.\\
Here the
$\Spin \times_{\Z_2^F}\Spin(10)$ structure imposes the spacetime and gauge bundle constraint
\bea
w_2(TM)=w_2(V_G)
\eea 
with $G=\Spin(10)/{\Z_2^F}=\SO(10)$. Moreover,
the Steenrod square $\Sq^1$ is an operation sending the second cohomology to the third cohomology class: $\H^2$ to $\H^3$,
which we can regard $\Sq^1 = \frac{1}{2} \delta$ with $\delta$ as a coboundary operator (see for example \cite{WanWang2018bns1812.11967}). 
Then, in the case $G=\SO(10)$, we can deduce another bundle constraint:
\bea
w_3(TM)+w_1(TM) w_2(TM)= \Sq^1 w_2(TM)= \Sq^1 w_2(V_G)= w_3(V_G). 
\eea 
On the orientable spacetime, the first Stiefel-Whitney class $w_1(TM)=0$, so
$$
w_3(TM)=w_3(V_G).
$$
Thus combining the above formulas, on the orientable $\Spin \times_{\Z_2^F}\Spin(10)$ structure, 
we derive that $w_2(TM)w_3(TM)=w_2(V_G)w_3(V_G)$ in \eq{eq:w2w3}, shorthand as $w_2w_3=w_2w_3(TM)=w_2w_3(V_G)$.
This derivation also works for other $G=\Spin(n)/{\Z_2^F}=\SO(n)$ for $n \geq 3$.
} 
\bea \label{eq:w2w3}
{\bf Z}(M^5)=(-1)^{\int_{M^5} w_2(TM) w_3(TM)}=(-1)^{\int_{M^5}  w_2(V_{\SO(10)}) w_3(V_{\SO(10)})}.
\eea
But this mod 2 anomaly is \emph{absent} and \emph{not} allowed on the $\Spin \times_{}\Spin(10)$ structure.
The difference between $\Spin \times_{\Z_2^F}\Spin(10)$ and $\Spin \times_{}\Spin(10)$
is the following: the fermion charge under $(-1)^F$ thus odd under ${\Z_2^F}$
must be in the $\Z_2$ normal subgroup of the center subgroup $Z(\Spin(10))=\Z_{4,X}$ so $(X)^2=(-1)^F$ in order to 
impose the spacetime-internal $\Spin \times_{\Z_2^F}\Spin(10)$ structure.
However, in contrast, the $\Spin \times_{}\Spin(10)$ allows other fermions to not obey the $(X)^2=(-1)^F$ relation.

As mentioned in \Refe{WangWen2018cai1809.11171, WangWenWitten2018qoy1810.00844} and footnote \ref{footnote:w2w3anomaly}, as $\Spin(10)  \supset \Spin(3) = \SU(2)$, so
\bea
\Spin \times_{\Z_2^F}\Spin(10) \supset \Spin \times_{\Z_2^F}\Spin(3) = \Spin \times_{\Z_2^F}\SU(2).
\eea
The $\Spin \times_{\Z_2^F}\Spin(10)$-structure is tightly related to the $\Spin \times_{\Z_2^F}\SU(2)$ also known as the $\Spin^h$-structure.
We can project the $\Spin \times_{\Z_2^F}\Spin(10)$-structure to the $\Spin^h$-structure. 
Then, in the $\Spin^h$-structure, 
because the fermionic wavefunction gains a $(-1)$ statistical sign under a $2\pi$ self rotation on a Spin manifold 
is identified with the $(-1)^F$ as the center $Z(\SU(2))=\Z_2^F$,
we can read that imposing the $\Spin^h$-structure \cite{WangWen2018cai1809.11171, WangWenWitten2018qoy1810.00844}:\\ 
$\bullet$ the fermions must be in the half-integer isospin representation 1/2, 3/2, $\dots$, etc. of SU(2) 
(namely, the even-dimensional representations ${\bf 2}, {\bf 4}, \dots,$ etc. of SU(2)).\\
$\bullet$  the bosons must be in the integer isospin representation 0, 1, 2, $\dots$, etc. of SU(2) 
(namely, the odd-dimensional representations ${\bf 0}, {\bf 1}, {\bf 3}, \dots,$ etc. of SU(2)).
\item The last but the most important comment above all,
is that in order to realize a possible continuous deconfined quantum phase transition,
we do require to use the $w_2w_3$ anomaly in \eq{eq:w2w3}, such that this anomaly occurs in
the phase transition between the GG and PS models in \Fig{fig:phase-schematic}.
So we do aim to impose the $\Spin \times_{\Z_2^F}\Spin(10)$-structure 
as in \eq{eq:spacetime-internal-embedding-1} in order to implement the $w_2w_3$ anomaly.  
In short, the readers can ask:%
\begin{multline}
\text{\emph{Why do we need the $w_2w_3$ anomaly near the criticality for establishing}}\cr
\text{\emph{a possible continuous quantum phase transition between the GG and PS models?}}\nn
\end{multline}
The answer is that:\\
$\bullet$ The GG and PS models are Landau-Ginzburg symmetry breaking type of phases (when we treat the internal symmetry as global symmetry) or the
gauge-symmetry breaking type of phases (when we treat the internal symmetry group as gauge group). The $w_2w_3$ anomaly is matched on two sides of
phases by GG and PS models via symmetry breaking. (In fact, no $w_2 w_3$ anomaly is allowed in GG and PS models.)\\
$\bullet$ But the $w_2 w_3$ anomaly can protect a gapless quantum phase transition (or a gapless intermediate quantum critical region) 
between the GG and PS models 
when the Spin(10) symmetry is restored at their phase transition.
Their phase transition can be protected to be Spin(10)-symmetry-preserving gapless due to the $w_2 w_3$ anomaly exists only in the enlarged Spin(10) internal symmetry group.

Because the conventional $so(10)$ GUT is free from the $w_2 w_3$ anomaly \cite{WangWen2018cai1809.11171, WangWenWitten2018qoy1810.00844},
we will need to explicitly introduce a new WZW-like term 
built out of GUT-Higgs field in the mother EFT, which allows the GUT-Higgs sector (beyond the SM sector) to saturate the $w_2 w_3$ anomaly. To this end, we will start from writing down a GUT-Higgs model in the context of $so(10)$ GUT, and then trying to modifying the GUT-Higgs model to
saturate the $w_2 w_3$ anomaly. (That mother EFT will be the main achievement later in \Sec{sec:EFTHiggs}.)

\end{enumerate}

\subsection{Branching Rule of SMs and GUTs, and a GUT-Higgs model}
\label{sec:SM-GUT-branch}

In the following, we motivate the GUT model with GUT-Higgs as the gauge symmetry breaking pattern to go to the lower energy EFT (such as SM).
Most of these  breaking patterns are well-established and overviewed in \cite{georgi2018liebook}. 
The additional new input is that we try to unify several models into a GUT-Higgs model with as minimum amount of GUT-Higgs as possible.
In Appendix \ref{sec:app-Branching-GUT-Higgs}, we try to go through the logic again, 
and carefully examine the consequences and possibilities of the types of required GUT-Higgs. 
Later we will motivate the possible Lagrangian of the GUT-Higgs potential.

\begin{figure}[!h] 
\centering
\hspace{-1.24cm}
\includegraphics[width=1.07\textwidth]{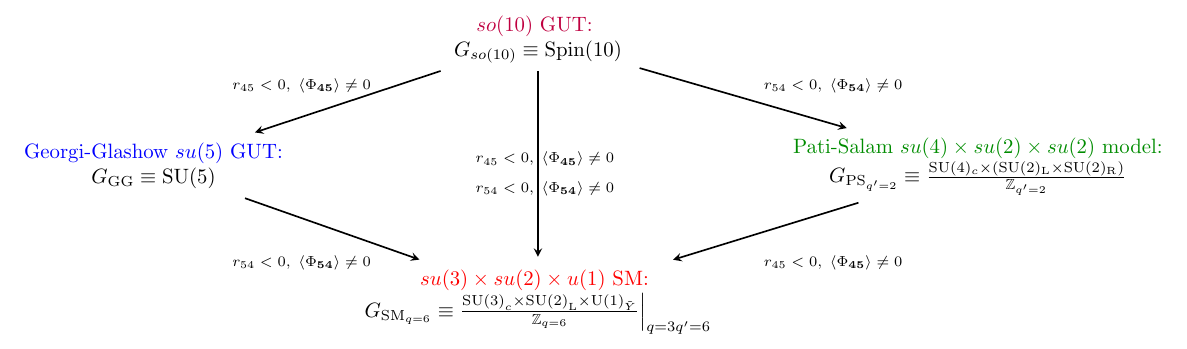}
\caption{Beware that the direction of the group symmetry breaking ``$\rightarrow$'' is the opposite direction to the group inclusion ``$\hookleftarrow$.''
(These colors are also designed to match the colors in \Fig{fig:SM-GUT-fig} to \Fig{fig:so10-GUT-fig}, and \Fig{fig:phase-schematic}).
}
\label{fig:GEQCembedGUTHiggs1}
\end{figure}

Here we summarize what we need from the analysis done in Appendix \ref{sec:app-Branching-GUT-Higgs}:
\begin{itemize}[leftmargin=.mm]
\item 
We can use
a Lorentz scalar boson with a 45-dimensional real representation of $so(10)$ or Spin(10):
\bea \label{eq:phi45}
\Phi_{so(10), \bf 45} \equiv
\Phi_{\bf 45}  \in \R.
\eea
to break the $\Spin(10)$ of $so(10)$ GUT to the $\SU(5)$ of GG model, 
also we can use this same $\Phi_{\bf 45}$ to break 
$G_{\PS_2}\equiv\frac{\Spin(6) \times \Spin(4)}{\Z_2}$ of PS model to the 
$G_{\SM_6}\equiv \frac{{\SU(3)}_c \times {\SU(2)}_{\rm L} \times \U(1)_Y}{\Z_6}$ of the SM.
\item
We can use
a Lorentz scalar boson with a 54-dimensional real representation of $so(10)$ or Spin(10):
\bea \label{eq:phi54}
\Phi_{so(10), \bf 54}\equiv
\Phi_{\bf 54}  \in \R,
\eea
to break the $\Spin(10)$ of $so(10)$ GUT to the 
$G_{\PS_2}\equiv\frac{\Spin(6) \times \Spin(4)}{\Z_2}$ of PS model, 
also we can use this same $\Phi_{\bf 54}$ to break 
$\SU(5)$ of GG model to the $G_{\SM_6}\equiv \frac{{\SU(3)}_c \times {\SU(2)}_{\rm L} \times \U(1)_Y}{\Z_6}$ of the SM.
\item The combinations of the two facts above is summarized in \Fig{fig:GEQCembedGUTHiggs1},
where we can use the $\Phi_{\bf 45}$ and $\Phi_{\bf 54}$ to write the GUT-Higgs model,
that can induce the qualitative phase diagram similar to \Fig{fig:phase-schematic}.
\end{itemize}
\begin{figure}[!h] 
\centering
\includegraphics[width=0.72\textwidth]{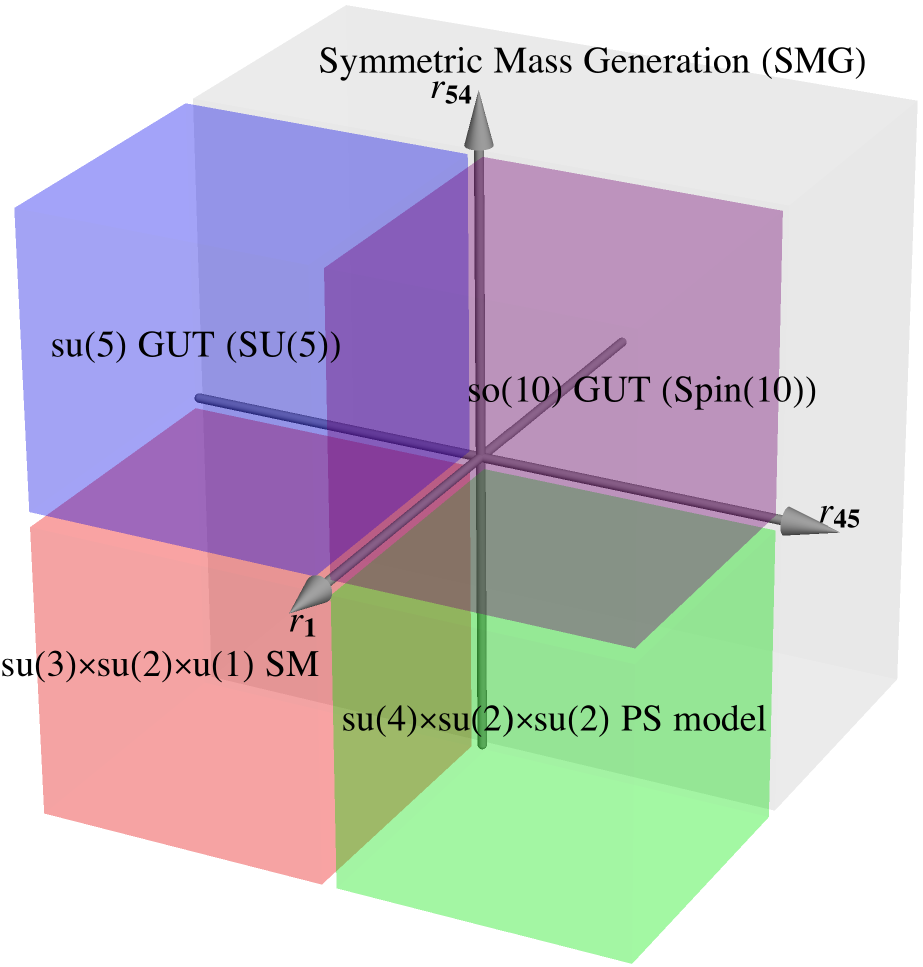} 
\caption{Schematic quantum phase diagram interpolating between the $so(10)$ GUT (Spin(10) group), 
the Georgi-Glashow $su(5)$ GUT (SU(5) group), 
the $su(4) \times su(2)_{\mathrm{L}} \times su(2)_{\mathrm{R}}$ Pati-Salam model (PS), and 
the $su(3)\times su(2)\times u(1)$ Standard Model (SM), and the symmetric mass generation (SMG). 
Here the real parameter $r_{\mathbf{R}} \in \R$ denotes the coefficient of the effective quadratic potential of $\Phi$ field in the representation ${\mathbf{R}}$. 
The corresponding GUT-Higgs $\Phi$ field will condense in the representation-${\mathbf{R}}$ if $r_{\mathbf{R}}<0$.
Relatively speaking, the infrared (IR) low energy is drawn with the \ccred{red} color (for SM), 
the intermediate neighbor phases are drawn with the \ccgreen{green} or \ccblue{blue} color (for PS or SU(5) models),
while the ultraviolet (UV) higher energy is drawn with the \ccpurple{violet purple} color (for Spin(10)).
These colors are also designed to match the colors of partitions of representations in \Fig{fig:SM-GUT-fig} to \Fig{fig:so10-GUT-fig}.
}
\label{fig:phase}
\end{figure}
Given the $so(10)$ GUT, to induce the three other models in \Fig{fig:GEQCembedGUTHiggs1},
we can add the GUT-Higgs potential ${\rm U}(\Phi_{{\mathbf{R}}})$ with $\Phi_{{\mathbf{R}}}$ of some representation ${\mathbf{R}}$.
The ${\rm U}(\Phi_{{\mathbf{R}}})$ is chosen to have positive $\Phi^4$ coefficients (thus $\lambda_{{\mathbf{45}}}, \lambda_{{\mathbf{54}}} >0$), 
while the $r_{{\mathbf{45}}}$ and $r_{{\mathbf{54}}}$
are real-number tunable parameters shown in \Fig{fig:phase-schematic} and \Fig{fig:phase}:
\bea \label{eq:UPhiR}
{\rm U}(\Phi_{{\mathbf{R}}}) =\Big(r_{{\mathbf{45}}} (\Phi_{{\mathbf{45}}})^2 +\lambda_{{\mathbf{45}}} (\Phi_{{\mathbf{45}}})^4\Big)+ 
\Big(r_{{\mathbf{54}}} (\Phi_{{\mathbf{54}}})^2 +\lambda_{{\mathbf{54}}} (\Phi_{{\mathbf{54}}})^4\Big).
\eea
A slice of \Fig{fig:phase} becomes the \Fig{fig:phase-schematic}. (Temporarily now we get rid of the GUT-Higgs $\Phi_{\mathbf{1}}$ thus get rid of $r_{{\mathbf{1}}}$ axis in  \Fig{fig:phase}. 
More on this $\Phi_{\mathbf{1}}$ later.)
We can use this ${\rm U}(\Phi_{{\mathbf{R}}})$ potential in \eq{eq:UPhiR}
to induce these interior parts of four phases (the $so(10)$ GUT, the $su(5)$ GUT, the PS model, and the SM).
\begin{itemize}
\item
If $\langle\Phi_{\mathbf{45}}\rangle$ condenses, namely if $r_{{\mathbf{45}}}<0$ so $\langle\Phi_{\mathbf{45}}\rangle\neq 0$, 
then the $so(10)$ GUT becomes Higgs down to the $su(5)$ GUT. 
\item
If $\langle\Phi_{\mathbf{54}}\rangle$ condenses, namely if $r_{{\mathbf{54}}}<0$ so $\langle\Phi_{\mathbf{54}}\rangle\neq 0$, 
then the $so(10)$ GUT becomes Higgs down to the PS model. 
\item
If $\langle\Phi_{\mathbf{45}}\rangle$ and $\langle\Phi_{\mathbf{54}}\rangle$ both condense, 
namely if $r_{{\mathbf{45}}}<0$ and $r_{{\mathbf{54}}}<0$ so that 
$\langle\Phi_{\mathbf{45}}\rangle\neq 0$ and $\langle\Phi_{\mathbf{54}}\rangle\neq 0$. 
The theory becomes Higgs down to the SM. 
\end{itemize}
All these above Higgs condensations induce {\bf \emph{continuous phase transitions}}.

The purpose of the next Section \ref{sec:EFTHiggs} is to design various EFT and to
explore the possible phase structures and phase transitions (of \Fig{fig:phase-schematic} and \Fig{fig:phase}).
In particular, we will write down a mother EFT such that it saturates the $w_2 w_3$ global anomaly
and it realizes an excotic quantum phase transition between the GG $su(5)$ GUT and the PS model. 

\newpage
\section{Mother Effective Field Theory with Competing GUT-Higgs fields}
\label{sec:EFTHiggs}

\subsection{Elementary GUT-Higgs model induces the SM}
\label{sec:ElementaryHiggs}

In Section \ref{sec:SM-GUT-more} (especially \Sec{sec:SM-GUT-branch}), we write down a GUT-Higgs potential ${\rm U}(\Phi_{{\mathbf{R}}})$ in \eq{eq:UPhiR} appending to the $so(10)$ GUT with 16n complex Weyl fermions $\psi_L$. 
Let us write down the full path integral $\bZ_{\GUT}$ of such $so(10)$ GUT plus ${\rm U}(\Phi_{{\mathbf{R}}})$, in a
Lorentzian signature, evaluated on a 4-manifold $M^4$:
\bea \label{eq:Zso10}
\bZ_{\GUT}\equiv
\int [{\cal D} {\psi_L}] [{\cal D}{\psi_L^\dagger}] [{\cal D} A][{\cal D} \Phi_{{\mathbf{R}}} ] \dots
\exp( \ii \left. S_{\text{GUT}}[\psi_L, {\psi_L^\dagger}, A, \Phi_{{\mathbf{R}}}, \dots ] \right  \rvert_{M^4}).
\eea
The action $S_{\text{GUT}}$ is:
\begin{multline} \label{eq:Sso10}
S_{\text{GUT}}= \int_{M^4} \Big(\Tr(F_{} \wedge \star F_{})
-\frac{\theta}{8 \pi^2} g^2
 \Tr( F_{} \wedge F_{}) \Big) 
+ \int_{M^4} \Big( {\psi}^\dagger_L  (\ii {\bar  \sigma}^\mu {D}_{\mu, A} ) \psi_L
 \\
+ | {D}_{\mu, A}\Phi_{{\mathbf{R}}} |^2 -{\rm U}(\Phi_{{\mathbf{R}}})
 -(  (\Phi_{{\mathbf{R}}}) ({\psi}^\dagger_L\cdots)({\psi}_L\cdots) +{\rm h.c.} )
 +  \dots
  \Big)\,\dd^4 x.
\end{multline}
The $S_{\text{YM}}=\int \Tr(F_{} \wedge \star F_{})$ part is the Yang-Mills gauge theory,
with Lie algebra valued field strength curvature 2-form $F = \dd A - \ii { g} A \wedge A$.
Here $({\psi}^\dagger_L\cdots)$ and $({\psi}_L\cdots)$ imply
indefinite multiple numbers of Weyl fermion fields, so as to properly match the representation $\mathbf{R}$ of the Higgs field $\Phi_{\mathbf{R}}$.
For the $so(10)$ GUT, we have to sum over the Spin(10) gauge bundle,  whose 1-form connection is the spin-1 Lorentz vector and Spin(10) gauge field, 
written as 
\bea
 A = 
 ( \sum_{\ra=1}^{45} {{\rm T}^\ra}   A_{{\Spin(10)},\mu}^\ra
)\dd x^\mu.
\eea
There are 45 of such Lie algebra generators, ${{\rm T}^\ra}$, with:\\ 
$\bullet$ rank-16 matrix representations that act on the quark-and-lepton matter representation ${\bf 16}^+$ of Spin(10).\\
$\bullet$ rank-45 matrix representations that  act on the $\Phi_{\mathbf{45}}$ as the ${\bf 45}$ of Spin(10).\\
$\bullet$ rank-54 matrix representations that act on the $\Phi_{\mathbf{54}}$ as the ${\bf 54}$ of Spin(10).

Locally the Spin(10) Lie algebra is the same as the $so(10)$ Lie algebra,
but globally we really need to define the principal Spin(10) gauge bundle $P_A$ to sum over.
So more precisely the path integral over the gauge field measure really means 
$\int [{\cal D} A] \dots \equiv \sum_{\text{gauge bundle $P_A$}} \int [{\cal D} \tilde A] \dots$,
where $\tilde A$ are  gauge connections over each specific gauge bundle choice $P_A$.
The $\theta$ term, $\theta \Tr( F_{} \wedge F_{})$, can be added or removed depending on the model.
In this work, we shall set $\theta=0$ or close to zero.

The $\psi_L$ is a 2-component spin-1/2 Weyl fermion ${\bf 2}_L$ of Spin(1,3).
The $\dagger$ is the standard complex conjugate transpose.
The $\bar{\sigma}^\mu=(\sigma^0,-\sigma^1,-\sigma^2,-\sigma^3)$ and ${\sigma}^\mu=(\sigma^0,\sigma^1,\sigma^2,\sigma^3)$ are the standard spacetime spinor rotational $su(2)$ Lie algebra generators for $L$ and $R$ Weyl spinors.
The action $S_{\text{GUT}}$ also includes the Weyl spinor kinetic term and GUT-Higgs kinetic term, coupling to gauge fields
via the covariant derivative operator 
${D}_{\mu, A } \equiv \nabla_{\mu} - \ii g  \,  A_\mu $.
The $\nabla_{\mu}$ can contain the curve-spacetime covariant derivative data such as Christoffel symbols or the spinor's spin-connection if needed.
The $\dots$ are possible extra deformation terms to be added later.

This subsection \Sec{sec:ElementaryHiggs} mostly treats the spin-0 Lorentz scalar Higgs field $\Phi_{\bf R}$ with some representation 
${\bf R}$ as the elementary Higgs field.
We will however \emph{fractionalize} this elementary Higgs field $\Phi_{\bf R}$ to other further elementary fermionic fields in the 
later \Sec{sec:CompositeHiggs} and \Sec{sec:FragmentaryHiggs}.

\subsubsection{Model I: Without Wess-Zumino-Witten term, and Symmetric Mass Generation}
\label{sec:ElementaryHiggs-noWZW}

Follow the choice in \Sec{sec:SM-GUT-branch} and in \eq{eq:UPhiR}, we can further adjust it to
\bea \label{eq:GUTHiggsU}
{\rm U}(\Phi_{{\mathbf{R}}}) =\Big(r_{{\mathbf{45}}} (\Phi_{{\mathbf{45}}})^2 +\lambda_{{\mathbf{45}}} (\Phi_{{\mathbf{45}}})^4\Big)+ 
\Big(r_{{\mathbf{54}}} (\Phi_{{\mathbf{54}}})^2 +\lambda_{{\mathbf{54}}} (\Phi_{{\mathbf{54}}})^4\Big)+
\Big(r_{{\mathbf{1}}} (\Phi_{{\mathbf{1}}})^2 +\lambda_{{\mathbf{1}}} (\Phi_{{\mathbf{1}}})^4\Big).
\eea
The property (whether
 $\langle\Phi_{\mathbf{45}}\rangle \neq 0$ or $\langle\Phi_{\mathbf{54}}\rangle \neq 0$ condenses, or both condense, 
namely whether $r_{{\mathbf{54}}}<0$ or $r_{{\mathbf{54}}}<0$) still follows \Sec{sec:SM-GUT-branch}.
The theory becomes Higgs down to the $su(5)$ GUT, or the PS model, or the SM, see \Fig{fig:GEQCembedGUTHiggs1}.
Here are some extra comments for adding $\Phi_{\mathbf{1}}$ or other $\Phi_{\mathbf{R}}$ terms to \Fig{fig:phase}:
\begin{itemize}
\item We can introduce
a Lorentz scalar boson with a 1-dimensional trivial but real representation of $so(10)$ or Spin(10):
\bea \label{eq:phi1}
\Phi_{so(10), \bf 1} \equiv
\Phi_{\bf 1}  \in \R.
\eea
\begin{itemize}
\item
If $\langle\Phi_{\mathbf{1}}\rangle =0$ does not condense, namely if $r_{{\mathbf{1}}}>0$, the theory remains in the $so(10)$ GUT. 
\item
If $\langle\Phi_{\mathbf{1}}\rangle \neq 0$ condenses, namely if $r_{{\mathbf{1}}}<0$, 
for a small $\langle\Phi_{\mathbf{1}}\rangle<\Phi_{{\mathbf{1}},c}$, the theory still remains in the $so(10)$ GUT (as $\langle\Phi_{\mathbf{1}}\rangle$ is an irrelevant perturbation). 
\item
However, not only $\langle\Phi_{\mathbf{1}}\rangle \neq 0$ condenses,
but when $\langle\Phi_{\mathbf{1}}\rangle > \Phi_{{\mathbf{1}},c}$ exceeds a critical value, it can drive to the Symmetric Mass Generation (SMG) phase 
and {gap out all fermions} while preserving the $G$-symmetry (if the theory is free from all 't Hooft anomalies in $G$).\footnote{The Symmetric Mass Generation (SMG) mechanism 
is explored in various references, for some selective examples,
by Fidkowski-Kitaev \cite{FidkowskifSPT2} in 0+1d, 
by Wang-Wen \cite{Wang2013ytaJW1307.7480,Wang2018ugfJW1807.05998} for gapping chiral fermions in 1+1d,
You-He-Xu-Vishwanath \cite{YouHeXuVishwanath1705.09313, YouHeVishwanathXu1711.00863} in 2+1d,
and notable examples in 3+1d by Eichten-Preskill \cite{Eichten1985ftPreskill1986},
Wen \cite{Wen2013ppa1305.1045}, 
You-BenTov-Xu \cite{You2014oaaYouBenTovXu1402.4151, YX14124784}, 
BenTov-Zee \cite{BenTov2015graZee1505.04312},
Kikukawa \cite{Kikukawa2017ngf1710.11618}, 
Wang-Wen \cite{WangWen2018cai1809.11171},
Catterall et al \cite{Catterall2020fep, CatterallTogaButt2101.01026},
Razamat-Tong \cite{RazamatTong2009.05037, Tong2104.03997}, etc.}
 \end{itemize}
 
How do we associate $\langle\Phi_{\mathbf{1}} \rangle > \Phi_{{\mathbf{1}},c}$ with the SMG effect?
First notice that the four of the spinor representation ${\bf 16}^+$ of Spin(10) can produce the tensor product decomposition 
\cite{1912.10969LieART}
\bea \label{eq:16161616branch}
&&{\bf 16} \otimes {\bf 16} \otimes {\bf 16} \otimes {\bf 16}
=
({\bf 10}_{} \oplus {\bf 120}_{}  \oplus \overline{\bf 126}_{} ) \otimes ({\bf 10}_{} \oplus {\bf 120}_{}  \oplus \overline{\bf 126}_{} )\cr
&&=({\bf 10} \otimes {\bf 10}) \oplus ({\bf 120} \otimes {\bf 120})  \oplus (\overline{\bf 126} \otimes \overline{\bf 126})
\oplus 2 ({\bf 10} \otimes {\bf 120})  \oplus 2 ({\bf 10} \otimes \overline{\bf 126}) \oplus 2 (  {\bf 120} \otimes \overline{\bf 126})   \cr
&&=({\bf 1}  \oplus {\bf 45} \oplus {\bf 54}) \oplus ({\bf 1}  \oplus {\bf 45} \oplus {\bf 54} \oplus 2 ({\bf 210}) \oplus {\bf 770}  \oplus {\bf 945}
\oplus {\bf 1050} \oplus \overline{\bf 1050} \oplus {\bf 4125} \oplus {\bf 5940}  )\cr
&& \;\quad \oplus ({\bf 54 \oplus 945 \oplus \overline{1050} \oplus \overline{2772} \oplus 4125 \oplus \overline{6930}})
\oplus 2 (   {\bf 45} \oplus {\bf 210} \oplus {\bf 945})
\oplus 2( {\bf 210} \oplus \overline{\bf 1050})\cr
&& \;\quad\oplus 2( {\bf 45} \oplus {\bf 210} \oplus {\bf 945} \oplus \overline{\bf 1050} \oplus {\bf 5940} \oplus \overline{\bf 6930})
\eea
More systematically, with the symmetric (S) or anti-symmetric (A) matrix representation subscript indicated on the right hand side:
\bea
{\bf 16} \otimes {\bf 16} &=& {\bf 10}_{\rm S} \oplus {\bf 120}_{\rm A}  \oplus \overline{\bf 126}_{\rm S}.\cr 
{\bf 10} \otimes {\bf 10} &=& {\bf 1}_{\rm S}  \oplus {\bf 45}_{\rm A} \oplus {\bf 54}_{\rm S}.\cr
{\bf 120} \otimes {\bf 120} &=& {\bf 1}_{\rm S}  \oplus {\bf 45}_{\rm A} \oplus {\bf 54}_{\rm S} 
\oplus  {\bf 210}_{\rm S} \oplus  {\bf 210}_{\rm A} \oplus {\bf 770}_{\rm S}  \oplus {\bf 945}_{\rm A}
\oplus {\bf 1050}_{\rm S} \oplus \overline{\bf 1050}_{\rm S} \oplus {\bf 4125}_{\rm S} \oplus {\bf 5940}_{\rm A}.  \cr
{\bf 126} \otimes {\bf 126} &=& {\bf 54_{\rm S} \oplus 945_{\rm A} \oplus 1050_{\rm S} \oplus 2772_{\rm S} \oplus 4125_{\rm S} \oplus 6930_{\rm A}}. \cr
{\bf 10} \otimes {\bf 120} &=&   {\bf 45} \oplus {\bf 210} \oplus {\bf 945}. \cr
{\bf 10} \otimes {\bf 126} &=&   {\bf 210} \oplus {\bf 1050}. \cr
{\bf 120} \otimes {\bf 126} &=&   {\bf 45} \oplus {\bf 210} \oplus {\bf 945} \oplus {\bf 1050} \oplus {\bf 5940} \oplus {\bf 6930}.
\eea
From \eq{eq:16161616branch},
{we learn that four of ${\bf 16}$ can produce two trivial representations ${\bf 1}$ of $so(10)$ or Spin(10), one from ${\bf 10}\otimes{\bf 10}$ and one from ${\bf 120}\otimes{\bf 120}$.}
Therefore, on the mean field level, we can deduce the expectation of the GUT-Higgs $\Phi_{\mathbf{1}}$ from 
some schematic effective four-fermion interactions of $\psi$ in ${\bf 16}$ of Spin(10):\footnote{Here 
fermions are anti-commuting Grassman variables, so this expression $\langle \psi \psi \psi \psi\rangle$ is only schematic.
The precise expression of $\langle \psi \psi \psi \psi\rangle$ includes 
additional spacetime-internal representation 
indices and also includes possible additional spacetime derivatives (for point-splitting the fermions to neighbor sites if writing them on a regularized lattice).}
\bea
\langle\Phi_{\mathbf{1}} \rangle \simeq \langle \psi \psi \psi \psi\rangle \neq 0.
\eea
But we do not wish to impose the ordinary Anderson-Higgs quadratic mass term induced by $\langle \psi \psi\rangle \neq 0$,
otherwise this $\langle \psi \psi\rangle \neq 0$ will lead to Spin(10) symmetry breaking, instead of the Spin(10) symmetry preserving SMG.
This means that we have to impose $\langle \psi \psi\rangle = 0$, so
\bea
\langle \psi \psi\rangle  \psi \psi =0, \quad \quad \text{no conventional mass due to $\langle \psi \psi\rangle =0$.} 
\eea
Thus the above argument implies that above a critical condensation value
$\langle\Phi_{\mathbf{1}} \rangle > \Phi_{{\mathbf{1}},c}$ as the interaction strength goes above a critical value, 
we do obtain the SMG effect in \Fig{fig:phase}!

To implement the SMG to gap out the {\bf \emph{16 Weyl fermions}} in ${\bf 16}$, a necessary check is that the fermions are 
{\bf \emph{free from all 't Hooft anomalies}} in the Spin(10),
or more precisely free from  all 't Hooft anomalies in the spacetime-internal
$\Spin \times_{\Z_2^F}\Spin(10)$ structure.
This is true based on \eq{eq:Spin10-anomaly}, because there is only a mod 2 class $w_2w_3$ global anomaly,
which the 16 Weyl fermions in ${\bf 16}$ do not carry any $w_2w_3$ global anomaly. So we are able to gap out the 
16 Weyl fermions  while preserving $\Spin \times_{\Z_2^F}\Spin(10)$-symmetry.

To strengthen and improve \Refe{Wen2013ppa1305.1045}'s argument, we may regard our 
$\Phi$ as a bivector of two 10-dimensional vector $\phi_{so(10),{\bf 10}}\equiv  \phi_{{\bf 10}} $ in ${\bf 10}$
(or regard $\Phi$ as a bivector of two 120-dimensional vector $\phi_{so(10),{\bf 120}} \equiv \phi_{{\bf 120}} $ in ${\bf 120}$).
Thus, schematically
\bea
\langle\Phi_{\mathbf{1}} \rangle \simeq 
\langle \phi_{{\bf 10}} \phi_{{\bf 10}} \rangle
+ \langle \phi_{{\bf 120}} \phi_{{\bf 120}} \rangle
+ \dots \simeq \langle \psi \psi \psi \psi\rangle + \dots \neq 0.
\eea
This $\langle\Phi_{\mathbf{1}} \rangle > \Phi_{{\mathbf{1}},c} \neq 0$ implies that
the bi-linear of vectors (bivector) condense: 
$\langle \phi_{{\bf 10}} \phi_{{\bf 10}} \rangle \neq 0$ and/or $\langle \phi_{{\bf 120}} \phi_{{\bf 120}} \rangle \neq 0$,
but the $\langle \phi_{{\bf 10}} \rangle = \langle \phi_{{\bf 120}} \rangle =0$. So \emph{no} ordinary quadratic fermion mass term is induced, but only the SMG  is induced.
The SMG causes the \emph{symmetry-preserving disordered mass}.

But one of the mother EFTs (Model II) 
that we will propose later in \Sec{sec:ElementaryHiggs-WZW}, indeed have {\bf \emph{an extra new bosonic sector}} 
{\bf \emph{carrying the mod 2 class $w_2w_3$ global anomaly}}. This bosonic sector include the WZW term built out of GUT-Higgs fields.
To reiterate, there is no conflict about gapping the 16 Weyl fermions, but having the extra bosonic sector carry another anomaly.
This simply implies that {\bf \emph{if we demand to preserve $\Spin \times_{\Z_2^F}\Spin(10)$-symmetry,
although we can gap out the Weyl fermions in ${\bf 16}$,
the extra GUT-Higgs WZW bosonic sectors will still induce 
additional symmetry-preserving gapless modes.}}

\item In the standard Anderson-Higgs  \emph{electroweak} symmetry breaking mechanism, Higgs coupling $( {\psi}^\dagger_L \Phi_{{\mathbf{R}}} ({\ii} \sigma^2 {\psi_L'}^*) +{\rm h.c.} )$ is introduced 
in order to give quadratic masses to Weyl fermions. In this work, we may need to introduce more general GUT-Higgs fields $\Phi_{{\mathbf{R}}}$ with various representations ${\mathbf{R}}$. For a generic representation ${\mathbf{R}}$, the Higgs field may couple to a product of even number (not limited to two) of fermion operators (e.g. $\psi^\dagger\psi^\dagger\psi\psi$ or $\psi\psi\psi\psi$), such that the fermion representation can combine to match the corresponding Higgs field representation. 
(We shall not get distracted to handle the Anderson-Higgs \emph{electroweak} symmetry breaking masses of Weyl fermions in this article, as this effect is well-studied.
But we make some comments in Appendix \ref{sec:app-Branching-GUT-Higgs}.)

{\item {\bf Scaling dimensions of tuning parameters} $r_\mathbf{R}$. Because the GUT-Higgs field $\Phi_\mathbf{45}$, $\Phi_\mathbf{54}$, and $\Phi_\mathbf{1}$ all couple to four fermion operators (e.g. $\psi^\dagger\psi^\dagger\psi\psi$ or $\psi\psi\psi\psi+\text{h.c.}$), the term $r_\mathbf{R}\Phi_\mathbf{R}^2$ that tunes the Higgs transition will correspond to a eight-fermion interaction. At the SM fixed point, the matter fermion $\psi$ has a scaling dimension $3/2$ . So the eight-fermion interaction that drives the Higgs transition will have a scaling dimension $3/2\times 8=12$, which is much higher than the space-time dimension $4$. For this reason, such interaction is often ignored in the existing study of the SM. Although such interaction is perturbatively irrelevant at the SM fixed point, strong enough interaction will lead to non-perturbative effect that modifies the tuning parameters $r_\mathbf{R}$ and eventually drives the Higgs transitions between the SM phase and its adjacent GUT phases (such as the PS and GG phases).}
\end{itemize}
So taking into account the GUT-Higgs condensation or non-condensation,
we obtain a qualitative phase diagram in \Fig{fig:phase}.

\subsubsection{Model II: With Wess-Zumino-Witten term, and Deconfined Quantum Criticality}
\label{sec:ElementaryHiggs-WZW}

Now we propose a new mother EFT path integral by modifying the action $S_{\text{GUT}}$ to $S_{\text{GUT}}^{\text{WZW}}$ via adding the WZW term and other terms,
in a Lorentzian signature path integral:
\bea \label{eq:Zso10-WZW}
\bZ_{\GUT}^{\text{WZW}}\equiv
\int [{\cal D} {\psi_L}] [{\cal D}{\psi_L^\dagger}] [{\cal D} A][{\cal D} \Phi_{{\mathbf{R}}} ][{\cal D} \Phi^{\rm{bi}} ][{\cal D} \phi ] \dots
\exp( \ii \left. S^{\text{WZW}}_{\text{GUT}}[\psi_L, {\psi_L^\dagger}, A, \Phi_{{\mathbf{R}}}, \Phi^{\rm{bi}}, \phi, \dots ] \right  \rvert_{M^4}).
\eea
\begin{multline} \label{eq:Sso10-WZW}
S_{\text{GUT}}^{\text{WZW}}\equiv \int_{M^4} \Tr(F_{} \wedge \star F_{}) 
+ \int_{M^4} \Big( {\psi}^\dagger_L  (\ii {\bar  \sigma}^\mu {D}_{\mu, A} ) \psi_L
+ | {D}_{\mu, A}\Phi_{{\mathbf{R}}} |^2 -{\rm U}(\Phi_{{\mathbf{R}}}) \\
+ \frac{1}{2}\phi^\intercal \Phi^{\rm{bi}} \phi
+ \frac{1}{2}\sum_{a=1}^{5}\big(\psi_L^\intercal\ii\sigma^2(\phi_{2a-1}\Gamma_{2a-1}-\ii\phi_{2a}\Gamma_{2a})\psi_L+\text{h.c.}\big)
  \Big)\,\dd^4 x
+S^\text{WZW}[\Phi^{\rm{bi}}].
\end{multline}
The purpose of the new discrete torsion class 4d WZW-like term (written on a 5d manifold with 4d boundary), 
that we will introduce in details later, is to saturate the $w_2 w_3$ global anomaly.
The mother EFT contains the following detailed ingredients:
\begin{enumerate}[leftmargin=.mm]
\item
There are 16n complex Weyl fermions, each $\psi_L$ is the ${\bf 16}$ of Spin(10) minimally coupled to $\Spin(10)$ gauge field in the covariant derivative. 
Properties of the Spin(10) gauge field $A$ and other familiar terms in $S_{\text{GUT}}$ had been explained in the earlier \Sec{sec:ElementaryHiggs}.

\item {An $\SO(10)$ real vector field $\phi \in \R$ is in ${\bf 10}$ of $so(10)$ also of Spin(10). 
To be explicit, $\phi$ contains one vector index, $\phi_a$ with $a\in \{1,2,\dots,10\}$.}

\item An $\SO(10)$ real bivector field $\Phi^{\rm{bi}}\in \R$  is obtained from the tensor product of the two $\phi$,
in the ${\bf 10} \otimes {\bf 10} = {\bf 1}_{\rm S}  \oplus {\bf 45}_{\rm A} \oplus {\bf 54}_{\rm S}$ of $so(10)$ also of Spin(10). 
To be explicit, $\Phi^{\rm{bi}}$ contains two vector indices, $\Phi^{\rm{bi}}_{ab}$ with $a,b\in \{1,2,\dots,10\}$.
We can arrange $\Phi^{\rm{bi}}_{ab}$ into three different representations ${\mathbf{R}}$ of $\Phi_{{\mathbf{R}}}$ 
as the three GUT-Higgs fields $\Phi_{{\mathbf{1}}}$, $\Phi_{{\mathbf{45}}}$ and $\Phi_{{\mathbf{54}}}$
(which appeared in \Sec{sec:ElementaryHiggs-noWZW}):
 \be\hspace{-2mm} \label{eq:Phi-bi-phi}
\Phi^{\rm{bi}}_{ab} = \phi_{a}\phi_{b} \text{ includes}\left\{\begin{array}{l}
\Tr\Phi^{\rm{bi}}=\sum_{a}\Phi^{\rm{bi}}_{aa} \text{ gives $\Phi_{{\mathbf{R}}}=\Phi_{{\mathbf{1}}}$ in ${\bf 1}_{\rm S}$}. \\ 
\hat{\Phi}^{\rm{bi}}\equiv
\Phi^{\rm{bi}}_{[a,b]} = \tfrac{1}{2}(\Phi^{\rm{bi}}_{ab}-\Phi^{\rm{bi}}_{ba})=\tfrac{1}{2}(\phi_a\phi_b-\phi_b\phi_a)=\tfrac{1}{2}[\phi_a,\phi_b] \text{ gives $\Phi_{{\mathbf{R}}}=\Phi_{{\mathbf{45}}}$ in ${\bf 45}_{\rm A}$} .\\
\tilde{\Phi}^{\rm{bi}}\equiv
\Phi^{\rm{bi}}_{\{a,b\}} = \tfrac{1}{2}(\Phi^{\rm{bi}}_{ab}+\Phi^{\rm{bi}}_{ba})=\tfrac{1}{2}(\phi_a\phi_b+\phi_b\phi_a)=\tfrac{1}{2}\{\phi_a,\phi_b\} \text{ gives $\Phi_{{\mathbf{R}}}=\Phi_{{\mathbf{54}}}$ in ${\bf 54}_{\rm S}$}.
\end{array}\right.
\ee
{For brevity, we also denote
the anti-symmetric bivector $\Phi^{\rm{bi}}_{[a,b]}$ or $\Phi_{{\mathbf{45}}}$ as $\hat{\Phi}^{\rm{bi}}$,
and denote the 
symmetric bivector $\Phi^{\rm{bi}}_{\{a,b\}}$ or $\Phi_{{\mathbf{54}}}$
as $\tilde{\Phi}^{\rm{bi}}$.
}
\item {\bf \emph{GUT-Higgs field kinetic term and covariant derivative}}: 
The kinetic term for the GUT-Higgs fields is written as 
 $| {D}_{\mu, A}\Phi_{{\mathbf{R}}} |^2 \equiv ( {D}_{A}^\mu \Phi_{{\mathbf{R}}})^\dagger ( {D}_{\mu, A}\Phi_{{\mathbf{R}}})$,
with the complex conjugate transpose written as dagger $\dagger$.

Moreover, we can also combine the kinetic terms for $\Phi_{{\mathbf{1}}}$, $\Phi_{{\mathbf{45}}}$ and $\Phi_{{\mathbf{54}}}$
in terms of the kinetic term for the bivector $\Phi^{\rm{bi}}_{}$. 
This kinetic term becomes $\Tr \big((D^{\mu}_{A} \Phi^{\rm{bi}})^\intercal (D_{\mu, A}\Phi^{\rm{bi}}) \big)$,
with the matrix transpose written as $\intercal$,
where the Trace Tr is over the 10-dimensional Lie algebra representation of $so(10)$.
We can write down the explicit form $(D_{\mu, A}\Phi^{\rm{bi} })_{ab} 
\equiv  \nabla_\mu \Phi^{\rm{bi} }_{ab}    - \ii g [A_\mu, \Phi^{\rm{bi} }]_{ab} 
=\nabla_\mu \Phi^{\rm{bi} }_{ab}    - \ii g (A_{\mu,ab} \Phi^{\rm{bi} }_{bc}- \Phi^{\rm{bi} }_{ab}  A_{\mu, bc} )$
with $a,b,c\in \{1,2,\dots,10\}$,\footnote{The reason that
$(D_{\mu, A}\Phi^{\rm{bi} })_{ab} \equiv  \nabla_\mu \Phi^{\rm{bi} }_{ab}   - \ii g [A_\mu, \Phi^{\rm{bi} }]_{ab}$ has 
a matrix commutator $[A_\mu, \Phi^{\rm{bi} }]$ in contrast with the familiar form
${D}_{\mu, A }\phi \equiv \nabla_{\mu} \phi - \ii g  \,  A_\mu \phi$, is due to the following fact:
The Lie group $G$ transformation for some $U \in G$
acts on the gauge field $A$ as $A \mapsto U (A+\frac{\ii}{g} \dd) U^\dagger$ (or $A \mapsto U (A+\frac{\ii}{g} \dd) U^{\intercal}$ when $U$ is real-valued).
However, the Lie group transformation acts on the vector field $\phi$ as $\phi \mapsto U \phi$,
while acts on the rank-10 matrix bivector field $\Phi^{\rm{bi} }$ as $\Phi^{\rm{bi} }_{ab} \mapsto U\Phi^{\rm{bi} }_{ab} U^\intercal$. 
} 
where $A_{\mu,ab} =\sum_{\alpha} A_{\mu}^{\alpha} {\rm T'}^{\alpha}_{ab}$ 
with another 45 pieces of the rank-10 matrix representation ${\rm T'}^{\alpha}$.

In general, the Lie algebra generator ${\rm T}^{\alpha}$ is hermitian.
In the case of the real representation ${\mathbf{10}}$, the ${\rm T'}^{\alpha}$ is not only hermitian, but also an imaginary and anti-symmetric matrix.

In summary, for our purpose, the two expressions of GUT-Higgs kinetic terms are both correct:
 $\sum_{{\mathbf{R}}={\bf 1}, {\bf 45}, {\bf 54}} 
 | {D}_{\mu, A}\Phi_{{\mathbf{R}}} |^2 \equiv 
 ( {D}_{A}^\mu \Phi_{{\mathbf{1}}})^\dagger ( {D}_{\mu, A}\Phi_{{\mathbf{1}}})
 + ( {D}_{A}^\mu \Phi_{{\mathbf{45}}})^\dagger ( {D}_{\mu, A}\Phi_{{\mathbf{45}}})+
  ( {D}_{A}^\mu \Phi_{{\mathbf{54}}})^\dagger ( {D}_{\mu, A}\Phi_{{\mathbf{54}}})$,
and the bi-vector field expression: $\Tr \big((D^{\mu}_{A} \Phi^{\rm{bi}})^\intercal (D_{\mu, A}\Phi^{\rm{bi}}) \big)$.

All these above GUT-Higgs fields (in the vector or bivector representations) 
also coupled to the $so(10)$ gauge fields in the standard way.

\item {\bf \emph{Yukawa-like coupling terms}}:
We also have several Yukawa-like coupling terms, \\
(i) between the GUT-Higgs bivectors $\Phi^{\rm{bi}}$ and the vectors $\phi$, 
 explicitly, $\phi^\intercal \Phi^{\rm{bi}} \phi \equiv \sum_{a,b} \phi_a^\intercal \Phi^{\rm{bi}}_{ab} \phi_b$.

(ii) between the GUT-Higgs  vectors $\phi$ and the Weyl spinor $\psi_L$,
the $\big(\psi_L^\intercal\ii\sigma^2(\phi_{2a-1}\Gamma_{2a-1}-\ii\phi_{2a}\Gamma_{2a})\psi_L+\text{h.c.}\big)$ is apparently a hermitian scalar.
The $\sigma^2$ matrix acts on the 2-component spacetime Weyl spinor $\psi_L$. $\Gamma_a$ (with $a\in \{1,2,\dots,10\}$) are ten rank-16 matrices satisfying $\{\Gamma_{2a-1},\Gamma_{2b-1}\}=2\delta_{ab}, \{\Gamma_{2a},\Gamma_{2b}\}=2\delta_{ab}, [\Gamma_{2a-1},\Gamma_{2b}]=0$ (for $a,b=1,2,\cdots,5$).

\item {\bf \emph{Mean-field approximation}}: If for a moment, we neglect the gauge field $A$ coupling in the covariant derivative, 
neglect the GUT-Higgs potential ${\rm U}(\Phi_{{\mathbf{R}}})$,
and neglect the possible WZW term $S^\text{WZW}[\Phi^{\rm{bi}}]$, 
then we only have the quadratic Lagrangian in between 
GUT-Higgs bivectors $\Phi^{\rm{bi}}$, vectors $\phi$, and the Weyl spinor $\psi_L$.
Then this quadratic Lagrangian,
$\frac{1}{2}\phi^\intercal \Phi^{\rm{bi}} \phi
+ \frac{1}{2}\sum_{a=1}^{5}\big(\psi_L^\intercal\ii\sigma^2(\phi_{2a-1}\Gamma_{2a-1}-\ii\phi_{2a}\Gamma_{2a})\psi_L+\text{h.c.}\big)$,
at the mean-field level, can be integrated out to impose constraints and relations between the 
bivectors $\Phi^{\rm{bi}}$, vectors $\phi$, and the Weyl spinor $\psi_L$.
In some sense, what is integrated out becomes a Lagrange multiplier to impose a constraint on the remained fields.
In this limit, we only need to regard the Weyl spinor $\psi_L$ as the elementary fields,
the vectors $\phi$ is the ${\bf 10}$ from the tensor product of two $\psi_L$ since
${\bf 16} \otimes {\bf 16} = ({\bf 10} \oplus {\bf 120}  \oplus \overline{\bf 126} )$.
Then the bivector $\Phi^{\rm{bi}}$ is from the tensor product of two $\phi$ as the ${\bf 10} \otimes {\bf 10}$, out of the quartic $\psi_L$'s
${\bf 16} \otimes {\bf 16} \otimes {\bf 16} \otimes {\bf 16}$. 

\item {\bf \emph{Wess-Zumino-Witten-like discrete torsion term:}} 
For now, we directly provide our endgame answer to WZW term, later we will backup and derive this WZW term in details from scratch in {\Sec{sec:HomotopyCohomology}}.

The schematic WZW action that we propose to match the mod 2 class $w_2w_3$ global anomaly is:
\bea \label{eq:BdCdeRham}
S^\text{WZW}[\Phi]=\pi \int_{M^5}B({\Phi})\wedge \dd B'({\Phi}),
\eea
in terms of differential form with mod 2 valued forms of $B$ and $B'$ fields, in the de Rham cohomology. 
The theory is defined on the 5d manifold ${M^5}$ whose boundary is the 4d space time $M^4=\partial {M^5}$.\footnote{Here we normalize the usual differential form
$B(\tilde{\Phi}^{\rm{bi}})/{\pi} \mapsto {B(\tilde{\Phi}^{\rm{bi}})}$ and
$B'(\hat{\Phi}^{\rm{bi}})/{\pi}  \mapsto B'(\hat{\Phi}^{\rm{bi}})$,
so the usual differential form partition function
$\exp(\ii \frac{2}{2\pi}  \int_{M^5}B(\tilde{\Phi}^{\rm{bi}})\wedge \dd B'(\hat{\Phi}^{\rm{bi}}) )$ 
maps to $\exp(\ii \pi \int_{M^5}B(\tilde{\Phi}^{\rm{bi}})\wedge \dd B'(\hat{\Phi}^{\rm{bi}}) ) $.
See a related discussion on the 5d $B \dd B'$ theory in \cite{KravecMcGreevySwingle1409.8339}.
{The quantization conditions on the closed cycles, also map from:
$\ointint B(\tilde{\Phi}^{\rm{bi}})$ or $\ointint B'(\hat{\Phi}^{\rm{bi}}) = n{\pi} \mod 2 \pi \mapsto \ointint {B(\tilde{\Phi}^{\rm{bi}})}$ or $\ointint B'(\hat{\Phi}^{\rm{bi}})  = n  \mod 2$.
It can be verified that this WZW has two properties: 
(1) invertible $|{\bf Z}(M^5)|=1$ on a closed 5-manifold, but on a specific manifold
${\bf Z}(M^5)=-1$ can possibly signature the underlying bulk 5d invertible TQFT $w_2w_3$. 
(2) this WZW term really is a 4d theory, having physical impacts only on the 4d $M^4$ --- it is a 4d boundary theory of the 5d bulk invertible TQFT
on the extended $M^5$.}
\label{ft:normalization-WZW}
}
The $B$ and $B'$ are constructed out of 
some GUT-Higgs field ${\Phi}$ 
{(such as the bivector $\tilde{\Phi}^{\rm{bi}}$ 
or $\hat{\Phi}^{\rm{bi}}$, for
$\Phi^{\rm{bi}}_{\{a,b\}}$ 
or $\Phi^{\rm{bi}}_{[a,b]}$ respectively, organized in \eq{eq:Phi-bi-phi})}. 
More precisely, the WZW term is written in the singular cohomology class of $B$ and $B'$ cochain fields: 
\bea  \label{eq:BdCsingular}
S^\text{WZW}[\Phi]{=\pi \int_{M^5}B(\tilde{\Phi}^{\rm{bi}})\smile \delta B'(\hat{\Phi}^{\rm{bi}})}
{= 2\pi \int_{M^5} B(\tilde{\Phi}^{\rm{bi}})\smile \frac{\delta}{2} B'(\hat{\Phi}^{\rm{bi}})}
{= 2\pi \int_{M^5} B(\tilde{\Phi}^{\rm{bi}})\smile \Sq^1 B'(\hat{\Phi}^{\rm{bi}})}.\quad
\eea
Here the 2-cochain fields are $\Z_2$-valued, they can be chosen as cohomology classes thus 
$B \in {\rm H}^2(M,\Z_2)$ and $B'  \in {\rm H}^2(M,\Z_2)$.
The $\delta$ is the coboundary operator, 
and the Steenrod square $\Sq^1 \equiv \frac{\delta}{2} \mod 2$ here maps the singular cohomology $\H^2(M,\Z_2) \mapsto \H^3(M,\Z_2)$,
on some triangulable manifold $M$.\footnote{
Generally, given a chain complex $\rC_{\bullet}$ and a short exact sequence of abelian groups:
$$
0\to \rA'\to \rA\to \rA''\to 0,
$$
we have a short exact sequence of cochain complexes:
$$
0\to\Hom(\rC_{\bullet},\rA')\to\Hom(\rC_{\bullet},\rA) \to\Hom(\rC_{\bullet},\rA'')\to0.
$$
Hence we can obtain a long exact sequence of cohomology groups:
$$
\cdots\to\H^n(\rC_{\bullet},\rA')\to\H^n(\rC_{\bullet},\rA)\to\H^n(\rC_{\bullet},\rA'')
\stackrel{\partial}{\to}\H^{n+1}(\rC_{\bullet},\rA')\to\cdots,
$$
the connecting homomorphism $\partial$ is called Bockstein homomorphism.
For instance,
$\beta_{(n,m)}:\H^*(-,\Z_{m})\to\H^{*+1}(-,\Z_{n})$ is the Bockstein homomorphism associated with the extension $\Z_n\stackrel{\cdot m}{\to}\Z_{nm}\to\Z_m$ where $\cdot m$ is the group homomorphism given by multiplication by $m$. 
Specifically, $\beta_{(2,2^n)}=\frac{1}{2^n}\delta\mod 2$, thus the 
Steenrod square obeys $\Sq^1 \equiv \beta_{(2,2)}  \equiv \frac{\delta}{2} \mod 2$.} 
The wedge product $\wedge$ of differential form in \eq{eq:BdCdeRham} becomes the
cup product $\smile$ of cochains or cohomology classes in \eq{eq:BdCsingular}.
Note that the triangulable manifold $M$ is always a smooth differentiable manifold, 
thus we can downgrade the singular cohomology result \eq{eq:BdCsingular} to
reproduce the de Rham cohomology expression \eq{eq:BdCdeRham}.

\item {\bf \emph{GUT-Higgs potential ${\rm U}(\Phi_{{\mathbf{R}}})$, and a relation to non-linear sigma model (NLSM)}}:
Mostly we shall simply choose the GUT-Higgs potential written in \eq{eq:GUTHiggsU},
$$
{\rm U}(\Phi_{{\mathbf{R}}}) =\Big(r_{{\mathbf{45}}} (\Phi_{{\mathbf{45}}})^2 +\lambda_{{\mathbf{45}}} (\Phi_{{\mathbf{45}}})^4\Big)+ 
\Big(r_{{\mathbf{54}}} (\Phi_{{\mathbf{54}}})^2 +\lambda_{{\mathbf{54}}} (\Phi_{{\mathbf{54}}})^4\Big)+
\Big(r_{{\mathbf{1}}} (\Phi_{{\mathbf{1}}})^2 +\lambda_{{\mathbf{1}}} (\Phi_{{\mathbf{1}}})^4\Big),
$$
which is sufficient for a continuum QFT description.
{Some lattice or condensed matter based theorists may wonder whether there is a non-linear sigma model (NLSM) description at a deeper UV.
One approach is to write down a potential with a NLSM constraint $(\Tr (\Phi^\intercal\Phi) - \mathrm{R}^2)$ with the norm of GUT-Higgs centered around a radius $\mathrm{R}$,
and introduce a Lagrange multiplier $\lambda$,
such that integrating out $\int [{\cal D} {\lambda}] \dots$ gives the fixed radius constraint at UV. With appropriate deformations, we anticipate a RG flow from UV to IR gives
the GUT-Higgs potential.
One reason to introduce a NLSM is that it is natural to adding the WZW term to NLSM.
However, an NLSM description turns out to be \emph{not necessary} for writing 
our WZW term.
}

\item {\bf \emph{Deconfined Quantum Criticality (DQC):}} 
The motivation to add this 4d $S^\text{WZW}[\Phi]$ into our 4d mother EFT is to induce the analogous phenomenon called 
the deconfined quantum criticality  \cite{SenthildQCP0311326}.
The original deconfined quantum criticality  \cite{SenthildQCP0311326} is proposed as a continuous quantum phase transition between
two kinds of Landau symmetry breaking orders: N\'eel anti-ferromagnet order and Valence-Bond Solid (VBS) order in 3d (namely, 2+1d).

Here in out gauge theory context in 4d (namely, 3+1d), between the GG  $su(5)$ GUT and the PS ${su(4) \times su(2)  \times su(2)}$ model,
we do not really have the conventional Landau symmetry breaking orders 
as both the $su(5)$ and ${su(4) \times su(2)  \times su(2)}$ are dynamically gauged as gauge theories.
But if we regard the $su(5)$ and ${su(4) \times su(2)  \times su(2)}$ are internal global symmetries that are not yet gauged,
then we are able to seek for a deconfined quantum criticality construction between the GG and PS models, as we will verify in the next \Sec{sec:HomotopyCohomology}.
\end{enumerate}

\subsection{Homotopy and Cohomology group arguments to induce a WZW term}
\label{sec:HomotopyCohomology}

We review the 3d WZW term construction in the familiar deconfined quantum criticality (dQCP) in 3d (namely, 2+1d)  \cite{SenthildQCP0311326},
in Appendix \ref{sec:HomotopyCohomology-NeelVBS}, 
based on more nonperturbative arguments from homotopy and cohomology groups, and anomaly classifications from cobordism. 
Here we proceed with the same logic, to construct the 4d WZW term in the new deconfined quantum criticality (DQC) 
in 4d (namely, 3+1d) to justify what we claimed in \eq{eq:BdCsingular}.

Below we write $G$ as the original larger symmetry group,
while ${G_{\text{sub}}}$ is the remained preserved unbroken symmetry in the corresponding order 
(i.e., N\'eel or VBS orders for 3d dQCP; the GG or PS for the 4d DQC we will propose).
Then we have the following fibration structure:
\bea \label{eq:fibration}
{G_{\text{sub}}} \lhook\joinrel\xrightarrow{\quad} G \longrightarrow \frac{G}{G_{\text{sub}}},
\eea
where the quotient space $\frac{G}{G_{\text{sub}}}$ is the base manifold (i.e., the orbit) as the \emph{symmetry-breaking order parameter space}.
The $G$ is the total space obtained from the fibration of the ${G_{\text{sub}}}$ fiber (i.e., the stabilizer) over the base $\frac{G}{G_{\text{sub}}}$. 
 
Now we follow the similar logic for the 3d dQCP summarized in Appendix \ref{sec:HomotopyCohomology-NeelVBS}, 
generalizing the idea to deal with our 4d DQC.

\subsubsection{Induce a 4d WZW term between Georgi-Glashow $su(5)$ and Pati-Salam $su(4) \times su(2) \times su(2)$ 
models on a 5d bulk $w_2(V_{\SO(10)})w_3(V_{\SO(10)})$}
\label{sec:HomotopyCohomology-GGPS}

Follow the principle in Appendix \ref{sec:HomotopyCohomology-NeelVBS}, 
we aim to {induce a 4d WZW term between Georgi-Glashow $su(5)$ and Pati-Salam $su(4) \times su(2) \times su(2)$ 
models on a 5d bulk $w_2(V_{\SO(10)})w_3(V_{\SO(10)})$}.
First we look at the order-parameter target manifold via the fibration structure \eq{eq:fibration}, formed by the \emph{bosonic} GUT-Higgs fields.
For the bosonic GUT-Higgs fields, we only have the internal SO(10) symmetry not the Spin(10) symmetry,
but we can include the orientation reversal which gives an $\O(10) = \SO(10) \rtimes \Z_2$ symmetry. 
Then the fibration \eq{eq:fibration} becomes:
\bea \label{eq:fibration-su5}
\text{GG $su(5)$ GUT: } \Big({G_{\text{sub}}}={\U(5)} \Big) \lhook\joinrel\xrightarrow{\quad} 
\Big(G={\O(10)}\Big) \longrightarrow \Big(\frac{G}{G_{\text{sub}}}=\frac{\O(10)}{\U(5)}\Big). 
\eea
Here we can keep the larger ${\U(5)}$ instead of SU(5) as the preserved internal symmetry of the $su(5)$ GUT.
\bea \hspace{-3mm}
\label{eq:fibration-su5}
\text{PS $su(4) \times su(2) \times su(2)$: }\Big({G_{\text{sub}}}={\O(6)\times\O(4)} \Big) \lhook\joinrel\xrightarrow{\quad} 
\Big(G={\O(10)}\Big) \longrightarrow \Big(\frac{G}{G_{\text{sub}}}=\frac{\O(10)}{\O(6)\times\O(4)}\Big). 
\eea
Recall that $su(4) \times su(2) \times su(2)$ has the same Lie algebra as $so(6) \times so(4)$.
Here we also keep the larger ${\O(6)\times\O(4)}$ instead of ${\SO(6)\times\SO(4)}$ 
as the preserved internal symmetry of the PS model.
{Homotopy groups for these target manifolds of GUT-Higgs fields are in the table:}
\bea \label{eq:HomotopyGGPS}
\begin{tabular}{lcccccc}
\hline
 & $\pi_0$ & $\pi_1$ & $\pi_2$ & $\pi_3$ & $\pi_4$ & $\pi_5$ \\
\hline
GG $\frac{\O(10)}{\U(5)}$ & $\Z_2$ & $0$ & $\Z$ & $0$ & $0$ & $0$ \\
\hline
PS $\frac{\O(10)}{\O(6)\times\O(4)}$ & $0$ & $\Z_2$ & $\Z_2$ & $0$ & $\Z^2$ & $\Z_2^2$ \\
\hline
$\O(10)$ & $\Z_2$ & $\Z_2$ & $0$ & $\Z$ & $0$ & $0$  \\
$\O(4)$ & $\Z_2$ & $\Z_2$ & $0$ & $\Z^2$ & $\Z_2^2$ & $\Z_2^2$  \\
$\O(6)$ & $\Z_2$ & $\Z_2$ & $0$ & $\Z$ & $0$ & $0$  \\
\hline
$\U(5)$ & $0$ & $\Z$ & $0$ & $\Z$ & $0$ & $\Z$  \\
\hline
$\SO(10)$ & $0$ & $\Z_2$ & $0$ & $\Z$ & $0$ & $0$  \\
$\SO(4)$ & $0$ & $\Z_2$ & $0$ & $\Z^2$ & $\Z_2^2$ & $\Z_2^2$  \\
$\SO(6)$ & $0$ & $\Z_2$ & $0$ & $\Z$ & $0$ & $0$  \\
\hline
$\SU(5)$ & $0$ & $0$ & $0$ & $\Z$ & $0$ & $\Z$  \\
\hline
\end{tabular}.
\eea
Let us comment about the construction of 4d WZW and its 4d 't Hooft anomaly, step by step,
\begin{enumerate}[leftmargin=.mm]
\item Start with the hint from homotopy groups, we need to find {\bf {\emph{topological defects trapped
in the order-parameter target manifold of bosonic GUT-Higgs fields}}}
 in the GG and PS models,\footnote{{\bf{\emph{Caveat}}}: 
We had emphasized again and again that here we are considering 
topological defects in the order-parameter target manifold of \emph{bosonic} GUT-Higgs fields.
We are \emph{not} talking about 
the topological objects of \emph{fermionic} sectors (quarks/leptons) or gauge theory sectors in GUTs or SMs. 
For example, there are 
magnetic monopoles in the GG and PS gauge theories from
$\pi_1(G_{\SM_6})=\pi_2(G_{\GG}/G_{\SM_6})=\pi_2(G_{\PS_2}/G_{\SM_6})=\Z$, also from 
$\pi_1(G_{\SM_3})=\pi_2(G_{\PS_1}/G_{\SM_3})=\Z$ or from any $\pi_1(G_{\SM_q})=\Z$ with $q=1,2,3,6$.
But we are talking about different topological objects in the order-parameter target manifold of 
\emph{bosonic} GUT-Higgs fields.}
classified by
$\pi_{n_{\GG}}(\frac{\O(10)}{\U(5)})$ and 
$\pi_{n_{\PS}}(\frac{\O(10)}{\O(6)\times\O(4)})$
such that the dimensionality
${n_{\GG}} + {n_{\PS}} = d$ where the $d$ is the total spacetime dimension thus $d=4$ (or one lower dimension compared with the 5d where the WZW is extended to put on).
This suggests that we take
$$
\pi_{2}(\frac{\O(10)}{\U(5)}) = \Z, \quad \pi_{2}(\frac{\O(10)}{\O(6)\times\O(4)}) = \Z_2, \quad {n_{\GG}} + {n_{\PS}} =2+2=4.
$$
Note that $(\frac{\O(m+n)}{\O(m)\times\O(n)}) \equiv {\rm Gr}(m, m+n)$ is a Grassmannian manifold.
Here we need ${\rm Gr}(6, 10)={\rm Gr}(4, 10)$.
\item We will use the cohomology construction of the WZW term, furnished by the hints of homotopy groups.
Then we need a relation between homotopy group and cohomology group.

In algebraic topology, an Eilenberg-MacLane space $K(G,n)$ is a topological space with a single nontrivial homotopy group, s.t.~$\pi_{n}(K(G,n))\cong G$ and $\pi_m(K(G,n))=0$ if $m\neq n$. It can be regarded as a building block for homotopy theory, 
also it provides a bridge between homotopy and cohomology. Let $X$ be a topological space or a manifold. 
The set $[X,K(G,n)]$ of based homotopy classes of based maps from $X$ to $K(G,n)$ is a natural bijection with the $n$-th singular cohomology group $\H^n(X,G)$. In particular, when $\pi_n(X)\cong G$,
\begin{equation}
\H^n(X,G)=\mathrm{Hom}(\pi_n(X), G)=\mathrm{Hom}(G,G).
\end{equation}
{There is a distinguished element $\omega\in \H^n(X,G)$, 
as the generator of the cohomology group $\H^n(X,G)$, 
corresponding to the identity morphism in $\mathrm{Hom}(G, G)$.} 
The morphism is realized as
\begin{equation}
\begin{split}
\omega:&\pi_n(X)\to G,
\quad\quad\quad
f\in\pi_n(X) \mapsto \int_{x\in S^n}\omega(f(x)) \in G.
\end{split}
\end{equation}

\item With the above homotopy group \eq{eq:HomotopyGGPS} in mind, 
we can use the Serre spectral sequence to derive the following:\footnote{We can answer in more general case $\O(2n)/\U(n)$. 
We will need the Universal Coefficient Theorem (UCT), so that
$\H^2(X,{\rm A})=\Hom(\H_2(X),{\rm A})\oplus {\rm Ext}(\H_1(X),{\rm A}),$ for some topological space $X$ and any abelian group coefficient ${\rm A}$.\\
The space $\O(2n)/\U(n)$ has two connected components, 
each of which is diffeomorphic to $\SO(2n)/\U(n)$, so $\H^k(\O(2n)/\U(n), {\rm A}) = \H^k(\SO(2n)/\U(n), {\rm A})\oplus \H^k(\SO(2n)/\U(n), {\rm A})$.\\
For $n > 1$, the space $\SO(2n)/\U(n)$ is simply connected with $\pi_2(\SO(2n)/\U(n)) = \mathbb{Z}$, 
so by the Hurewicz Theorem we have $\H_1(\SO(2n)/\U(n), \mathbb{Z}) = 0$ and $\H_2(\SO(2n)/\U(n), \mathbb{Z}) = \mathbb{Z}$. 
Therefore by UCT,
so we have
$\H^2(\SO(2n)/\U(n), {\rm A}) = \Hom(\mathbb{Z}, {\rm A})\oplus\operatorname{Ext}(0, {\rm A}) = {\rm A}$. 
Thus, $\H^2(\O(2n)/\U(n), {\rm A}) = {\rm A}^2$.
}
\bea
\H^2(\O(10)/\U(5), \Z) =\Z^2.\quad\quad\quad
\H^2(\O(10)/\U(5), \Z_2)=\Z_2^2.
\eea
In fact, we just need one of the two components from $\SO(10)/\U(5)$, whose cohomology group:
\bea
\H^2(\SO(10)/\U(5), \Z) =\Z.\quad\quad\quad
\H^2(\SO(10)/\U(5), \Z_2)=\Z_2.
\eea
\item We can also derive
\bea
\H^2(\O(10)/(\O(6)\times\O(4)), \Z) =\Z_2.\quad\quad\quad
\H^2(\O(10)/(\O(6)\times\O(4)), \Z_2)=\Z_2^2.
\eea
The mod 2 cohomology of real Grassmannian manifold is well-known from the theory of Stiefel-Whitney characteristic classes.
The integral cohomology is trickier but it can be worked out.
\item
We now take a $\Z_2$ cohomology class called $B(\tilde{\Phi}^{\rm{bi}})$ out of
\bea
B(\tilde{\Phi}^{\rm{bi}}) \in \H^2(\O(10)/(\O(6)\times\O(4)), \Z_2),
\eea
and another $\Z_2$ cohomology class called $B'(\hat{\Phi}^{\rm{bi}})$ out of
\bea
B'(\hat{\Phi}^{\rm{bi}}) \in \H^2(\O(10)/\U(5), \Z_2).
\eea
$\bullet$ The $B(\tilde{\Phi}^{\rm{bi}})$-field as a second cohomology class, 
can be constructed out of the GUT-Higgs field $\Phi_{{\mathbf{54}}}$ in the ${\mathbf{54}}$ representation of $so(10)$.
In particular, we can also write $\Phi_{{\mathbf{54}}}$ as 
a bivector GUT-Higgs field {\bf symmetric} representation, ${\bf 54}_{\rm S}$ out of 
${\bf 10} \otimes {\bf 10}$, called $\tilde{\Phi}^{\rm{bi}}$ that we detail in \Sec{sec:CompositeHiggs}.\\
$\bullet$ The $B'(\hat{\Phi}^{\rm{bi}})$-field as a second cohomology class, 
can be constructed out of the GUT-Higgs field $\Phi_{{\mathbf{45}}}$ in the ${\mathbf{45}}$ representation of $so(10)$.
In particular, we can also write $\Phi_{{\mathbf{45}}}$ as 
a bivector GUT-Higgs field {\bf anti-symmetric} representation, ${\bf 45}_{\rm A}$ out of 
${\bf 10} \otimes {\bf 10}$, called $\hat{\Phi}^{\rm{bi}}$ that we detail in \Sec{sec:CompositeHiggs}.

Similar to the familiar 3d dQCP in Appendix \ref{sec:HomotopyCohomology-NeelVBS}, 
we can also provide the physical intuitions on the link invariants between various topological defects:
between the  {\bf {\emph{ charged objects}}} and the  {\bf {\emph{charge operators}}} constructed from homotopy groups and cohomology groups. 
For example,
\begin{enumerate}[leftmargin=2.mm, label=\textcolor{blue}{(\roman*)}., ref={(\roman*)}]
\item  
{\bf {\emph{Georgi-Glashow GUT-Higgs target manifold and topological defects}}}:\\
The $B'(\hat{\Phi}^{\rm{bi}}) \in \H^2(\O(10)/\U(5), \Z_2)$ can be placed on a 2-surface called $\hat\varrho^2$, as a {\bf {\emph{charge operator}}}
$\exp( \ii \pi \ointint_{\hat \varrho^2}   B'(\hat{\Phi}^{\rm{bi}})) = 
\exp( \ii \pi \ointint_{\hat \varrho^2}   c_1(V_{\U(5)}))$ (i.e., symmetry generator)
measures the charge of a preserved U(5) symmetry in the topological defect 
trapped in the target manifold $\O(10)/\U(5)$.
The first Chern class $c_1(V_{\U(5)})$ of the associated vector bundle of U(5) 
 evaluates a magnetic flux mod 2 on this 2-surface ${\hat \varrho^2}$.
There is a topological defect line along a 1d loop called $\varsigma^1_{\GG}$, 
paired up with a 1-connection called $\hat v$ gives a 1d line operator $\exp(\ii \pi \oint_{\varsigma^1_{\GG}} \hat v )$
as a {\bf {\emph{ charged object}}}.
The {charge operator} 2-surface ${\hat \varrho^2}$ can be linked with a {charged} 1d loop $\varsigma^1_{\GG}$ in the 4d spacetime.
Follow the generalized higher global symmetry language \cite{Gaiotto2014kfa1412.5148},
 this nontrivial linking number Lk implies a measurement of U(5) symmetry on the topological defect.
 Precisely, the linking number Lk, manifested as a statistical Berry phase,
is evaluated via the expectation value of path integral:
\bea  \label{eq:Lk-2d-1d-GG}
\langle \exp(\ii \pi \ointint_{\hat \varrho^2} B'(\hat{\Phi}^{\rm{bi}}) ) \cdot 
\exp(\ii \pi \oint_{\varsigma^1_{\GG}} \hat v )\rangle = (-1)^{{\rm Lk}(\hat \varrho^2, {\varsigma^1_{\GG}} )} \Big\vert_{M^4}.
\eea
Related descriptions of link invariants of QFTs can be found in \cite{Putrov2016qdo1612.09298PWY,GuoJW1812.11959} and references therein.
\item 
{\bf {\emph{Pati-Salam GUT-Higgs target manifold and topological defects}}}:\\
The $B(\tilde{\Phi}^{\rm{bi}}) \in \H^2(\O(10)/(\O(6)\times\O(4)), \Z_2)$ can be placed on a 2-surface called $\tilde\varrho^2$, as a {\bf {\emph{charge operator}}}
$\exp( \ii \pi \ointint_{\tilde \varrho^2}   B(\tilde{\Phi}^{\rm{bi}}) ) = 
\exp( \ii \pi \ointint_{\tilde \varrho^2}   w_2(V_{(\O(6)\times\O(4))}))$\footnote{Note that the second Stiefel-Whitney class of associated vector bundle of
the product of orthogonal groups satisfies $w_2(V_{(\O(n)\times\O(m))})=w_2(V_{\O(n)})+w_2(V_{\O(m)})+w_1(V_{\O(n)})w_1(V_{\O(m)})$.
} 
(i.e., symmetry generator)
measures the charge of a preserved $(\O(6)\times\O(4))$ symmetry in the topological defect 
trapped in the target manifold $\O(10)/(\O(6)\times\O(4))$.
There is a topological defect line along a 1d loop called $\varsigma^1_{\PS}$, 
paired up with a 1-connection called $\tilde v$ gives a 1d line operator $\exp(\ii \pi \oint_{\varsigma^1_{\PS}} \tilde v )$
as a {\bf {\emph{ charged object}}}.
The {charge operator} 2-surface ${\tilde \varrho^2}$ can be linked with a {charged} 1d loop $\varsigma^1_{\PS}$ in the 4d spacetime.
Follow the generalized higher global symmetry language \cite{Gaiotto2014kfa1412.5148},
 this nontrivial linking number Lk implies a measurement of $(\O(6)\times\O(4))$ symmetry on the topological defect.
 Precisely, the linking number Lk, manifested as a statistical Berry phase,
is evaluated via the expectation value of path integral:
\bea \label{eq:Lk-2d-1d-PS}
\langle \exp(\ii \pi \ointint_{\tilde \varrho^2} B(\tilde{\Phi}^{\rm{bi}}) ) \cdot \exp(\ii \pi \oint_{\varsigma^1_{\PS}} \tilde v)\rangle = 
(-1)^{{\rm Lk}(\tilde \varrho^2, {\varsigma^1_{\PS}} )}
\Big\vert_{M^3}.
\eea
\item
If we extend the 4d spacetime $t,x,y,z$ to an extra 5th dimension $\varpi$,
the previous 1d loop $\varsigma^1_{\GG}$ trajectory can be a 2d pseudo-worldsheet $\varsigma'^2_{\GG}$ in the 5d $M^5$.
Similarly, the previous 1d loop $\varsigma^1_{\PS}$ trajectory can be a 2d pseudo-worldsheet $\varsigma'^2_{\PS}$ in the 5d $M^5$.
Such two 2d configurations can be linked in 5d, with a linking number:
$$
{{\rm Lk}({\varsigma'^2_{\GG}}, {\varsigma'^2_{\PS}})} \Big\vert_{M^5}.
$$
This describes the link in the extended 5d spacetime of two \emph{charged objects}, 
charged under U(5) and $(\O(6)\times\O(4))$ respectively.\\
\item 
In a parallel story, the \emph{charge operators} (of the above charged objects) are
the 2d 
$B'(\hat{\Phi}^{\rm{bi}})$ operator on $\hat\varrho^2$, 
and 2d 
$B(\tilde{\Phi}^{\rm{bi}})$ surface operator on $\tilde\varrho^2$.
Such two configurations can be linked in 5d, with a linking number:
$$
{{\rm Lk}(B'(\hat{\Phi}^{\rm{bi}}) \text{ on } \hat \varrho^2, B(\tilde{\Phi}^{\rm{bi}}) \text{ on } \tilde\varrho^2)} \Big\vert_{M^5}.
$$
This describes the link in the extended 5d spacetime of two \emph{charge operators}.

{$\bullet$ If we open up the closed 
$ \ointint_{\tilde \varrho^2} B(\tilde{\Phi}^{\rm{bi}}) $ on ${\tilde \varrho^2}$ 
with an open end on the 4d boundary ${M^4}$ of the bulk ${M^5}$, then this open end carries 
a closed 1d loop 
$\oint_{\varsigma^1_{\GG}} \hat v$. 
Their link configuration in 4d corresponds to the earlier \eq{eq:Lk-2d-1d-GG}:
$$
{{\rm Lk}(\hat \varrho^2, {\varsigma^1_{\GG}} )} \Big\vert_{M^4}.
$$
$\bullet$ If we open up the closed $\ointint_{\hat \varrho^2} B'(\hat{\Phi}^{\rm{bi}})$ on ${\hat \varrho^2}$ with an open end 
on the 4d boundary ${M^4}$ of the bulk ${M^5}$, then this open end carries a closed 1d loop 
$\oint_{\varsigma^1_{\PS}} \tilde v$.  Their link configuration in 4d corresponds to the earlier \eq{eq:Lk-2d-1d-PS}:
$$
{{\rm Lk}(\tilde \varrho^2, {\varsigma^1_{\PS}} )} \Big\vert_{M^4}.
$$
}
\end{enumerate}
We leave more of these picturesque discussions and imaginative figures, in a companion work. 
\item
Based on the above observations about the link invariants,
follow Appendix \ref{sec:HomotopyCohomology-NeelVBS}'s logic,
our 4d DQC construction is valid if we introduce a mod 2 class 4d WZW term,
 defined on a 4d boundary $M^4$ of a 5d manifold $M^5$,
schematically in a differential form or de Rham cohomology,
\bea \label{eq:BdCdeRham-bi}
\exp(\ii S^\text{WZW}[\Phi])=\exp(\ii \pi \int_{M^5}B(\tilde{\Phi}^{\rm{bi}})\wedge \dd B'(\hat{\Phi}^{\rm{bi}}) )  \Big\vert_{{M^4=\prt {M^5}}}.
\eea
Recall the footnote \ref{ft:normalization-WZW} about our normalizations of differential forms and cohomology classes.
More precisely, we can improve this to construct WZW in the singular cohomology class:
\bea  \hspace{-5mm}
\label{eq:BdCsingular-bi}
\exp(\ii S^\text{WZW}[\Phi])
{=\exp(\ii \pi \int_{M^5}B(\tilde{\Phi}^{\rm{bi}}) \smile \delta B'(\hat{\Phi}^{\rm{bi}}) )  \Big\vert_{{M^4=\prt {M^5}}}}
{=\exp(\ii  2\pi \int_{M^5}B(\tilde{\Phi}^{\rm{bi}})\smile \Sq^1 B'(\hat{\Phi}^{\rm{bi}}) )  \Big\vert_{{M^4=\prt {M^5}}} }.\;\;\;
\eea
We thus succeed to verify our claims in \eq{eq:BdCdeRham} and \eq{eq:BdCsingular}, while all notations here follow there in \Sec{sec:ElementaryHiggs-WZW}.
\item
Our 4d DQC construction will be supported by a 4d 't Hooft anomaly in the spacetime-internal global symmetry 
$(\Spin \times_{\Z_2^F}\Spin(10))$ on a 4-manifold $M^4$, 
captured by a 5d bulk invertible TQFT \cite{WangWen2018cai1809.11171, WangWenWitten2018qoy1810.00844} living on a 5-manifold $M^5$ with $\prt M^5= M^4$:
\bea \label{eq:w2w3-2}
{\exp(\ii \pi \int_{M^5} w_2(TM)w_3(TM))=}\exp(\ii \pi \int_{M^5} w_2(V_{\SO(10)})w_3(V_{\SO(10)})).
\eea
This 4d 't Hooft anomaly is a mod 2 class global anomaly,
mentioned already in \eq{eq:Spin10-anomaly} and \eq{eq:w2w3}. 
We comment more about the cobordism group data on 
{perturbative local and nonperturbative global anomalies}
in various SMs and GUTs in Appendix \ref{app:Anomalies via Cobordism}.
\end{enumerate}
These conclude our derivation of 4d WZW and 't Hooft anomaly for a candidate 4d DQC for GG-PS GUT transition.

\subsection{Composite GUT-Higgs model within the SM}
\label{sec:CompositeHiggs}

Before analyzing the effect of the 4d WZW term, we will first review how $so(10)$ GUT, GG, PS, and SM can be unified in the same quantum phase diagram by the different condensation pattern of the $\SO(10)$ bivector GUT-Higgs field. 
Follow \Sec{sec:SM-GUT-branch}, for this discussion, we will first turn off the WZW term, assuming that the theory has no additional $w_2w_3$ anomaly.  Starting from the $so(10)$ GUT phase, which has the largest internal symmetry group Spin(10), the GUT-Higgs field can be unified as an $\SO(10)$ bivector field
\bea \label{eq:Phi-bi-phi-2}
\Phi_{ab}^\text{bi}\sim \phi_{a}\phi_{b} \quad\text{(for $a,b=1,2,\cdots,10$)},
\eea
which can be considered as a composition of two $\SO(10)$ vector fields $\phi_a$, where the $\SO(10)$ vector $\phi_a$ can be further considered as a composition of two Weyl fermions $\psi$ 
\bea
\phi_{2a-1}\sim \frac{1}{2}(\psi^\intercal\ii\sigma^2\Gamma_{2a-1}\psi+\text{h.c.}),\quad
\phi_{2a}\sim \frac{1}{2\ii}(\psi^\intercal\ii\sigma^2\Gamma_{2a}\psi-\text{h.c.}),\quad \text{(for $a=1,2,\cdots,5$).}
\eea
Here when two quantum fields $\Phi_A$ and $\Phi_B$ are linearly coupled with each other in the field theory (as source and original fields), 
we denote them in this notation $\Phi_A\sim \Phi_B$, 
such that they are ``dual'' to each other and share exactly the same symmetry properties. There are $16\times 16$ real symmetric matrices $\Gamma_a$ acting in the fermion flavor space, which are determined by the following algebraic relations (for $a,b=1,2,\cdots,5$):
\bea
\{\Gamma_{2a-1},\Gamma_{2b-1}\}=2\delta_{ab},\quad \{\Gamma_{2a},\Gamma_{2b}\}=2\delta_{ab},\quad [\Gamma_{2a-1},\Gamma_{2b}]=0.
\eea
In view of  the above composite construction, we refer to the bivector representation $\Phi^\text{bi}$ as the composite GUT-Higgs field. 

The composite Higgs field contains elementary Higgs components of both $\Phi_{\bf 45}$ and $\Phi_{\bf 54}$, since 
$\mathbf{10}\otimes\mathbf{10}=\mathbf{1}\oplus\mathbf{45}_{\rm A}\oplus\mathbf{54}_{\rm S}$. 
{Follow \eq{eq:Phi-bi-phi},} we introduce the following notations to denote different irreducible representations of the composite GUT-Higgs field (in terms of $\SO(10)$ vector bilinears):
\begin{itemize}
\item $\Tr\Phi^\text{bi}\sim\sum_{a}\phi_a\phi_a$ is equivalent to $\Phi_{\bf 1}$ as the ${\bf 1}_{\rm S}$ of $\SO(10)$.
\item $\hat{\Phi}^\text{bi}:\Phi^\text{bi}_{[a,b]}\sim\frac{1}{2}[\phi_a,\phi_b]$  is equivalent to $\Phi_{\bf 45}$ as the ${\bf 45}_{\rm A}$, 
antisymmetric (A) part of $\mathbf{10}\otimes\mathbf{10}$, of $\SO(10)$.
\item $\tilde{\Phi}^\text{bi}:\Phi^\text{bi}_{\{a,b\}}-\frac{1}{10}\Tr\Phi^\text{bi}\delta_{ab}\sim\frac{1}{2}\{\phi_a,\phi_b\}-\frac{1}{10}\sum_{c}\phi_c\phi_c\delta_{ab}$  is equivalent to $\Phi_{\bf 54}$ as the ${\bf 54}_{\rm S}$, symmetric (S) part of $\mathbf{10}\otimes\mathbf{10}$, of $\SO(10)$.
\end{itemize}

The competition between $\tilde{\Phi}^\text{bi}$ and $\hat{\Phi}^\text{bi}$ condensation leads to different GUT or SM phases in the phase diagram. We enumerate all the symmetry breaking patterns (below ``$\to$'' means ``breaking to'')
as follows:
\begin{enumerate}
\item
$\Spin(10)\to\frac{\Spin(6)\times\Spin(4)}{\Z_2}=\frac{\SU(4)\times\SU(2)_{\rm L}\times\SU(2)_{\rm R}}{\Z_2}$ by condensing $\tilde{\Phi}^\text{bi}$ 
(the $\mathbf{54}_{\rm S}$ symmetric representation) to the following specific configuration
{in the symmetric rank-10 bi-vector matrix form:}
\bea\label{eq: 54config}
\langle\tilde{\Phi}^\text{bi}\rangle=\Big(-3 \sum_{a=1}^{4}+2 \sum_{a=5}^{10}\Big)\Phi^\text{bi}_{\{a,a\}}=\phi^\intercal\bpm -3\cdot\one_{2\times 2} & \\  &2\cdot\one_{3\times 3}\epm\otimes \sigma^0\phi\in\frac{\O(10)}{\O(6)\times\O(4)}.
\eea
The GUT-Higgs field $\tilde{\Phi}$ discriminates the $\SO(4)$ vector $(\phi_1,\phi_2,\phi_3,\phi_4)$ from the $\SO(6)$ vector $(\phi_5,\phi_6,\phi_7,\phi_8,\phi_9,\phi_{10})$, which breaks $\Spin(10)$ down to $\frac{\Spin(6)\times\Spin(4)}{\Z_2}$ realizing the Pati-Salam symmetry $G_{\mathrm{PS}_2}$. The 16 Weyl fermions split as
$\mathbf{16}\sim(\mathbf{4},\mathbf{2},\mathbf{1})_{\rm L} \oplus (\overline{\mathbf{4}},\mathbf{1},\mathbf{2})_{\rm R}$ 
under $\frac{\SU(4)\times\SU(2)_{\rm L}\times\SU(2)_{\rm R}}{\Z_2}$.\footnote{Recall 
in footnote \ref{footnote:LR}, about the left or right spinors, 
the $\rm L$/$\rm R$ notations here are for the internal-symmetry's spinors, 
while the $L/R$ notations are for the spacetime-symmetry's Weyl spinors.}
The $\rm L$/$\rm R$ sectors are distinguished by the operator 
\bea\label{eq: chi}
\chi=\psi^\dagger (\prod_{a=1}^{10}\Gamma_a)\psi=\pm1.
\eea
Let $\rho_{[a,b]}=\frac{1}{2\ii}[\phi_a,\phi_b]$ be the $\SU(4)$ generators (for $a,b=5,6,\cdots,10$). Using algebraic relations, we can check that in the ${\rm L}$ sector, 
$\SU(4)$ acts as $\psi_{\rm L} \mapsto e^{\ii \rho_{[a,b]}}\psi_{\rm L} e^{-\ii \rho_{[a,b]}}$, matching the $\mathbf{4}$ representation; and in the ${\rm R}$ sector, 
$\SU(4)$ acts as {$\psi_{\rm R}\mapsto  e^{-\ii \rho_{[a,b]}^*}\psi_{\rm R} e^{\ii \rho_{[a,b]}^*}$}, matching the $\overline{\mathbf{4}}$ representation.

\item
$\Spin(10)\to\SU(5)\times \Z_{4,X}$ by condensing $\hat{\Phi}^\text{bi}$ (the $\mathbf{45}_{\rm A}$ antisymmetric representation) to the following specific configuration
{in the antisymmetric rank-10 bi-vector matrix form:}
\bea\label{eq: 45config}
\langle\hat{\Phi}^\text{bi}\rangle=\sum_{a=1}^{5}\Phi^\text{bi}_{[2a-1,2a]}=-\frac{1}{2}\phi^\intercal\one_{5\times 5}\otimes\ii\sigma^2\phi\in\frac{\O(10)}{\U(5)}.
\eea
If we combine the $\SO(10)$ vector 
{$\phi_b$ (for $b=1,2,\cdots,10$)}
into a 5-component complex vector $\varphi_a=(\phi_{2a-1}+\ii\phi_{2a})/\sqrt{2}$ (for $a=1,2,\cdots,5$), $\varphi$ would transform as the 
{$\mathbf{5}_1$ under $\U(5)=\frac{\SU(5)\times \U(1)}{\Z_5}$}\footnote{
{\Refe{WangYou2111GEQC, HAHSIV} points out the subtle differences between different non-isomorphic versions of
U(5) Lie groups (and their corresponding gauge theories) that we should refine and redefine them
as several $\U(5)_{\hat q}$ with ${\hat q} \in \Z$:
\bea \label{eq:U5q}
\U(5)_{\hat q} \equiv \frac{\SU(5) \times \U(1)_{\hat q} }{\Z_5} \equiv
\{ (g, \e^{\ii \theta}) \in \SU(5) \times \U(1) \big\vert  ( \e^{\ii \frac{2 \pi n}{5}} \mathbb{I}, 1) \sim ( \mathbb{I}, \e^{\ii \frac{2 \pi n {\hat q}}{5}} ), n\in \Z_5 \}
\eea
where we use two data $(g, \e^{\ii \theta})$ to label the $\SU(5) \times \U(1)$ group elements respectively,
while we identify $( \e^{\ii \frac{2 \pi n}{5}} \mathbb{I}, 1) \sim ( \mathbb{I}, \e^{\ii \frac{2 \pi n {\hat q}}{5}} )$ for $n\in \Z_5$,
with a rank-5 identity matrix $\mathbb{I}$. They have the group isomorphisms between different $\hat q$ as
$$\U(5)_{\hat q}  \cong \U(5)_{-\hat q}  \cong \U(5)_{5 m \pm \hat q}.$$ 
See further discussions in footnote \ref{ft:U(5)q-II}. Whenever we mention $\U(5) \subset \SO(10)$, we really require $\U(5)_{ \hat q=1,4} \subset \SO(10)$.
In contrast, whenever we mention $\U(5) \subset \Spin(10)$, we really require $\U(5)_{ \hat q=2,3} \subset \Spin(10)$.}
\label{ft:U(5)q-I}
}
in $\SO(10)$. 
The GUT-Higgs field $\hat{\Phi}^\text{bi}=\sum_{a=1}^{5}\varphi_a^\dagger \varphi_a$ itself defines the generator of the $\U(1)_X$ group, whose $\Z_4$ subgroup defines $\Z_{4,X}$. The 16 Weyl fermions split as
$\mathbf{16}\sim \overline{\mathbf{5}}_1\oplus \mathbf{10}_1\oplus \mathbf{1}_1$
under $\SU(5)\times \Z_{4,X}$. The $\Z_{4,X}$ generator in the $\Spin(10)$ spinor representation is given by
\bea\label{eq:q''}
q_X=\sum_{a=1}^{5}\psi^\dagger \ii\Gamma_{2a-1}\Gamma_{2a}\psi.
\eea
By diagonalizing $q_X$ operator, we indeed found five-fold eigenvalues of $-3$, ten-fold eigenvalues of $1$ and a one-fold eigenvalue of $5$. After mod 4, they all correspond to charge $1$ under $\Z_{4,X}$. Further investigate the representation of $\SU(5)$ generators in each $q_X$-charge sectors, we can confirm that the $q_X=-3$ sector is indeed in the anti-fundamental representation $\overline{\mathbf{5}}$ and 
{so on to form
$\mathbf{16}\sim \overline{\mathbf{5}}_{-3}\oplus \mathbf{10}_1\oplus \mathbf{1}_5$.}

\item $\Spin(10)\to\frac{\SU(3)\times\SU(2)\times\U(1)_{\tilde{Y}}}{\Z_6}\times\Z_{4,X}$ by simultaneously condensing $\tilde{\Phi}^\text{bi}$ and $\hat{\Phi}^\text{bi}$ 
(both $\mathbf{54}_{\rm S}$ and $\mathbf{45}_{\rm A}$ representations) 
to configurations specified in \Eqn{eq: 54config} and \Eq{eq: 45config}. The unbroken symmetry group is generated by the sub-algebra of $so(10)$ that commute with  both GUT-Higgs condensates $\langle\tilde{\Phi}^\text{bi}\rangle$ and $\langle\hat{\Phi}^\text{bi}\rangle$, which must take the form of
\bea
\phi^\intercal\bpm\ii {\rm A}_{2\times 2} & \\ & \ii {\rm A}_{3\times 3}\epm\otimes\sigma^0\phi \text{ or }
\phi^\intercal\bpm {\rm S}_{2\times 2} & \\ & {\rm S}_{3\times 3}\epm\otimes\sigma^2\phi,
\eea
where ${\rm A}_{n\times n}=-{\rm A}_{n\times n}^\intercal\in\R_{n\times n}$ are real antisymmetric matrices and 
${\rm S}_{n\times n}={\rm S}_{n\times n}^\intercal\in\R_{n\times n}$ are real symmetric matrices. They can be combined in the complex representation as
\bea
\varphi^\dagger \bpm {\rm S}_{2\times 2}+\ii {\rm A}_{2\times 2} &\\ & {\rm S}_{3\times 3}+\ii {\rm A}_{3\times 3}\epm\varphi = \varphi^\dagger\bpm {\rm H}_{2\times 2} &\\ & {\rm H}_{3\times 3}\epm\varphi,
\eea
such that ${\rm H}_{n\times n}={\rm H}_{n\times n}^\dagger \in \C_{n\times n}$ are complex Hermitian matrices. 
There is no traceless condition imposed on ${\rm H}_{3\times 3}$ and ${\rm H}_{2\times 2}$ 
and they act independently in each subspace, 
so they generate the $\U(3)\times\U(2)$ subgroup of $\U(5)$, 
which is further a subgroup of $\SO(10)$. The two $\U(1)$ subgroups of $\U(3)$ and $\U(2)$ are generated by $\sum_{a=3}^{5}\varphi_a^\dagger \varphi_a$ and  $\sum_{a=1}^{2}\varphi_a^\dagger \varphi_a$ respectively. Since the $\U(1)_X$ (or $\Z_{4,X}$) generator has already been identified  as $\sum_{a=1}^{5}\varphi_a^\dagger \varphi_a$, so the $\U(1)_{\tilde{Y}}$ generator must be given by the remaining $\U(1)$ generator $\frac{1}{2}(-3\sum_{a=1}^{2}+2\sum_{a=3}^{5})\varphi_a^\dagger \varphi_a$, which is represented in the $\Spin(10)$ spinor representation as
\bea\label{eq:q'}
q_{\tilde{Y}}=\frac{1}{2}\Big(-3\sum_{a=1}^{2}+2\sum_{a=3}^{5}\Big)\psi^\dagger \ii\Gamma_{2a-1}\Gamma_{2a}\psi.
\eea

By diagonalizing  $\chi$, $q_{\tilde{Y}}$ and $q_{X}$ operators jointly (defined in Eqns.\,\Eq{eq: chi}, \Eq{eq:q'}, \Eq{eq:q''}), 
we can classify the 16 Weyl fermions $\psi$ {(actually they are all in the 
left-handed spacetime Weyl spinor $\psi_L$ basis)}
by the quantum numbers as follows
\bea
\begin{array}{cccccc}
\hline
\U(1)_{\tilde{Y}} & \U(1)_{X} & \text{internal } {\rm L}/{\rm R} & \SU(2)_{\rm L}^z & \SU(2)_{\rm R}^z & \psi\\
\hline
2 & -3 & {\rm R} & 0 & 1 & \bar{d}_R\\
-3 & -3 & {\rm L} & 1 & 0 & \nu_L\\
-3 & -3 & {\rm L} & -1 & 0 & e_L\\
1 & 1 & {\rm L} & 1 & 0 & u_L\\
1 & 1 & {\rm L} & -1 & 0 & d_L\\
-4 & 1 & {\rm R} & 0 & -1 & \bar{u}_R\\
6 & 1 & {\rm R} & 0 & 1 & \bar{e}_R\\
0 & 5 & {\rm R} & 0 & -1 & \bar{\nu}_R\\
\hline
\end{array},
\eea
matching all the fermion contents in the SM (see \Table{table:SMfermion}).
\end{enumerate}

{\bf No bilinear mass generation by bivector GUT-Higgs}: Unlike the SM-Higgs that generates a bilinear mass for SM Weyl fermions, the GUT-Higgs 
in {\bf 45} and {\bf 54} do not generate a bilinear mass for SM Weyl fermions. Because the $\SO(10)$ bivector GUT-Higgs field ${\Phi}^\text{bi}$ corresponds to four-fermion operators, which is supposed to be perturbatively irrelevant. Even if it condenses, it is not expected to gap out the Weyl fermions if its vaccum expectation value is small (but it will Higgs down the gauge group), so the theory remains gapless in the fermion sector in all phases. However, sufficiently strong Higgs condensation of $\Tr\Phi^\text{bi}$ (or $\Phi_{\bf 1}$ equivalently) can lead to symmetric mass generation (SMG)\cite{FidkowskifSPT2,Wang2013ytaJW1307.7480,Wang2018ugfJW1807.05998,YouHeXuVishwanath1705.09313, YouHeVishwanathXu1711.00863,Eichten1985ftPreskill1986,Wen2013ppa1305.1045,You2014oaaYouBenTovXu1402.4151, YX14124784,BenTov2015graZee1505.04312,Kikukawa2017ngf1710.11618,Catterall2020fep, CatterallTogaButt2101.01026,RazamatTong2009.05037, Tong2104.03997} as discussed previously.

\subsection{Fragmentary
GUT-Higgs Liquid model beyond the SM}
\label{sec:FragmentaryHiggs} 

\subsubsection{Low-energy descriptions for the WZW theory}

The WZW term and its associated $w_2w_3$ global anomaly can significantly modify  the dynamics in the GUT-Higgs sector. There are several possibilities for the low-energy fate of the WZW theory:
\begin{enumerate}
\item {\bf Spontaneous symmetry breaking (SSB).} 
The $\SO(10)$ internal symmetry 
of WZW term (or Spin(10) for the full modified $so(10)$ GUT)
is spontaneously broken by GUT-Higgs condensation. Within this scenario, there are a few different symmetry breaking patterns relevant to our discussion (recall \Sec{sec:SM-GUT-branch}):
\begin{itemize}
\item $\langle\Phi_{\bf 45}\rangle\neq 0$, the $so(10)$ GUT is Higgs down to the $su(5)$ GUT.
\item $\langle\Phi_{\bf 54}\rangle\neq 0$, the $so(10)$ GUT is Higgs down to the PS model.
\item $\langle\Phi_{\bf 45}\rangle\neq 0$ and $\langle\Phi_{\bf 54}\rangle\neq 0$, the $so(10)$ GUT is Higgs down to the SM.
\end{itemize}
{In all three cases, the $w_2w_3(V_{\SO(10)})$ anomaly is matched by symmetry breaking the Spin(10) down to the GG, PS and 
SM groups.\footnote{However, the $\Z_2$ class $w_2w_3(V_{\SO(10)})$ anomaly of SO(10) bundle is split to different kinds of $w_2w_3$ anomalies of SO(6) and SO(4) bundles in
the PS symmetry group: More precisely,
see Appendix \ref{app:Anomalies via Cobordism} in detail,
$w_2(V_{\SO(10)})w_3(V_{\SO(10)})
=w_2(V_{\SO(6)})w_3(V_{\SO(6)})
+ w_2(V_{\SO(4)})w_3(V_{\SO(4)}) + 
w_2(V_{\SO(6)})w_3(V_{\SO(4)})
+ w_2(V_{\SO(4)})w_3(V_{\SO(6)})
 \mod 2$, where the crossing term 
 $w_2(V_{\SO(6)})w_3(V_{\SO(4)})
+ w_2(V_{\SO(4)})w_3(V_{\SO(6)})$ may or may not survive depending on whether we include additional time-reversal $T$ or $CP$ type of discrete symmetries protection or not.
}}
The resulting vacua is in the same quantum phase as the corresponding vacua in the absence of the WZW term. 

\item The $\SO(10)$ symmetry remains unbroken, and the $w_2w_3$ anomaly persists to low-energy. The low-energy effective theory must saturate the anomaly requirement, which further leads to several different possibilities:
\begin{enumerate}
\item {\bf WZW conformal field theory (CFT):} The WZW theory flows to a non-trivial CFT fixed point, where the GUT-Higgs field $\Phi$ remains gapless and disordered (not condensing), and also does not deconfine into fragmented excitations.
\item {\bf Deconfined quantum criticality (DQC):} The GUT-Higgs field $\Phi$ deconfines into fragmented excitations: partons and emergent gauge fields, which are new particles beyond the SM. The low-energy physics will be described by new quantum electrodynamics (QED$'$) or quantum chromodynamics (QCD$'$) sectors. In any case, the total gauge group must be enlarged to include the emergent gauge structure of partons, which is a phenomenon called gauge enhanced quantum criticality (GEQC) \cite{WangYouZheng1910.14664}. This can be viewed as the generalization of the deconfined quantum criticality (DQC)\cite{SenthildQCP0311326,Vishwanath2013Physics,Sandvik2007Evidence,WangSenthildQCP1703.02426} to gauge-Higgs models. Possible field theory descriptions of the DQC can be classified by the parton statistics as:
\begin{itemize}
\item {\bf Fermionic parton} theory, where the fractionalized particles in the emergent matter sector are fermions, which is the focus of our following work.
\item {\bf Bosonic parton} theory, where the fractionalized particles in the emergent matter sector are bosons.
\end{itemize}
It is possible that two seemly different descriptions (e.g. fermionic  v.s. bosonic parton theories) may be related by dualities, as discussed in\cite{Zou2018Symmetry,WangSenthildQCP1703.02426}. In this scenario, the $w_2w_3$ anomaly should be matched either by the anomalous fermionic matter or by a non-trivial $\theta$-term of the emergent gauge field.

\item {\bf Topological order with low-energy non-invertible TQFT}: The $w_2w_3$ anomaly could also be matched by a certain 4d topological order. A simplest possibility is the $\Z_2$-gauge theory topological order (more precisely, generated by \emph{dynamical spin structures}), 
which can be considered as a descendent of the DQC when the emergent gauge group is reduced to $\Z_2$ by some further Higgsing.

\end{enumerate}

\end{enumerate}

Among the above possibilities: 1. The SSB scenario in the WZW theory has no substantial difference with our previous discussions without the WZW term, which will not be repeated here. 2.(a) The WZW CFT is a non-trivial possibility, which the authors are not aware of suitable theoretical tools to study it, which will thus be left for future exploration. 2.(b) The DQC scenario will be the focus of the following discussion. In particular, we will consider a {\bf QED$'_4$ theory with fermionic partons} as the effective field theory description. The WZW theory could potentially admits dual bosonic parton descriptions as well, but we will also leave this possibility for future study. 2.(c) The topological order scenario could be derived from the DQC scenario, which will also be left for future study.

\subsubsection{{Dirac} Fermionic Parton Theory and a Double-Spin structure DSpin within a modified $so(10)$ GUT}
\label{sec:QED4Parton}

{Here we propose a fermionic parton construction for the WZW term in \Sec{sec:HomotopyCohomology}. 
We propose that WZW term \Eqn{eq:BdCdeRham} can also be viewed as a low-energy description of this  {Dirac} fermionic parton theory}
with an action: 
\begin{equation}\label{eq: QED4}
S_{\text{QED}'_4}{[\xi,\bar{\xi},a,\Phi]}=\int_{M^4} \bar{\xi}(\ii\gamma^\mu D_\mu-\tilde{\Phi}^\text{bi}-\ii {\gamma^{\text{FIVE}}}\hat{\Phi}^\text{bi})\xi \; \dd^4x.
\end{equation}
{We will soon argue that importantly the fermion parity $\Z_2^{F'}$ 
of this  {Dirac} fermionic parton $\xi$ requires to be different from the original fermion parity $\Z_2^{F}$ of the standard model or GUT fermions $\psi$.
Namely, we will soon introduce a new kind of spin structure with two distinct fermion parities, which we name it formally a double spin structure: 
\bea
\DSpin \equiv (\Z_2^{F} \times \Z_2^{F'}) \rtimes \SO.
\eea}
The theory contains the following ingredients:
\begin{enumerate}[leftmargin=.mm]
\item There are 10 Dirac fermions $\xi$ forming the $\mathbf{10}$  (vector representation) of $\SO(10)$. Here $\gamma^\mu$ are the standard rank-4 $\gamma$ matrices of 4-component Dirac fermions with 
 {$\gamma^{\text{FIVE}}=\ii \gamma^0\gamma^1\gamma^2\gamma^3$} and  $\bar{\xi}=\xi^\dagger\gamma^0$.
\item The covariant derivative $D_\mu=\nabla_\mu-\ii a_\mu-\ii g A_\mu$ contains the minimal coupling of the fermionic parton $\xi$ to a new emergent dynamical 
{$\U(1)'$} gauge field $a_\mu$, as well as
the minimal coupling to the $\SO(10)$ gauge field $A_\mu$ 
{(which is part of the Spin(10) gauge field in the conventional $so(10)$ GUT in \Sec{sec:SM-GUT-branch})}. 
We may treat the $\SO(10)$ gauge field $A_\mu$ as a background  field for now, and discuss how it can be gauged later.
\item The GUT-Higgs field $\Phi$ is written as its $10\times10$ matrix representation $\Phi^\text{bi}$ of the $\SO(10)$ bivector form. It couples to the fermionic partons by taking its traceless symmetric component $\tilde{\Phi}^\text{bi}$ (the ${\bf 54}$ of $\SO(10)$) as the vector mass of $\xi$ and its antisymmetric component $\hat{\Phi}^\text{bi}$ (the ${\bf 45}$ of $\SO(10)$) as the axial mass of $\xi$. In this way, the $\SO(10)$ bivector GUT-Higgs boson effectively deconfines into two $\SO(10)$ vector fermions: $\Phi^\text{bi}_{ab}\sim\xi^\dagger_a\xi_b$.\footnote{ {If this theory has 't Hooft anomaly in $G$, 
it cannot be trivially gapped by preserving the $G$-symmetry. Since we like to construct fermion parton theory 
${\text{QED}'_4}$ \eq{eq: QED4} to saturate the $w_2 w_3$ anomaly of $\SO(10)$ symmetry (or 
$\Spin  \times_{\Z_2^F}\Spin(10)$ symmetry), we should forbid the \eq{eq: QED4} 
to get any quadratic mass term that preserves the $\SO(10)$. It turns out that the ${\text{QED}'_4}$ have
$\U(1)'$, CP$'$, and T$'$ symmetries that can forbid any $\SO(10)$ symmetric quadratic mass term:\\
(i) The $\U(1)'$ symmetry: $\xi\to e^{\ii \theta}\xi$ forbids any Majorana mass of the form $\xi_\text{L/R}^{\rm T}\ii\sigma^2\xi_\text{L/R}$ that potentially gaps out the Dirac fermion 
(written as two Weyl fermions: $\xi = \xi_\text{L} + \xi_\text{R}$).\\ 
(ii) The CP$'$ symmetry $\Z_2^{\text{CP}'}$: $\xi(t,\vec{x})\to\gamma^0\gamma^\text{FIVE}\xi^*(t,-\vec{x})$
forbids the vector $\bar{\xi}\xi$ mass:
$\bar{\xi}\xi\to-\bar{\xi}\xi$. \\
(iii) The T$'$ symmetry $\Z_2^{\text{T}'}$: 
$\xi(t,\vec{x})\to\mathcal{K} \gamma^0\gamma^\text{FIVE}\xi(-t,\vec{x})$
forbids the axial $\ii \bar{\xi}\gamma^\text{FIVE}\xi$ mass:
$\ii \bar{\xi}\gamma^\text{FIVE}\xi\to-\ii \bar{\xi}\gamma^\text{FIVE}\xi$. }}
\item 
In the ${\text{QED}'_4}$ theory $S_{{\text{QED}'_4}}$, the GUT-Higgs field fractionalizes into gapless fermionic partons with emergent $\U(1)'$ gauge interactions. The situation is similar to the $\U(1)$ Dirac spin liquid\cite{Zhou2017Quantum,Savary2017Quantum} discussed in the condensed matter physics context. Therefore we may also call this QED$'_4$ theory as the Fragmentary GUT-Higgs Liquid model.\footnote{ {Because the order-parameter target manifold in our construction
involves a Grassmannian manifold $(\frac{\O(m+n)}{\O(m)\times\O(n)}) \equiv {\rm Gr}(m, m+n)$, the corresponding GUT-Higgs Liquid may also be called Grassmannian Liquid by some condensed matter people.}}

\item  {The name of ``fragmentary'' GUT-Higgs liquid (\Sec{sec:QED4Parton}) 
is meant to distinguish and emphasize the fractionalization of bivector field as
$\Phi_{ab} \sim \xi_a^\dagger \xi_b$ of fermionic partons in \eq{eq: QED4}, instead of 
$\Phi_{ab}^\text{bi}\sim \phi_{a}\phi_{b}$
of the bosonic partons in \eq{eq:Phi-bi-phi} and \eq{eq:Phi-bi-phi-2} for the ``composite'' GUT-Higgs model (\Sec{sec:CompositeHiggs}).}
\end{enumerate}

We first argue that the QED$'_4$ theory (without a $\theta$-term) in \Eqn{eq: QED4} saturates the same $w_2w_3$ anomaly as the WZW term  in \Sec{sec:HomotopyCohomology}. 
The starting point is to identify that the spacetime-internal symmetry (here ${\Spin'\times_{\Z_2^{F'}}\U(1)'}$) and the gauge group (here $\SO(10)$) of the fermionic parton theory is
\begin{equation}
\label{eq:GQED4}
{G_\text{QED$'_4$}\equiv {{\Spin' \times_{[\Z_2^{F'}]}[\U(1)']}\times {  \SO(10)} } \equiv \Spin^{c'} \times\SO(10),}
\end{equation}
with fermions in the ${\bf 10}_1$ representation of $\SO(10)$ and $\U(1)'$. 
{Notice that we use the prime notation to indicate that
those groups contain the new fermion parity ${\Z_2^{F'}}$.
Such that $\U(1)'\supset {\Z_2^{F'}}$, $\Spin' \supset {\Z_2^{F'}}$, and $\Spin^{c'} \supset {\Z_2^{F'}}$.
Here we use the bracket notation around $[\U(1)']$ to indicate that this $\U(1)'$ is \emph{dynamically gauged} 
eventually in terms of the emergent gauge fields near the quantum criticality.
In other words, the new fermion parity ${\Z_2^{F'}}$ must also be dynamically gauged because $[\U(1)']  \supset [{\Z_2^{F'}}]$.}

{How do we reconcile the Spin structure (of the familiar SM and GUT in \Sec{sec:EFTHiggs}) and the $\Spin'$ structure (of this new fermion parton theory \eq{eq: QED4}) in the full theory? After all, we have to place a full theory on some curved spacetime with a single unified geometric structure. 
The full spacetime-internal structure of this modified $so(10)$-GUT, that we require to include 
$\Spin \times_{\Z_2^F}\Spin(10)$ of \eq{eq:spacetime-internal-embedding-1} and $\Spin^{c'} \times\SO(10)$ of \eq{eq:GQED4} as subgroups, 
turns out to be:\footnote{
{Again we use the bracket notation around $[\U(1)']$ and $[\Z_2^{F'}]$ to 
indicate that they must be \emph{dynamically gauged}. 
Although the Spin(10) is also \emph{dynamically gauged} in the GUT,
the Spin(10) may still be treated as a \emph{global symmetry} in the context of quantum criticality of the internal flavor symmetry
of fermions in the condensed matter system.
However, the $[\U(1)']$ and $[\Z_2^{F'}]$ must be \emph{dynamically gauged} due to their roles at quantum criticality, 
regardless whether the Spin(10) is gauged or not.
In summary, there is a hierarchy of gauging: the brackets $[..]$ implies those degrees of freedom have a higher priority to be gauged.
}}
\bea \label{eq:Gmodifiedso10}
G^{\text{modified}}_\text{$so(10)$-GUT}\equiv
(\DSpin  \times_{\Z_2^F}\Spin(10)) \times_{[\Z_2^{F'}]}[\U(1)'],
\eea
where
we implement the early advertised double spin structure 
$\DSpin \equiv (\Z_2^{F} \times \Z_2^{F'}) \rtimes \SO$ structure. 
We leave the detail construction of this full spacetime-internal
$G^{\text{modified}}_\text{$so(10)$-GUT}$ symmetry based on the group extension in the footnote remark\footnote{
{
Here are some comments about our construction of spacetime-internal symmetry. More details are in Appendix \ref{sec:DSpin}.\\
First, the $\psi$ fermion  in the ${\bf 16}$ of $\Spin(10)$ requires a fermion parity ${\Z_2^F}$,
while the $\xi$ fermion in the ${\bf 10}$ of $\SO(10)$ requires another new fermion parity ${\Z_2^{F'}}$.
Next, both $\psi$ and $\xi$ fermions require the common $\SO \times \SO(10)$ structure (as the quotient group of the total symmetry group),
because they share the same bosonic part of spacetime rotational special orthogonal symmetry group SO, and their SO(10) gauge fields are the same.
However, the $\psi$ fermion requires a total structure $\Spin \times_{\Z_2^F}\Spin(10)$ under the short exact sequence:
$1 \to \Z_2^F \to \Spin \times_{\Z_2^F}\Spin(10) \to \SO \times \SO(10) \to 1$;
the $\xi$ fermion requires a different total structure ${{\Spin' \times\SO(10)}}$ under the short exact sequence:
$1 \to \Z_2^{F'} \to {{\Spin' \times\SO(10)}} \to \SO \times \SO(10) \to 1$. Their structures cannot be compatible under the same fermion parity,
thus we require to introduce two fermion parities
with the $\DSpin \equiv (\Z_2^{F} \times \Z_2^{F'}) \rtimes \SO$ structure under
$1 \to \Z_2^F \times  \Z_2^{F'} \to \DSpin \to \SO \to 1$ such that $\DSpin \supset \Spin = \Z_2^{F} \rtimes \SO$ and 
$\DSpin \supset \Spin' = \Z_2^{F'} \rtimes \SO$.
The above short exact sequences can be combined into the following 
group extensions:
\begin{equation}
\label{eq:extension-web}
\xymatrix{
&  &1 \ar[d] & 1 \ar[d]\\
&  &{{ \mathbb{Z}_{2}^{F'}}}  \ar[d]  & {{ \mathbb{Z}_{2}^{F'}}}  \ar[d]\\
1\ar[r]&  \Z_2^F \ar[r]  & {{{(\DSpin  \times_{\Z_2^F}\Spin(10))} }} \ar[d] \ar[r] & 
 {{{\Spin' \times\SO(10)}}} \ar[r] \ar[d] &1\\
1\ar[r]& \Z_2^F \ar[r] & \Spin \times_{\Z_2^F}\Spin(10) \ar[r]^{}  \ar[d]& \SO \times \SO(10) \ar[r] \ar[d] &1\\
&  &1 & 1
}.
\end{equation}
This total extended spacetime-internal ${(\DSpin  \times_{\Z_2^F}\Spin(10))}$ group is compatible with both fermionic spectrum restrictions for $\psi$ and $\xi$.
By modifying the ${{ \mathbb{Z}_{2}^{F'}}}$ into $\U(1)'$ in the web of \eq{eq:extension-web},
we thus obtain the
$G^{\text{modified}}_\text{$so(10)$-GUT}\equiv
(\DSpin  \times_{\Z_2^F}\Spin(10)) \times_{\Z_2^{F'}}\U(1)'$ in \eq{eq:Gmodifiedso10}.\\
Related to the DSpin structure, by including an extra discrete symmetry such as a time-reversal symmetry,
the literatures also discover the structures known as DPin \cite{Kaidi2019tyfJulioTachikawa1911.11780} and EPin \cite{Wan2019sooWWZHAHSII1912.13504} structures, 
see also an interpretation via the regularized quantum many-body model \cite{PrakashJW2011.13921}. 
See more elaborations in Appendix \ref{sec:DSpin}.}
}
and the Appendix \ref{sec:DSpin}.}

The $\U(1)'$ group is free of anomaly, which is consistent with the fact that this emergent $\U(1)'$ structure can be gauged. 
Gauging $\U(1)'$ out of $\Spin^{c'} \times\SO(10)$
removes the spin structure of the fermion theory, allowing the gauge theory to be placed on non-spin manifolds. So the resulting theory is a bosonic theory with 
{an $\SO \times \SO(10)$ symmetry. It is expected that the spacetime $\SO$ group should carry the $w_2w_3$ anomaly}, and 
 {the anomaly could only originate from the fermionic partons in the QED$'_4$ theory.} 

To check the anomaly in the fermion sector, we first turn off the Higgs coupling (as it does not affect the anomaly analysis), such that the theory becomes as simple as $\int_{M^4}\bar{\xi}\gamma^\mu D_\mu\xi \,\dd^4 x$. Without coupling to the GUT-Higgs field, the theory has an enlarged 
{$\SU(2)'$} gauge group, generated by $\xi^\dagger \xi$, $\Re \xi^\intercal\gamma^5\xi$, $\Im \xi^\intercal\gamma^5\xi$, among which $\xi^\dagger \xi$ generates the $\U(1)'$ gauge group as a subgroup of $\SU(2)'$. With the enlarged $\SU(2)'$ gauge group, the fermionic parton theory is promoted from a QED$'_4$ theory to a QCD$'_4$ theory (without enlarging the fermion content), whose group structure is\footnote{
{Similar
to \eq{eq:extension-web},
by modifying the ${{ \mathbb{Z}_{2}^{F'}}}$ into $\SU(2)'$ in the web,
we thus obtain a modification on \eq{eq:Gmodifiedso10} into
\bea
G^{\text{modified}}_\text{$so(10)$-GUT}\equiv
(\DSpin  \times_{\Z_2^F}\Spin(10)) \times_{[\Z_2^{F'}]}[\SU(2)'],
\eea  
that has a quotient group $G_\text{QCD$'_4$}  \equiv \Spin^{h'} \times\SO(10)$ in \eq{eq:GQCD4}.
See more elaborations in Appendix \ref{sec:DSpin}.}}
\begin{equation}
\label{eq:GQCD4}
G_\text{QCD$'_4$}={{\Spin' \times_{[\Z_2^{F'}]}[\SU(2)']}\times \SO(10)} \equiv \Spin^{h'} \times\SO(10),
\end{equation}
 {the original Dirac fermion $\xi$ is in ${\bf 2}_{L} \oplus {\bf 2}_{R}$ of Spin(1,3) and $({1},{\bf 10})$ of $\U(1)'\times\SO(10)$,
while now the fermion $\xi$ becomes in ${\bf 2}_{L}$ of Spin(1,3) and
in the $({\bf 2},{\bf 10})$ representation of $\SU(2)'\times\SO(10)$.}
{Again we use the bracket notation around $[\SU(2)']$ and $[\Z_2^{F'}]$ to indicate that they must be dynamically gauged near the criticality.}
This QED$'_4$ to QCD$'_4$ promotion does not change the anomaly structure, 
because the $\SU(2)'$ group is still anomaly-free.
{Namely,
there are only two possible combinations of nonperturbative global anomalies out of the cobordism classification for ${\Spin' \times_{\Z_2^{F'}}\SU(2)'}$ symmetry
given by $\TP_5({\Spin' \times_{\Z_2^{F'}}\SU(2)'})=\Z_2^2$
\cite{WangWen2018cai1809.11171, WangWenWitten2018qoy1810.00844, WanWang2018bns1812.11967}: 
\\[-10mm]
\begin{enumerate}[leftmargin=.mm]
\item
No Witten SU(2)$'$ anomaly \cite{Witten1982fp}: 
Given that there are even number (ten) of fundamental fermions ${\bf 2}$ of $\SU(2)'$, so $10 \mod 2 = 0$. 
\item No new SU(2)$'$ anomaly \cite{WangWen2018cai1809.11171}: 
Given that there is no ${\bf 4}$ of $\SU(2)'$ fermions, so $0 \mod 2 = 0$. 
\end{enumerate}}
So the anomaly is still contained in the $\SO(10)$ group out of $G_\text{QCD$'_4$}= \Spin^{h'} \times\SO(10)$. 
To match the $w_2w_3$ anomaly, we make a connection to the recently discovered new $\SU(2)$ anomaly 
\cite{WangWenWitten2018qoy1810.00844} by the following trick on the $\SO \times \SO(10)$ sector: we first embed $\SU(2)' \times\SO(10)$ in $\Sp(10)$ and use a sequence of maximal 
{\emph{special} (S) \emph{or regular} (R) \emph{Lie subalgebra} \cite{1912.10969LieART}} decomposition $\Sp(10)\hookleftarrow\Sp(2)\times\Sp(8)\hookleftarrow\SU(2)''\times\Sp(8)$ to show that a different $\SU(2)''$ subgroup carries the $w_2w_3$ anomaly. Under the embedding, the representation of the fermionic parton $\xi$ splits as\footnote{{Here we apply the symplectic group notation under 
$\Sp(n)=\USp(2n)=\Sp(2n, \C) \cup \U(2n)$,
such that $\Sp(1)={\rm USp}(2)=\SU(2)=\Spin(3)$ and 
$\Sp(2)={\rm USp}(4)=\Spin(5)$. 
The $G_1 \hookrightarrow G_2$ means that the inclusion $G_1 \subset G_2$ as a subgroup.
The representations on two sides of $``\sim''$ shows their decomposition relation.}}
\begin{equation} \label{eq:anomaly-decomposition}
\begin{array}{ccccccccc}
\U(1)'\times\SO(10) & \hookrightarrow
&\SU(2)'\times\SO(10) &  \hookrightarrow
& \Sp(10) & \hookleftarrow
 & \Sp(2)\times\Sp(8) &  \hookleftarrow & \SU(2)''\times\Sp(8)\\
{\bf 10}_1 & & ({\bf 2},{\bf 10}) & \overset{\rm S}{\sim} & {\bf 20} & \overset{\rm R}{\sim} & ({\bf 4},{\bf 1}) \oplus ({\bf 1},{\bf 16}) & \overset{\rm S}{\sim}  & ({\bf 4},{\bf 1}) \oplus ({\bf 1},{\bf 16}).
\end{array}
\end{equation}
Some comments on \eq{eq:anomaly-decomposition}: 
\begin{itemize}[leftmargin=.mm]
\item
{The $({\bf 1},{\bf 16})$ is free from both the old Witten's $\SU(2)'$ and the new $\SU(2)'$ anomaly, 
but the $({\bf 4},{\bf 1})$ has the new $\SU(2)''$ anomaly {$w_2w_3(V_{\SO(3)''})$} \cite{WangWenWitten2018qoy1810.00844}.} 
\item Since we have argued that $({\bf 2},{\bf 10})$ in $\SU(2)'\times\SO(10)$ has no Witten or the new $\SU(2)'$ anomalies in the $\SU(2)'$ sector,
so the new-$\SU(2)''$ anomaly must come from
the 
{remained $\SO(10)$, or more precisely the remained $\SO \times \SO(10)$ out of the full $\Spin^{h'} \times\SO(10)$ in \eq{eq:GQCD4}}. 
{According to \cite{WanWang2018bns1812.11967, WW2019fxh1910.14668},
 the classification of 't Hooft anomaly of $\SO \times \SO(10)$ symmetry is generated respectively by the cobordism group:
\bea
 \TP_5({\SO \times \SO(10)})=\Z_2^2, \quad
\left\{
\begin{array}{l}
\text{$(-1)^{\int w_2w_3(TM)}$ out of the tangent bundle $TM$ of $\SO$ ,}\\
\text{$(-1)^{\int w_2 w_3(V_{\SO(10)})}$ out of the associated vector bundle of $\SO(10)$.}
\end{array}
\right.
\eea
Therefore, 
we claim that the new-$\SU(2)''$ anomaly 
can be identified by $w_2w_3(V_{\SO(10)})$, 
come from the remained $\SO(10)$ out of the $\Spin^{h'} \times\SO(10)$. }
\item 
{We can further extend the $\Spin^{h'} \times\SO(10)$ structure of the fermionic parton theory {QCD$'_4$}
to the full $(\DSpin  \times_{\Z_2^F}\Spin(10)) \times_{[\Z_2^{F'}]}[\SU(2)']$ structure of the modified $so(10)$ GUT,
under the pullback:
\bea
1 \to \Z_2^F \to (\DSpin  \times_{\Z_2^F}\Spin(10)) \times_{[\Z_2^{F'}]}[\SU(2)'] \to \Spin^{h'} \times\SO(10) \to 1.
\eea
In terms of the interpretation of the anomaly (we can gauge the anomaly-free $\SU(2)'$), we are left with
\bea
1 \to \Z_2^F \to  \Spin  \times_{\Z_2^F}\Spin(10) \to \SO \times \SO(10) \to 1.
\eea
The two $w_2w_3(TM)$ and $w_2w_3(V_{\SO(10)})$ anomalies in the $ \TP_5({\SO \times \SO(10)})=\Z_2^2$ 
becomes identified as the same anomaly in the 
$\TP_5(\Spin \times_{\Z_2^F}\Spin(10))=\Z_2$ of \eq{eq:Spin10-anomaly}.
Thus, 
of course, 
now we can also interpret as the gauge anomaly $w_2 w_3(V_{\SO(10)})$
as the gravitational anomaly $w_2w_3(TM)$ due to the relation $(-1)^{\int w_2w_3(TM)}=(-1)^{\int w_2 w_3(V_{\SO(10)}) }$ as mentioned before.} 
The analysis establishes that the proposed QED$'_4$ or QCD$'_4$ theory in \Eqn{eq: QED4} at least 
has the same 4d nonperturbative global mixed gauge-gravitational $w_2 w_3$ anomaly as the {proposed 4d WZW term in \eq{eq:BdCsingular}}.
\end{itemize}
To reproduce the WZW term more explicitly, we extend the QED$'_4$ theory to the 5d bulk 
\begin{equation}
S_\text{QED$'_5$}[\xi,\bar{\xi},a,\Phi]=\int_{M^5} 
\bar{\xi}(\ii\gamma^\mu D_\mu -m-\gamma^5\tilde{\Phi}^\text{bi}-\gamma^6\ii\hat{\Phi}^\text{bi})\xi \; \dd^5 x,
\end{equation}
where $\xi$ still forms the ${\bf 10}_1$ under $\U(1)'\times\SO(10)$. 
Note that in 5d, each Dirac fermion already defines five gamma matrices $\gamma^0,\gamma^1,\gamma^2,\gamma^3,\gamma^4$, which are rank-4 matrices. 
By doubling the fermion content ( {which means we need two sets of 5d Dirac fermions in {\bf 10}, 
thus there are $2 \times {\bf 10}$ Dirac fermions in 5d}), 
we are able to introduce two more gamma matrices, denoted $\gamma^5$ and $\gamma^6$, such that all seven gamma matrices $\gamma^0,\cdots ,\gamma^6$ are 
rank-8 matrices satisfying the Clifford algebra relation $\{\gamma^\mu,\gamma^\nu\}=2\delta^{\mu\nu}$. The bulk fermions are gapped by the mass term $m$. The boundary QED$'_4$ theory (with massless fermions) is reduced from the bulk QED$'_5$ theory (with massive fermions) as the effective domain wall theory, which lives on the 4d domain wall separating the $m>0$ and $m<0$ phases in 5d.\footnote{ {The 5d
theory has the $2 \times {\bf 10}$ Dirac fermions of 4 complex components (alternatively, ${\bf 10}$ of 8 complex components), 
while the domain wall theory in 4d has ${\bf 10}$ Dirac fermions of 4 complex components, in one lower dimension.
The 4d domain wall fermion has only half of degrees of freedom of the 5d bulk.}
}

To show that the QED$'_4$ theory is equivalent to the WZW theory,  we only need to show that the bulk QED$'_5$ theory 
can reproduce the WZW term \eq{eq:BdCsingular}. 
For this purpose, we introduce two 2-form $\R$ gauge fields $\CB=\CB_{\mu\nu}\dd x^{\mu}\wedge\dd x^{\nu}$ and 
$\CB'=\CB'_{\mu\nu}\dd x^{\mu}\wedge\dd x^{\nu}$ that couple to the fermionic parton as
\begin{equation}
S_\text{QED$'_5$}[\xi,\bar{\xi},a,\Phi,\CB,\CB']=\int_{M^5}\bar{\xi}(\ii\gamma^\mu D_\mu -m-\gamma^5\tilde{\Phi}^\text{bi}-\gamma^6\ii\hat{\Phi}^\text{bi}-\ii\gamma^5\gamma^\mu\gamma^\nu \CB_{\mu\nu}-\ii\gamma^6\gamma^\mu\gamma^\nu \CB'_{\mu\nu})\xi\;\dd^5 x.
\end{equation}
Integrating out the massive fermion $\xi$, 
we obtain the BF 5-form term with 2-form $\CB$ and $\CB'$ fields:
\begin{equation}
S_{\text{BF}_5}[\CB,\CB']=\frac{1}{\pi}\int_{M^5} \CB\wedge\dd \CB',
\end{equation}
with the constraint that the 2-form gauge fields $\CB$ and $\CB'$ 
are locked to the cohomology classes that measure the topological defects in $\tilde{\Phi}^\text{bi}$ and $\hat{\Phi}^\text{bi}$ respectively
\begin{equation} \label{eq:BCnormalization}
B(\tilde{\Phi}^\text{bi})= \frac{\CB}{\pi}= \frac{\CB(\tilde{\Phi}^\text{bi})}{\pi}\in {\rm H}^2(\O(10)/(\O(6)\times\O(4)),\Z_2),\quad
B'(\hat{\Phi}^\text{bi})= \frac{\CB'}{\pi}= \frac{\CB'(\hat{\Phi}^\text{bi})}{\pi}\in {\rm H}^2(\O(10)/\U(5),\Z_2).
\end{equation}
The emergent $\U(1)'$ gauge field $a$ decouples from the GUT-Higgs field $\Phi$ and the 2-form gauge fields $\CB,\CB'$, which can be integrated out independently. Further integrate out the 2-form gauge fields $\CB,\CB'$, we obtain an action for $\Phi$ (simply by substituting the constraint),
$
S_\text{WZW}[\Phi]=\frac{1}{\pi}\int_{M^5} \CB(\tilde{\Phi}^\text{bi})\wedge\dd \CB'(\hat{\Phi}^\text{bi}).
$ 
Recall the footnote \ref{ft:normalization-WZW} about our normalizations of differential forms and cohomology classes.
This leads to the proposed WZW term in \Eqn{eq:BdCsingular} 
\begin{equation}
S_\text{WZW}[\Phi]={\pi}\int_{M^5} B(\tilde{\Phi}^\text{bi})\smile \delta B'(\hat{\Phi}^\text{bi}),
\end{equation}
which is expected to be placed on the 5d manifold $M^5$ whose boundary is the 4d spacetime $M^4=\partial M^5$.

\subsubsection{Color-Flavor Separation and Dark Gauge Sector: 4d Deconfined Quantum Criticality}

The QED$'_4$ theory describes the DQC scenario of the 4d WZW-term like theory at low-energy. In this scenario, the GUT-Higgs field deconfines into fragmentary excitations, 
which are new 0d particles beyond the SM:
\begin{itemize}
\item 10 new fermions $\xi$ in the ${\bf 10}_1$ of $\U(1)'\times\SO(10)$, as {\bf fermionic partons} that fractionalize the GUT-Higgs field;
\item a new $\U(1)'$ photon $a_\mu$ in the ${\bf 1}_0$ of $\U(1)'\times\SO(10)$, which mediates a new gauge force that exists between and only between fermionic partons. It does not couple to 
{any particle in the SM sector}, hence appears dark to us. Therefore, we will call it the {\bf dark photon}.
\end{itemize}
The GUT-Higgs boson can be considered as the bound state of two fermionic partons (of opposite emergent $\U(1)'$ gauge charges) bind together by the the emergent $\U(1)'$ gauge force mediated by dark photons.\\ 
$\bullet$ From particle physic perspective, the fermionic partons and dark photons are more fundamental constituents of the GUT-Higgs bosons.\\ 
$\bullet$ From condensed matter physics perspective, these fragmentary excitations are 
{emergent collective modes} of the GUT-Higgs field instead. \\
The two complementary viewpoints are a matter of culture. The readers can take whichever interpretation that is more favorable to their mindset.

Because the QED$'_4$ theory 
{is deconfined} in 4d, the fragmentary GUT-Higgs liquid is expected to be a stable phase in the phase diagram Fig.\,\ref{fig:phase-schematic}.  It covers the quantum critical region (critical in the sense that excitations are gapless), and may possibly extend into the modified $so(10)$ GUT phase (as long as fermionic partons remain deconfined). Starting from the fragmentary GUT-Higgs liquid phase, we can access the adjacent phases by GUT-Higgs condensation.
\begin{itemize}
\item $\langle\tilde{\Phi}^\text{bi}\rangle\neq 0$, the system enters the PS GUT phase, where fermionic partons are fully gapped by the vector mass.
\item $\langle\hat{\Phi}^\text{bi}\rangle\neq 0$, the system enters the $su(5)$ GUT phase, where fermionic partons are fully gapped by the axial mass.
\item $\langle\tilde{\Phi}^\text{bi}\rangle\neq 0$ and $\langle\hat{\Phi}^\text{bi}\rangle\neq 0$, 
the system enters the SM phase, where fermionic partons are fully 
{gapped by both vector and axial masses}.
\end{itemize}
In all phases, the dark photon will remain gapless and decoupled from all the other particles, which provides a new candidate for the light dark matter.

A substantial difference of fermionic partons $\xi$ in the fragmentary GUT-Higgs liquid from quarks and leptons $\psi$ in the SM, lies in their distinct assignment of quantum numbers. 
For the spacetime symmetry representation, the Dirac fermion partons $\xi$ is in 
the complex ${\bf 2}_L  \oplus {\bf 2}_R$ of ${\Spin(1,3)}$;
the SM's Weyl fermion is in the complex ${\bf 2}_L $ of ${\Spin(1,3)}$.

For the internal symmetry representation,
consider entering the SM phase from the fragmentary GUT-Higgs liquid, the  {Dirac} fermionic partons, apart from the gap opening, also has its representation split from ${\bf 10}_1$ under 
$\U(1)'\times\SO(10)$ to\footnote{
{Here we use the branching rule of the Lie algebra representations 
for the following inclusion:
$so(10) \hookleftarrow su(5) \times u(1)_X$ (R regular subalgebra), so that
${\bf 10} \sim {\bf 5}_{-2} \oplus \overline{\bf 5}_{2}$; and also the 
$su(5) \hookleftarrow su(3) \times su(2) \times u(1)_{\tilde Y}$ (R regular subalgebra)
so that
${\bf 5} \sim (1, {\bf 2})_{3} \oplus ({\bf 3}, 1)_{-2}$
and
$\overline{\bf 5} \sim (1, {\bf 2})_{-3} \oplus (\overline{\bf 3}, 1)_{2}$.
}} 
%
$$
({\bf 1},{\bf 2})_{1,3,-2}\oplus ({\bf 3},{\bf 1})_{1,-2,-2}\oplus ({\bf 1},{\bf 2})_{1,-3,2}\oplus ({\bf \bar 3},{\bf 1})_{1,2,2}
\text{ under } \SU(3)_{c}\times\SU(2)_{\rm L}\times {\U(1)'}_{\text{gauge}}^{\text{dark}}\times\U(1)_{\tilde Y}\times\U(1)_X
$$
of the SM. The weak $\SU(2)$ flavor and the strong $\SU(3)$ color quantum numbers separate to different fermions, called \emph{flavoron} and \emph{coloron}, 
denoted by the 
{$f$ and $c$  {Dirac} fermions as Grassmann numbers} 
respectively, as summarized in 
Table \ref{tab: parton}. 
We shall name this phenomenon as {\bf color-flavor separation}, as it is analogous to the spin-charge separation 
\cite{Tomonaga1950Remarks,Luttinger1963An-Exactly,Anderson2000Spin-charge} in condensed matter physics.

\begin{table}[htp]
\begin{center}
\begin{tabular}{c|c|ccccc}
\hline
 & ${\U(1)'}_{\text{gauge}}^{\text{dark}}$ & $\SU(3)_{c,\text{color}}$ & $\SU(2)_{\rm{L},\text{flavor}}$ & $\U(1)_{\tilde Y}$ & $\U(1)_X$ & $\U(1)_\text{EM}$ \\
\hline
$f$ & $1$ & ${\bf 1}$ & ${\bf 2}$ & $3$ & $-2$ & $1$ or $0$\\
$c$ & $1$ & ${\bf 3}$ & ${\bf 1}$ & $-2$ & $-2$ & $-1/3$\\
${f}'$ & $1$ & ${\bf 1}$ & ${\bf 2}$ & $-3$ & $2$ & $0$ or $-1$\\
${c}'$ & $1$ & ${\bf \bar 3}$ & ${\bf 1}$ & $2$ & $2$ & $1/3$\\
\hline
\end{tabular}
\caption{The  {Dirac} fermionic parton $\xi$ contains flavorons $f$ and colorons $c$ as Grassmann numbers. Please beware that the
${\U(1)'}_{\text{gauge}}^{\text{dark}}$ is for the Dark Gauge (dark photon) sector, which is \emph{totally distinct} from the $\U(1)_\text{EM}$.
The $\U(1)_\text{EM}$ is from the electroweak Higgs symmetry breaking of the $\SU(2)_{\rm{L},\text{flavor}} \times \U(1)_{\tilde Y}$ down to a subgroup $\U(1)_\text{EM}$.
}
\label{tab: parton}
\end{center}
\end{table}

The flavoron can participate SU(2) weak interaction but not SU(3) strong interaction. 
On the contrary, the coloron can participate SU(3) strong interaction but not SU(2) electroweak interaction. 
Many of them also carry electromagnetic charge, such that they can also participate electromagnetic interaction. Beyond the SM interactions, the flavoron and coloron also interact among themselves by the emergent $\U(1)'$ gauge force mediated by the dark photon. 
{Note that there exist a flavoron 
(in the $f_L$ sector) which do not participate SU(3) strong and electromagnetic interactions. 
It only participate SU(2) weak interaction (like left-handed neutrinos) and dark gauge interaction (unlike neutrinos), 
which makes it especially a potential better candidate for heavy dark matter.}

\newpage
\section{Conclusion: Mother Effective Field Theory for BSM Gauge Enhanced Quantum Criticality}

\label{sec:EFTBSMGEQC}

\subsection{Summary of Main Results:\\
EFT for Internal Spin(10) Global Symmetry or Dynamical Gauge Theory}

To conclude, here  in \Table{tab:field}, we summarize the quantum field content 
of the mother effective field theory of the 4d $so(10)$ GUT + GUT-Higgs potential + with or without WZW term.
we summarize our physical findings.
on the various quantum vacua of mother effective field theory

\begin{table}[htp]
\begin{center}
\begin{tabular}{c|c|ccccc}
 \hline
Field content & $\Spin \equiv {\Spin(1,3)}$ & $\Spin(10)$ & $\Z_2^F$ & $\Z_2^{F'}$ & ${\U(1)'}_{\text{gauge}}^{\text{dark}}$ \\
\hline
  \multicolumn{7}{c}{Model I}\\
\hline
$\psi$ & ${\bf 2}_L$   & ${\bf 16}$ & $1$ & $0$ & $0$\\
\arrayrulecolor{mygray}\hline
$A$ & ${\bf 4}$ & ${45}_{\text{adj.}}$ & $0$ & $0$ & $0$ \\
\arrayrulecolor{mygray}\hline
$\begin{array}{c}
{\Phi}^\text{bi}=\\
{\Phi}_{\bf 1} \oplus \hat{\Phi}^\text{bi} \oplus \tilde{\Phi}^\text{bi} 
\end{array}$ &  
 ${\bf 1}$ &
$\begin{array}{c}
{\bf 10} \otimes {\bf 10}={\bf 100}=\\
{\bf 1} \oplus {\bf 45} \oplus{\bf 54}
\end{array}$ 
&
  $0$ & $0$ & $0$ &\\
\arrayrulecolor{mygray}\hline
$\phi$ & ${\bf 1}$ & ${\bf 10}$ & $0$ & $0$ & $0$ & \\
\arrayrulecolor{black}\hline
  \multicolumn{7}{c}{Model II (include Model I's above + extra below)}\\
\arrayrulecolor{black}\hline
$\xi$ & ${\bf 2}_L  \oplus {\bf 2}_R$ & ${\bf 10}$ & $0$ & $1$ & $1$ \\
\arrayrulecolor{mygray}\hline
$a$ & ${\bf 4}$ & ${\bf 1}$ & $0$ & $0$ & $1_{\text{adj.}}$ \\
\arrayrulecolor{black}\hline
\end{tabular}
\caption{Quantum field representations (reps) for two Toy Models.
Model I contains the Weyl spacetime-spinor $\psi$,
the Spin(10) gauge field $A$ (45 Lie algebra generators denoted as ${45}_{\text{adj.}}$, but not the ${\bf 45}$ rep),
the SO(10)-bivector spacetime-scalar ${\Phi}^\text{bi}$,
and 
the SO(10)-vector spacetime-scalar ${\phi}$ as an auxiliary field (Lagrange multiplier with no dynamics).
Model II contains all the field contents of Model I, in addition, Model II contains extra fields:
the 4d WZW term $\pi \int_{M^5}B(\tilde{\Phi}^{\rm{bi}}) \smile \delta B'(\hat{\Phi}^{\rm{bi}})$ lives on the boundary of a 5d bulk
can induce a candidate low energy QED$'_4$ theory with 
a Dirac spacetime-spinor $\xi$ (as a fermionic parton)
and a ${\U(1)'}$ emergent dark gauge field $a$
(1 Lie algebra generator denoted as ${1}_{\text{adj.}}$, which carries no ${\U(1)'}$ charge).
The rep of fermionic parton $\xi$ in $su(3) \times su(2) \times u(1)_{\tilde Y}  \times u(1)_{X}$ is given in \Table{tab: parton}.
There are two types of fermion parities in a double Spin structure $\DSpin \equiv (\Z_2^{F} \times \Z_2^{F'}) \rtimes \SO$.
}
\label{tab:field}
\end{center}
\end{table}
 
Based on three binary conditions:
\begin{enumerate}[leftmargin=2.mm, label=\textcolor{blue}{(\alph*)}.]
\item {\bf Without or with the GUT-Higgs potential ${\rm U}(\Phi_{{\mathbf{R}}})$ and GUT-Higgs condensation $\langle \Phi_{{\mathbf{R}}}\rangle \neq 0$
of \Eqn{eq:GUTHiggsU}}: 
(i) Whether we stays in the Spin(10) group of $so(10)$ GUT, or (ii) add
the GUT-Higgs potential to Higgs down the Spin(10) deforming it to $G_{\GG}$, $G_{\PS}$, and $G_{\SM}$.
\item {\bf Without or with the WZW term 
$S^\text{WZW}[\Phi]{=\pi \int_{M^5}B(\tilde{\Phi}^{\rm{bi}})\smile \delta B'(\hat{\Phi}^{\rm{bi}})}$ 
of \Eqn{eq:BdCsingular}}: Namely 
(i) Whether we stays in the Model I --- an $so(10)$ GUT without the $w_2w_3$ anomaly, 
or (ii) the Model II --- a modified so(10) GUT + WZW matches the $w_2w_3$ anomaly.
\item {\bf Without or with the dynamically gauged internal symmetry group $G=G_{\text{internal}}$}: (i) Whether we keep the $[G_{\text{internal}}]$ symmetry as a global symmetry, 
or (ii) we gauge the $[G_{\text{internal}}]$,\footnote{We may use the bracket notation on a group $[G_{\text{internal}}]$ to emphasize that group is dynamically gauged.} 
namely gauging  $[\Spin(10)], [G_{\GG}]$, $[G_{\PS}]$, and $[G_{\SM}]$.
\end{enumerate}
The three binary conditions enumerate totally eight possibilities 
(where below we can use 3-bits, ``???'', each bit ``?''
labels a ``x'' or ``o'' to specify
without or with that binary condition holds), 
which we enlist their physics interpretations, one by one:
\begin{enumerate}[leftmargin=2.mm, label=\textcolor{blue}{\arabic*}.]
\item xxx - {\bf Without ${\rm U}(\Phi_{{\mathbf{R}}})$, without WZW, without gauged $[G_{\text{internal}}]$}:\\
We stay in the Landau-Ginzburg phase of the Spin(10) global symmetry.
\item  oxx - {\bf With ${\rm U}(\Phi_{{\mathbf{R}}})$, without WZW, without gauged $[G_{\text{internal}}]$}:\\
We stay in the Landau-Ginzburg phases, but the ${\rm U}(\Phi_{{\mathbf{R}}})$ potentially 
breaks the Spin(10) global symmetry to
other continuous Lie group global symmetries $G_{\GG}$, $G_{\PS}$, and $G_{\SM}$,
via spontaneous global symmetry breaking.
There are 45, 24, 21, and 12 Lie algebra generators for each of these groups.
So there are corresponding numbers of the low energy Nambu-Goldstone modes,
matching the number of the broken Lie algebra generators based on the Goldstone's theorem.

In principle, 
because there is no 't Hooft anomaly for the 16n chiral fermions with these 
$G_{\text{internal}}$ internal global symmetries, 
we can gap out all chiral fermions while preserving $G_{\text{internal}}$
via a \emph{symmetric mass generation} through appropriate interactions \cite{WangWen2018cai1809.11171,RazamatTong2009.05037}.

\item xxo  - {\bf Without ${\rm U}(\Phi_{{\mathbf{R}}})$, without WZW, with gauged $[G_{\text{internal}}]$}:\\
We obtain the familiar $so(10)$ GUT with the $[\Spin(10)]$ gauged. 
At a deep UV higher energy, there shows the \emph{asymptotic freedom} of 16n Weyl fermions 
(quarks and leptons are liberated with a weaker coupling at a shorter distance for such a non-abelian Lie group gauge force \cite{Gross1973id, Politzer1973fx}).
At an IR lower energy, the Spin(10) gauge fields confine the 16n Weyl fermions,
which is a strongly coupled gauge theory 
with all fermions can gain an energy gap (i.e., ``mass'' due to the confinement).

\item  oxo - {\bf With ${\rm U}(\Phi_{{\mathbf{R}}})$, without WZW, with gauged $[G_{\text{internal}}]$}:\\
Then we are in the dynamical gauge theory phases but with gauge symmetry breaking.
The ${\rm U}(\Phi_{{\mathbf{R}}})$ potentially 
breaks the Spin(10) gauge group to
other continuous Lie gauge group $G_{\GG}$, $G_{\PS}$, and $G_{\SM}$,
via Anderson-Higgs mechanism of \emph{spontaneous gauge symmetry breaking}.
There are 45, 24, 21, and 12 Lie algebra generators for each of these groups.
Recall in the global symmetry story, 
there are corresponding numbers of the low energy Nambu-Goldstone modes,
matching the number of the broken Lie algebra generators based on the Goldstone's theorem.
But now some massless gauge fields can "eats" the degrees of freedom of Goldstone bosons,
so to become the massive gauge field with extra degrees of freedom.

Note that again, 
at a deep UV higher energy, 
there shows the \emph{asymptotic freedom} of Weyl fermions;
while at an IR lower energy, 
the non-abelian Lie gauge forces 
of $G_{\GG}$, $G_{\PS}$, and $G_{\SM}$)
can \emph{confine} some of the Weyl fermions.
In this strongly coupled gauge theory, 
some fermions can gain an energy gap (i.e., ``mass'') due to the confinement.
But we do still have the electroweak-Higgs causing spontaneous gauge symmetry breaking  
$su(2)_{\rm L}  \times u(1)_Y \to u(1)_{\EM}$.
The $u(1)_{\EM}$ stays \emph{deconfined} and propagate the gaplss electromagnetic waves in our vacuum.
Here the fermion mass can come from a combination of mechanism from:
the confinement mass, the Anderson-Higgs (gauge-)symmetry-breaking mass, or
the gauge theory analog of the {symmetric mass generation}.
\item xox - {\bf Without ${\rm U}(\Phi_{{\mathbf{R}}})$, with WZW, without gauged $[G_{\text{internal}}]$}: \label{item:xox}\\
We stay in the Landau-Ginzburg phase of the Spin(10) global symmetry,
but the 4d WZW term causes the 4d \emph{deconfined quantum criticality} (DQC) with
fractionalized fragmentary excitations.

This DQC is also a \emph{gauge-enhanced criticality} (GEQC) because
we have a new gauge force (that we call Dark Gauge force with 
${\U(1)'}_{\text{gauge}}^{\text{dark}}$ 
Dark Photons) emergent near the criticality. 
The fractionalized fragmentary excitations carry the ${\U(1)'}_{\text{gauge}}^{\text{dark}}$ gauge charge.
If the ${\U(1)'}_{\text{gauge}}^{\text{dark}}$ dark photons stay gapless dynamically at deep IR,
then it is due to the protection of $w_2w_3$ anomaly.
\item oox - {\bf With ${\rm U}(\Phi_{{\mathbf{R}}})$, with WZW, without gauged $[G_{\text{internal}}]$}:\\
We stay in the Landau-Ginzburg phases, but the ${\rm U}(\Phi_{{\mathbf{R}}})$ potentially 
breaks the Spin(10) global symmetry to
other continuous Lie group global symmetries $G_{\GG}$, $G_{\PS}$, and $G_{\SM}$,
via spontaneous global symmetry breaking.
Other than the low energy Nambu-Goldstone modes
matching the number of the broken Lie algebra generators
in the neighbor phases, we still have the fractionalized fragmentary excitations
that also carries ${\U(1)'}_{\text{gauge}}^{\text{dark}}$ gauge charge,
with ${\U(1)'}_{\text{gauge}}^{\text{dark}}$ Dark Photons.
\item xoo - {\bf Without ${\rm U}(\Phi_{{\mathbf{R}}})$, with WZW, with gauged $[G_{\text{internal}}]$}: \label{item:xoo}\\
We obtain the modified $so(10)$ GUT + WZW with the $[\Spin(10)]$ gauged. 
At a deep UV higher energy, 
the GUT-Higgs potential + WZW term may affect the renormalizability of EFT;
however, what we concern is the EFT that works below certain energy cutoff scale 
such as GUT scale ${M_{\rm GUT}}$
or the 5d bulk invertible TQFT energy gap ${\Delta_{\rm iTQFT}}$.
Other than the DQC and GEQC phenomena described above in the scenario \ref{item:xox}, the theory shows:\\ 
$\bullet$ The Spin(10) gauge bosons can propagate or leak to the 5d bulk.\\
$\bullet$ The 16n Weyl fermions are gappable (because there is no anomaly protection for these 16n fermions).\\
$\bullet$  We have again the 10 fractionalized fragmentary fermions, gauge charged under ${\U(1)'}_{\text{gauge}}^{\text{dark}}$ Dark Photon.
Furthermore, the 10 fractionalized fragmentary fermions carry also the strong SU(3)$_c$ gauge charge,
and the weak SU(2)$_{\rm L}$ gauge charge, recall from Table \ref{tab: parton}.\\
$\bullet$ Here we are doing the {Fragmentary GUT-Higgs Liquid model beyond the SM} 
(with 10 fractionalized fragmentary fermions coupled to ${\U(1)'}_{\text{gauge}}^{\text{dark}}$ Dark Photon)
of \Sec{sec:FragmentaryHiggs} that can match the $w_2w_3$ anomaly.
In contrast, we are not thinking of the 10 gauge neutral bosons from {Composite GUT-Higgs model within the SM}
of \Sec{sec:CompositeHiggs} that does not have the $w_2w_3$ anomaly.

\item 
{ooo - {\bf With ${\rm U}(\Phi_{{\mathbf{R}}})$, with WZW, with gauged $[G_{\text{internal}}]$}:\\
This scenario follows directly from the scenario \ref{item:xoo}, but with a GUT-Higgs potential triggering (gauge-)symmetry-breaking.
All statements in the scenario \ref{item:xoo} follow also here. Moreover,\\
$\bullet$ There is a sequence of various possibilities at various energy scales from the UV to the IR dynamical fates of this QFT.
We do not know the definite answer of quantum {dynamics}. 
Here we only enlist the possibilities of quantum dynamical fates of 
{\bf the modified $so(10)$ GUT + 4d WZW term} (with 16n Weyl fermions)
 based on the $w_2w_3$ anomaly matching constraints:\\
i).\,Spin(10) gauge group can be broken down to contain an SU(2) gauge subgroup 
such that there is a new SU(2) anomaly of mixed gauge-gravity type $w_2w_3(TM)=w_2w_3(V_{\rm SO(3)})$ within 
the $\Spin \times_{\Z_2^F}\SU(2) \equiv \Spin^h$ symmetry \cite{WangWenWitten2018qoy1810.00844},
again dynamically gauging SU(2) makes the SU(2) gauge bosons can propagate to the 5d bulk.\\
ii).\,The gauge group can be broken down to contain a U(1) gauge subgroup
which can also have a pure gravitational $w_2w_3(TM)$ anomaly if the theory is all-fermion U(1) gauge theory 
\cite{WangPotterSenthil1306.3238, KravecMcGreevySwingle1409.8339}. 
The $\Spin \times_{\Z_2^F}\U(1) \equiv \Spin^c$ structure trivializes the $w_2w_3(TM)$ anomaly. \\
iii).\,The gauge group can be broken down to contain a  $\Z_2$ gauge subgroup
which can also have a pure gravitational $w_2w_3(TM)$ anomaly if the theory has fermionic strings  \cite{Thorngren1404.4385, ChenHsin2110.14644, FidkowskiHaahHastings2110.14654, WanWangWen2112.12148}.
The Spin structure trivializes the $w_2w_3(TM)$ anomaly. \\
\\
$\bullet$ However, the WZW dynamics in the quantum critical region 
that we propose in \Sec{sec:QED4Parton} shows none of the above.
Instead, we suggest a different IR low energy fate of WZW theory: the Spin(10) symmetry can be fully preserved,
while the mixed gauge-gravity anomaly $w_2w_3(TM)=w_2w_3(V_{\rm SO(10)})$
is matched by a
{{Dirac} fermionic parton theory QED$'_4$ with emergent ${\U(1)'}$ dark gauge force and with a DSpin structure.}
\Fig{fig:DQC} shows a schematic phase diagram. 
For Model I, without a WZW term,  there is no deconfined QED$'_4$ within the dashed circle region.
For Model II, with a WZW term,  there is a deconfined QED$'_4$ within the dashed circle region.
 }
\end{enumerate}

\begin{figure}[htbp]
\begin{center}
\includegraphics[width=0.35\columnwidth]{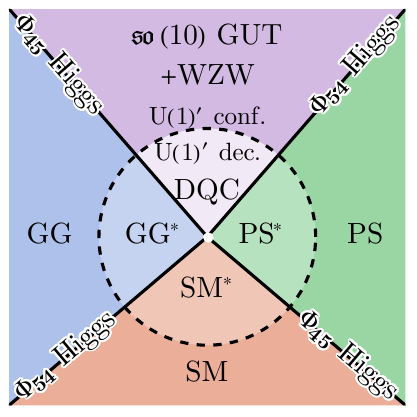}
\caption{Follow \Fig{fig:phase-schematic},
here we show the same phase diagram in the presence of the WZW term if its low energy consequence 
is the fermionic parton theory QED$'_4$ (\Sec{sec:QED4Parton}). 
The dashed circle denotes the confine-deconfine phase transition of the emergent $\U(1)'$ gauge field.
The solid-line phase boundaries between two neighbor phases all are described by 
GUT-Higgs condensation continuous phase transitions.
The SM$^*$ phase means a modification of SM plus additional BSM fields due to QED$'_4$,
within the $\U(1)'$ deconfined region inside the dashed circle. Similar situations for GG$^*$ and PS$^*$.}
\label{fig:DQC}
\end{center}
\end{figure}

\newpage
{\bf \emph{4d boundary criticality and a 5d bulk bosonic invertible TQFT}}: 
Notice that we can interpret the above 4d criticality as a 
{\bf\emph{ boundary criticality}} with the $w_2w_3$  anomaly on the 5d bulk 
of a mod 2 class invertible TQFT. 
The 4d WZW, that can be built from the GUT-Higgs fields, can saturate 4d $w_2w_3(TM)=w_2w_3(V_{\rm SO(10)})$ anomaly. 
So we only require the 5d bulk as some 5d invertible topological order
or symmetry-protected topological states (SPTs) if we require an onsite Spin(10) symmetry on the 4d boundary and on the 5d bulk; 
see an overview of modern quantum matter terminology and definitions in \cite{Wen2016ddy1610.03911, Senthil1405.4015}.

{\bf \emph{Bosonic UV completion}}: 
For this 16n Weyl fermion models, once the $[\Spin(10)]\supset [\Z_2^F]$ is dynamically gauged,
the whole UV completion of the full 4d and 5d system requires only the bosons, as the local onsite Hilbert space with gauge-invariant bosonic operators.

Although above we focus on the 16n-Weyl-fermion SMs or GUTs,
we can consider the 15n-Weyl-fermion models, especially for {\bf the $su(5)$ GUT and the SM + 4d WZW term},
see \Sec{sec:16nvs15nWeylfermions}.

\subsection{16n  vs 15n Weyl fermions: Give ``mass'' to ``right-handed sterile'' neutrinos, canceling mod 2 and mod 16 anomalies,
and topological quantum criticality}
\label{sec:16nvs15nWeylfermions}

Although we mostly focus on the 16n-Weyl-fermion SMs or GUTs in this work,
here we comment about several ways to obtain the low-energy 15n-Weyl-fermion models 
(since the real-world experiments only observed the 15n-Weyl-fermion so far)
by giving a large mass to the 16th Weyl fermions, 
the so-called ``right-handed sterile'' neutrinos (in any of the 3 generation of leptons).\footnote{Note that the
``right-handed sterile $\nu_R$'' neutrino is just the conventional name used in the HEP phenomenology.
We would mostly write this $\nu_R$ in the left-handed Weyl fermion basis.
Also the $\nu_R$ although is sterile to the $G_{\SM}$ and SU(5), the $\nu_R$ is not sterile to Spin(10) and $\Z_{4,X}$. }

What are examples of conventional ways \cite{georgi2018liebook}
to give a large (Anderson-Higgs type quadratic) mass to the 16th Weyl fermions?
We can pair Weyl fermion to itself (i.e., Majorana mass) or to another Weyl fermion (e.g., Dirac mass):
\begin{enumerate}
\item Introduce a Higgs $\Phi_{so(10), \bf 126}$ which can be paired with $ \overline{\bf 126}$ out of two Weyl fermions in
${\bf 16} \otimes {\bf 16} = {\bf 10} \oplus {\bf 120} \oplus \overline{\bf 126}$.
\item Introduce a Higgs $\Phi_{so(10), \bf 16}$ and add an extra Weyl fermion (17th Weyl fermion) singlet ${\bf 1}$ under Spin(10).
This works only if some of the following holds: 
\begin{enumerate}
\item
The 17th Weyl fermion is \emph{not} charged under the $\Z_{4,X}$-symmetry, so we have the $\Z_{16}$-anomaly cancelled already by 16n Weyl fermions.
This is likely to be true because this 17th Weyl fermion is singlet ${\bf 1}$ under Spin(10), thus is also not acted by the center $Z(\Spin(10))=\Z_{4,X}$.
\item 
If the 17th Weyl fermion is also charged under the $\Z_{4,X}$-symmetry,
then we require the $\Z_{4,X}$-symmetry is broken (thus the $\Z_{16}$-anomaly is removed), or the $\Z_{4,X}$-symmetry is preserved but 
$17 \mod 16$-anomaly is cancelled again by additional new sectors with $-1 \mod 16$-anomaly.
\end{enumerate}
\end{enumerate}
What are other new ways to leave only the observed 15n Weyl fermions at low energy, but the $\Z_{16}$ global anomaly can still be cancelled in the full quantum system?
To begin with, to characterize the full 4d anomaly of this 15n SMs or GUTs,
we should combine the two types of anomalies:
First, a potential global $\Z_2$ anomaly, the $w_2 w_3$ for our 4d WZW term, such as in the {Fragmentary GUT-Higgs Liquid model} in \Sec{sec:FragmentaryHiggs}.
Second, the $\Z_{16}$ global anomaly captured by a 5d version of Atiyah-Patodi-Singer (APS) eta invariant for the $\Spin \times_{\Z_2^F} \Z_{4,X}$-structure
 from $\TP_5(\Spin \times_{\Z_2^F} \Z_{4,X})=\Z_{16}$.
We can write that 5d APS invariant in terms of the 4d APS invariant of $\Pin^+$-structure from $\TP_4(\Pin^+)=\Z_{16}$.
The two combined invertible TQFT, labeled by ${\rm p} \in \Z_2$ and $\upnu\in \Z_{16}$, has a partition function ${\bf Z}$ on $M^5$,
which together labels a deformation class of SM  \cite{WangWanYou2112.14765}:
\begin{multline}\hspace{-4mm}
{{\bf Z}_{{\text{$5$d-iTQFT}}}^{({\rm p}, \upnu)} }
\equiv
\exp(\ii \pi\cdot {\rm p}\cdot \int_{M^5} w_2 w_3) 
\cdot
\exp(\frac{2\pi \ii}{16} \cdot\upnu \cdot \eta(\text{PD}(\CA_{{\Z_2}})) \bigg\rvert_{M^5}), \\ 
\text{ with } 
{\rm p} \in \Z_2, \quad
\text{a 4d Atiyah-Patodi-Singer $\eta$ invariant} \equiv \eta_{\text{Pin}^+} \in \Z_{16}, \quad \upnu\in \Z_{16}
.\;
\end{multline}
The cohomology classes of background gauge field
 $\CA_{{\Z_2}} \in \H^1(M,\frac{\Z_{4,X}}{\Z_2^F})$
is defined on a ${\Spin \times_{\Z_2^F} {\Z_{4,X}}}$-manifold $M$ obeys a constraint:
$w_2(TM) = \CA_{{\Z_2}}^2$.

Inspired by highly-entangled interacting quantum matter recent developments (see reviews in \cite{Wen2016ddy1610.03911, Senthil1405.4015}),
\Refe{JW2006.16996, JW2008.06499, JW2012.15860} proposed additional new sectors to cancel the anomalies, for example,
\begin{enumerate} \setcounter{enumi}{2}
\item Symmetry-preserving anomalous gapped 4d TQFT.
\item Symmetric-preserving 5d invertible TQFT in the extra dimension.
\item Symmetry-breaking gapped phase of Landau-Ginzburg kinds.
\item Symmetry-preserving (or breaking) 5d topological gravity theory.
\item Symmetry-preserving or symmetry-breaking gapless phase, 
e.g., extra massless theories, free or interacting conformal field theories (CFTs). 
The interacting CFT can also be related to unparticle physics \cite{Georgi200703260Unparticle} in the high-energy phenomenology community.
\end{enumerate}
The heavy gapped new sectors above can be {\bf   \emph{heavy Dark Matter}} candidates.
The interesting constraints from mod 2 and mod 16 global anomalies on our 4d DQC model are:
\begin{itemize}[leftmargin=.mm]
\item {\bf \emph{$\Z_{16}$ anomaly constraints on the GG and SM of 15n Weyl fermions}}:
On the Georgi-Glashow $su(5)$ GUT and the Standard Model SM${}_{q=6}$ side, we can have 15n Weyl fermions, plus additional new sectors enlisted 
(above and in \cite{JW2006.16996, JW2008.06499, JW2012.15860}) to match the $\Z_{16}$ anomaly.

\item {\bf \emph{$\Z_{2}$ $w_2w_3$ anomaly constraints on the $so(10)$ GUT and PS of 16n Weyl fermions}}:
On the $so(10)$ GUT and the Pati-Salam model sides, there are various types of  $\Z_{2}$ class $w_2w_3$ anomalies,
of the $\SO(10)$, $\SO(6)$, or $\SO(4)$  bundles.
The $\Z_{2}$ $w_2w_3$ anomaly is meant to be cancelled by our 4d WZW term.

\item At the vicinity of the 4d DQC we have proposed, there can be another interplay between the 15n Weyl fermions (GG and SM) to 16n Weyl fermions (the $so(10)$ GUT and PS),
such that the DQC becomes a  {\bf \emph{topological quantum phase transition}} or  {\bf \emph{topological quantum criticality}}.

\end{itemize}

{\bf \emph{4d boundary criticality to a 5d bulk criticality}}: 
Compare with the phase diagram in \Fig{fig:phase-schematic}.
Notice that we can interpret the above 4d criticality as a 
{\bf \emph{boundary criticality}} --- 
\begin{itemize}[leftmargin=.mm]
\item 
 On the modified $so(10)$ GUT and the PS model + WZW term side with 16n Weyl fermions in \Fig{fig:phase-schematic}:
with the $w_2w_3$ $\Z_2$-class anomaly on the 5d bulk  of a mod 2 class invertible TQFT.
\item  
On the modified $su(5)$ GUT and the SM + WZW term side with 15n Weyl fermions  in \Fig{fig:phase-schematic}:
with the $\eta(\text{PD}(\CA_{{\Z_2}}))$ $\Z_{16}$-class anomaly on the 5d bulk  of a mod 2 class invertible TQFT. 
\end{itemize}
Once the $[\Spin(10)]$ is dynamically gauged, 
\begin{itemize}[leftmargin=.mm]
\item 
The 5d bulk on the modified $so(10)$ GUT and the PS model side (16n Weyl fermions): 
The $[\Spin(10)]$ dynamical gauge fields can propagate and leak to the 5d bulk are \emph{deconfined and gapless}.
\item The 5d bulk on the modified $su(5)$ GUT and the SM side (15n Weyl fermions): 
Only the $[\Z_{4,X}]$ subgroup ($Z(\Spin(10))=\Z_{4,X}$)
are dynamically gauged in the 5d bulk of the original fermionic invertible TQFT $\eta(\text{PD}(\CA_{{\Z_2}}))$. 
Gauging $[\Z_{4,X}]$ turns the 5d fermionic bulk to a 5d bosonic bulk TQFT 
(with long-range entanglement, gapped topological order, and described by gauged cohomology, gauged cobordism, or higher category theory).
The 5d bulk can remain to be \emph{gapped}. 
\end{itemize}
Thus there is a phase transition between the {\bf\emph{deconfined and gapless}} 5d bulk to another side of
{\bf\emph{gapped}} 5d bulk. This phase transition can be interpreted as a {\bf \emph{5d bulk topological quantum criticality}}.


\section{Acknowledgements} 
JW thanks Pierre Deligne, Lena Funcke, Jun Hou Fung, Chang-Tse Hsieh, Pavel Putrov, Subir Sachdev, Luuk Stehouwer, David Tong, Zheyan Wan,
and Yunqin Zheng for helpful comments.
YZY thanks John McGreevy for helpful comments.
YZY is supported by a startup fund at UCSD.
This work is supported by 
NSF Grant DMS-1607871 ``Analysis, Geometry and Mathematical Physics'' 
and Center for Mathematical Sciences and Applications at Harvard University.

\appendix

\newpage
\section{Quantum Numbers and Representations of SMs and GUTs in Tables}
\label{sec:SM-GUT-table}

Here we summarize the representations of ``elementary'' chiral fermionic particles of quarks and leptons of SMs and GUTs in Tables.
\paragraph{Spacetime symmetry representation}
{Here 
Weyl fermions are spacetime Weyl spinors, which we prefer to write all Weyl fermions as
\bea
{\bf 2}_L \text{ of } {\Spin(1,3)}=\rm{SL}(2,\C)
\eea 
with a complex representation in the 4d Lorentz signature.
On the other hand,
the Weyl spinor is 
\bea
{\bf 2}_L \text{ of } \Spin(4)=\SU(2)_L \times \SU(2)_R
\eea
with a {pseudoreal representation}
in the 4d Euclidean signature.}

\paragraph{Internal symmetry representation}

Below we provide two Tables, \ref{table:SMfermion} and \ref{table:PSfermionAll},
to organize the internal symmetry representations of particle contents of
the SM, the $su(5)$ GUT, the Pati-Salam model, the $so(10)$ GUT.

\subsection{Embed the SM into the $su(5)$ GUT, then into the $so(10)$ GUT}
There is a QFT embedding, 
the $so(10)$ GUT $\supset$ the $su(5)$ GUT $\supset$ the SM$_6$ only for  $G_{\SM_{q=6}}$ 
via an internal symmetry group embedding: 
\bea
{\Spin(10)} \supset G_{\GG}\equiv \SU(5) \supset   G_{\SM_{6}} \equiv \frac{{\SU(3)}_c \times {\SU(2)}_{\rm L} \times \U(1)_{\tilde{Y}}}{\Z_6}.
\eea
The representations of quarks and leptons for these models are organized in \Table{table:SMfermion}.
There are two versions of electroweak hypercharge normalization listed in \Table{table:SMfermion},
such that the charge of $\U(1)_{Y}$ is $\frac{1}{6}$ of the charge of  $\U(1)_{\tilde Y }$.
\begin{table}[!h]
$\hspace{-16.9mm}
  \begin{tabular}{ccccccc c c c c c c c c}
    \hline
    $\begin{array}{c}
    \textbf{SM}\\ 
   \textbf{fermion}\\
   \textbf{spinor}\\ 
   \textbf{field}
       \end{array}$
   & ${\SU(3)}$& ${\SU(2)}$ & $\U(1)_{Y}$  
   & $\U(1)_{\tilde Y }$
   & $\U(1)_{\rm{EM}}$ 
    & $\U(1)_{{ \mathbf{B}-  \mathbf{L}}}$  & $\U(1)_{X}$
     & 
    $\Z_{5,X}$
 & 
    $\Z_{4,X}$
    & $\Z_{2}^F$  & SU(5) & Spin(10) \\
        \hline\\[-4mm]
    $\bar{d}_R$& $\overline{\mathbf{3}}$& $\mathbf{1}$ & 1/3  & 2 & 1/3 & $-1/3$ & $-3$ & $-3$ &1   &   1 &  \multirow{2}{*}{
     $\overline{\bf{5}}$}  &  \multirow{6}{*}{
     ${\bf{16}}$} \\
\cline{1-11} $l_L$& $\mathbf{1}$& $\mathbf{2}$& $-1/2$  & $-3$ & 0 or $-1$ & $-1$ & $-3$& $-3$  &1  &   1 \\
\cline{1-12}  $q_L$& ${{\mathbf{3}}}$& $\mathbf{2}$& 1/6  &1 & 2/3 or $-1/3$ &  1/3  &1&1 & 1 &    1 & \multirow{3}{*}{
     ${\bf{10}}$} \\
\cline{1-11} $\bar{u}_R$& $\overline{\mathbf{3}}$& $\mathbf{1}$& $-2/3$  & $-4$ & $-2/3$ & $-1/3$ &1 & $1$  &1 &   1 \\
\cline{1-11} $\bar{e}_R= e_L^+$& $\mathbf{1}$& $\mathbf{1}$& 1 & 6 & 1 & 1  &1 &1 &   1 &    1\\
\cline{1-12} $\bar{\nu}_R= {\nu}_L $& $\mathbf{1}$& $\mathbf{1}$& 0  & 0 & 0 & 1  & 5 &0 &   1 &   1 & \;\;${\bf{1}}$ &\\
    \hline\end{tabular}
$
\caption{{\bf Embed the $su(3) \times su(2) \times u(1)$ SM into the Georgi-Glashow $su(5)$ GUT, then into the $so(10)$ GUT.} 
We show the quantum numbers of $15 +1 = 16$ left-handed Weyl fermion (spacetime spinors ${\bf 2}_L$ in $\Spin(1,3)$) in each of three generations of matter fields in SM. 
The $15$ of 16 Weyl fermion are $ \overline{\bf 5} \oplus {\bf 10}$  of SU(5); namely,
$(\overline{\bf 3},{\bf 1},1/3)_L \oplus ({\bf 1},{\bf 2},-1/2)_L \sim \overline{\bf 5}$
and $({\bf 3},{\bf 2}, 1/6)_L \oplus (\overline{\bf 3},{\bf 1}, -2/3)_L \oplus ({\bf 1},{\bf 1},1)_L  \sim {\bf 10}$  of SU(5).
The $1$ of 16 is presented neither in the standard GSW SM nor in the $su(5)$ GUT, but it is within  ${\bf 16}$ of the $so(10)$ GUT.
The numbers in the Table entries indicate the quantum numbers associated with the representation of the groups given in the top row. 
We show a generation of SM fermion matter fields in  Table \ref{table:SMfermion}. 
There are 3 generations, triplicating Table \ref{table:SMfermion}, in SM.
All fermions have the fermion parity  $\Z_{2}^F$ representation charge 1.
{In the $su(5)$ GUT, by including the $\U(1)_{X}$,
we have the $(\SU(5) \times \U(1)_{X})/\Z_5=\U(5)_{\hat q=2}$ structure described in \Refe{WangYou2111GEQC, HAHSIV}.}
Here  $\U(1)_{X} \supset \Z_{4,X} \supset \Z_{2}^F$ and $\SU(5) \supset  \U(1)_Y$. 
Both $\U(1)_{X}$ and $\U(1)_{{ \mathbf{B}-  \mathbf{L}}}$ are outside the $\SU(5)$.
}
 \label{table:SMfermion}
\end{table}

\subsection{Embed the SM into the Left-Right and Pati-Salam models, into the $so(10)$ GUT}

There are two version of internal symmetry groups for Pati-Salam (PS) model \cite{Pati1974yyPatiSalamLeptonNumberastheFourthColor}:
$$
G_{\PS_{q'}}\equiv \frac{\SU(4)\times \SU(2)_{\rm L} \times \SU(2)_{\rm R}}{\Z_{q'}}
\equiv \frac{\Spin(6)\times \Spin(4)}{\Z_{q'}}$$ with $q'=1,2$.
There are two version of internal symmetry groups for 
Senjanovic-Mohapatra's Left-Right (LR) model \cite{SenjanovicMohapatra1975},
$$
G_{\LR_{q'}} \equiv \frac{{\SU(3)}_c \times {\SU(2)}_{\rm L} \times  {\SU(2)}_{\rm R} \times \U(1)_{\frac{ \mathbf{B}-  \mathbf{L}}{2}} }{\Z_{3 q'}}
$$ with $q'=1,2$.
In general, there is a QFT embedding, the PS model $\supset$ the LR model $\supset$ the SM for both $q'=1,2$ 
via the internal symmetry group embedding: 
\bea
G_{\PS_{q'}} \supset  G_{\LR_{q'}} \supset 
G_{\SM_{q=3q'}}  \equiv \frac{{\SU(3)}_c \times {\SU(2)}_{\rm L} \times \U(1)_{\tilde{Y}}}{\Z_{q=3q'}}.
\eea
Namely, when $q'=1$, we have
\bea
G_{\PS_{1}} \supset  G_{\LR_{1}} \supset G_{\SM_{3}}.
\eea
Furthermore, only when $q'=2$, we can have the whole embedded into the Spin(10) for the $so(10)$ GUT:
\bea
{\Spin(10)} \supset G_{\PS_{2}} \supset  G_{\LR_{2}} \supset G_{\SM_{6}}.
\eea
The representations of quarks and leptons for these models are organized in \Table{table:PSfermionAll}.

 \begin{table}[!h]
$\hspace{-12mm}
  \begin{tabular}{ccccccc c c c c c  }
    \hline
    $\begin{array}{c}
    \textbf{SM}\\ 
   \textbf{fermion}\\
   \textbf{spinor}\\ 
   \textbf{field}
       \end{array}$
   & ${\SU(3)}$& ${\SU(2)}_{\rm L}$& ${\SU(2)}_{\rm R}$ 
     & $\U(1)_{\frac{ \mathbf{B}-  \mathbf{L}}{2}}$  
   & $\U(1)_Y$  
   & $\U(1)_{Y_R}$   
   & $\U(1)_{\rm{EM}}$ 
   & $\U(1)_{X}$ & 
    $\Z_{4,X}$
    & $\Z_{2}^F$  & Spin(10)\\
        \hline\\[-4mm]
$u_L$& ${{\mathbf{3}}}$& 
\multicolumn{1}{c}{\multirow{2}{*}{
    {$q_L:\mathbf{2}$} 
     }
     }
&$\mathbf{1}$
& 1/6 
& 1/6  
& 2/3 & 2/3   
 &1 & 1 &    1 &   \multirow{8}{*}{
     ${\bf{16}}$}
   \\
 \cline{1-2} \cline{4-11}   
$d_L$& ${{\mathbf{3}}}$&  &$\mathbf{1}$
& 1/6 
& 1/6 
& $-1/3$ &   $-1/3$ 
 &1 & 1 &    1 & 
   \\
    \cline{1-11}  
 $\nu_L$& $\mathbf{1}$& 
\multicolumn{1}{c}{\multirow{2}{*}{
    {$l_L:\mathbf{2}$} 
    }}
 & $\mathbf{1}$ 
  & $-1/2$
 & $-1/2$ 
 &  0 & 0   
 & $-3$  &1  &   1
    \\
 \cline{1-2} \cline{4-11}   
 $e_L$& $\mathbf{1}$& & $\mathbf{1}$ 
 & $-1/2$
 & $-1/2$
 &  $-1$  &  $-1$ 
  & $-3$  &1  &   1 \\
 \cline{1-11}  
 $\bar{u}_R$& $\bar{\mathbf{3}}$& $\mathbf{1}$ & 
\multicolumn{1}{c}{\multirow{2}{*}{
    {$q_R:\mathbf{2}$} 
     }
     }
     & $-1/6$
& $-2/3$ & 
$-1/6$
& $-2/3$   & $1$  &1 &   1 \\
     \cline{1-3} \cline{5-11}   
$\bar{d}_R$& $\bar{\mathbf{3}}$& $\mathbf{1}$ & 
& $-1/6$
& 1/3  
& $-1/6$
& 1/3    & $-3$ &1   &   1 &
  \\
 \cline{1-11}    
 $\bar{\nu}_R= {\nu}_L $& $\mathbf{1}$& $\mathbf{1}$& 
\multicolumn{1}{c}{\multirow{2}{*}{
    {$l_R:\mathbf{2}$} 
     }  }
& 1/2 
& 0  &  1/2 & 0  & 5 &   1 &   1 & \\
\cline{1-3} \cline{5-11}   
$\bar{e}_R= e_L^+$& $\mathbf{1}$& $\mathbf{1}$& 
 & 1/2
& 1  &
1/2  & 1   &1 &   1 &    1\\
    \hline\end{tabular}
$
   \caption{{\bf Embed the $su(3) \times su(2) \times u(1)$ SM into the Pati-Salam ${su(4) \times su(2)  \times su(2)}$, then into the $so(10)$ GUT.}
   We have
   $T_{3,\rm{L}}+ Y=Q_{\rm{EM}}$,
  the Lie algebra linear combination ${\SU(2)}_{\rm L}$ (the third generator) and  $\U(1)_Y$ gives the $\U(1)_{\rm{EM}}$ charge.
 We have $T_{3,\rm{R}}+ Y=\frac{ \mathbf{B}-  \mathbf{L}}{2}$,
  the Lie algebra linear combination of ${\SU(2)}_{\rm R}$ (the third generator) and  $\U(1)_Y$ gives the $\U(1)_{ \mathbf{B}-  \mathbf{L}}$.
  We choose the right-handed anti-particle to be in {$\mathbf{2}$} of ${\SU(2)}_{\rm R}$
  (so its right-handed particle to be in {$\bar{\mathbf{2}}$} of ${\SU(2)}_{\rm R}$) that makes a specific assignment on the $\pm$ sign of its $T_{3,{\rm R}}$ charge.
  So we have the formula, $T_{3,{\rm L}}-T_{3,{\rm R}}=Q_{\rm{EM}}-\frac{ \mathbf{B}-  \mathbf{L}}{2}$.}
 \label{table:PSfermionAll}
\end{table}

\section{Representation and Branching Rule for GUT-Higgs symmetry breaking}
\label{sec:app-Branching-GUT-Higgs}

Here are we organize the set of branching rules of representations following the symmetry breaking pattern of various GUTs to SM 
(these rules are used in \Sec{sec:SM-GUT-group-branch}):
\begin{enumerate}
\item $\Spin(10) \hookleftarrow \SU(5) \hookleftarrow \frac{{\SU(3)}_c \times {\SU(2)}_{\rm L} \times \U(1)_Y}{\Z_6}$ branching rules:\\
$\diamond$ For $\Spin(10) \hookleftarrow \SU(5)$, also for
$\SO(10) \hookleftarrow \U(5)_{\hat q=1}= \frac{\SU(5) \times \U(1)_{\hat q=1}}{\Z_5}$
or $\Spin(10) \hookleftarrow \U(5)_{\hat q=2}= \frac{\SU(5) \times \U(1)_{\hat q=2}}{\Z_5}$ 
(or in terms of Lie algebra $so(10) \hookleftarrow {su(5) \times u(1)}$ with a regular Lie subalgebra in 
\cite{1912.10969LieART}),\footnote{
{
Follow footnote \ref{ft:U(5)q-I}
different non-isomorphic versions of
U(5) Lie groups defined as $\U(5)_{\hat q} \equiv \frac{\SU(5) \times \U(1)_{\hat q} }{\Z_5} \equiv
\{ (g, \e^{\ii \theta}) \in \SU(5) \times \U(1) \big\vert  ( \e^{\ii \frac{2 \pi n}{5}} \mathbb{I}, 1) \sim ( \mathbb{I}, \e^{\ii \frac{2 \pi n {\hat q}}{5}} ), n\in \Z_5 \}$,
the Lie group embedding shows (the proof is given in \cite{WangYou2111GEQC, HAHSIV})
$$\text{$\Spin(10) \supset \SU(5)$ and $\Spin(10) \supset \U(5)_{\hat q=2,3}$, but $\Spin(10) \not\supset \U(5)_{\hat q=1,4}$,}$$ 
while
$$\text{$\SO(10) \supset \SU(5)$ and $\SO(10) \supset \U(5)_{\hat q=1,4}$, but $\SO(10) \not\supset \U(5)_{\hat q=2,3}$.}$$ 
The embedding $\SO(10) \supset \U(5)_{\hat q=1,4}$ cannot be lifted to Spin(10) thus $\Spin(10) \not\supset \U(5)_{\hat q=1,4}$; 
but $\Spin(10) \supset \U(5)_{\hat q=2,3}$.}
\label{ft:U(5)q-II}
}
the branching rule says:
%
%
%
%
%
\bea \label{eq:Spin10-SU5}
\left\{\begin{array}{l c l}
{\bf 10} \sim {\bf 5} \oplus {\overline{\bf 5}}
& \text{or} &  {\bf 10} \sim {\bf 5}_2 \oplus {\overline{\bf 5}}_{-2}.\\
{\bf 16} \sim {\bf 1} \oplus {\overline{\bf 5}} \oplus {\bf 10} & \text{or} & {\bf 16} \sim {\bf 1}_{-5} \oplus {\overline{\bf 5}}_3 \oplus {\bf 10}_{-1}.\\
{\bf 45} \sim {\bf 1} \oplus {\bf 10} \oplus {\overline{\bf 10}} \oplus {\bf 24} & \text{or} & {\bf 45} \sim {\bf 1}_0 \oplus {\bf 10}_4 \oplus {\overline{\bf 10}}_{-4} \oplus {\bf 24}_0.\\
{\bf 54} \sim {\bf 15} \oplus {\overline{\bf 15}} \oplus {\bf 24}  & \text{or} & {\bf 54} \sim  {\bf 15}_4 \oplus {\overline{\bf 15}}_{-4} \oplus {\bf 24}_0.\\
{\bf 120} \sim   {\bf 5}\oplus \overline{\bf 5}\oplus {\bf 10}\oplus \overline{\bf 10}\oplus {\bf 45} \oplus \overline{\bf 45}
& \text{or} & {\bf 5}_2\oplus \overline{\bf 5}_{-2}\oplus {\bf 10}_{-6}\oplus \overline{\bf 10}_{6}\oplus {\bf 45}_{2} \oplus \overline{\bf 45}_{-2}.\\
{\bf 126} \sim   {\bf 1}\oplus {\bf 5}\oplus \overline{\bf 10}\oplus {\bf 15}\oplus \overline{\bf 45} \oplus {\bf 50}
& \text{or} & {\bf 1}_{10}\oplus {\bf 5}_{2}\oplus \overline{\bf 10}_{6}\oplus {\bf 15}_{-6}\oplus \overline{\bf 45}_{-2}\oplus {\bf 50}_{2}  .
\end{array}
\right.
\eea

$\diamond$ For $\SU(5) \hookleftarrow \frac{{\SU(3)}_c \times {\SU(2)}_{\rm L} \times \U(1)_Y}{\Z_6}$
(or in terms of Lie algebra $su(5) \hookleftarrow {su(3) \times su(2) \times u(1)}$ with a regular Lie subalgebra in \cite{1912.10969LieART}),
the branching rule says:
%
%
%
%
%
%
\bea \label{eq:SU5-SU3SU2U1}
\left\{\begin{array}{l }
{\bf 5} \sim ({\bf 1},{\bf 2})_{-3} \oplus  ({\bf 3},{\bf 1})_{2}.\\
{\bf 10} \sim ({\bf 1},{\bf 1})_{-6} \oplus (\overline{\bf 3},{\bf 1})_{4} \oplus ({\bf 3},{\bf 2})_{-1}.\\
{\bf 15} \sim ({\bf 1},{\bf 3})_{-6} \oplus ({\bf 3},{\bf 2})_{-1} \oplus ({\bf 6},{\bf 1})_{4}.\\
{\bf 24} \sim ({\bf 1},{\bf 1})_0 \oplus ({\bf 1},{\bf 3})_0 \oplus ({\bf 3},{\bf 2})_5 \oplus 
(\overline{\bf 3},{\bf 2})_{-5}  \oplus ({\bf 8},{\bf 1})_0.\\
\dots\\
{\bf 45} \sim ({\bf 1},{\bf 2})_{-3} \oplus  ({\bf 3},{\bf 1})_{2} 
\oplus  (\overline{\bf 3},{\bf 1})_{-8} \oplus (\overline{\bf 3},{\bf 2})_{7} \oplus  ({\bf 3},{\bf 3})_{2} 
\oplus  (\overline{\bf 6},{\bf 1})_{2} \oplus  ({\bf 8},{\bf 2})_{-3} .\\
{\bf 50} \sim ({\bf 1},{\bf 1})_{12} \oplus  ({\bf 3},{\bf 1})_{2} 
\oplus  (\overline{\bf 3},{\bf 2})_{7} \oplus ({\bf 6},{\bf 1})_{-8} \oplus  (\overline{\bf 6},{\bf 3})_{2} 
 \oplus  ({\bf 8},{\bf 2})_{-3} .\\
\end{array}
\right.
\eea
(1) First, in order to break the Spin(10) or SO(10) down to SU(5), we take the representation whose branching rule in \eq{eq:Spin10-SU5} 
contains the ${\bf 1}$ of SU(5) or ${\bf 1}_0$ of U(5)
on the right-handed side so that SU(5) or U(5) is left unbroken. This means that we may take a GUT-Higgs ${\bf 45}$ that we name it as \eq{eq:phi45}:
\bea \label{eq:phi45-app}
\Phi_{so(10), \bf 45} \equiv
\Phi_{\bf 45}.
\eea
(2) Second, in order to break SU(5) further down to $G_{\SM_6} \equiv \frac{{\SU(3)}_c \times {\SU(2)}_{\rm L} \times \U(1)_Y}{\Z_6}$,
we take the representation whose branching rule in \eq{eq:SU5-SU3SU2U1}
contains the $ ({\bf 1},{\bf 1})_0$ of $G_{\SM_6}$. This means that we can take the ${\bf 24}$ of SU(5) as the second GUT-Higgs called $\Phi_{su(5), \bf 24}$.
But if we want to obtain this second GUT-Higgs from a higher-energy $so(10)$ GUT, it turns out that
we can find $\Phi_{su(5), \bf 24}$ within \eq{eq:phi54}:
\bea \label{eq:phi54-app}
\Phi_{so(10), \bf 54}\equiv
\Phi_{\bf 54},
\eea
from \eq{eq:Spin10-SU5} more naturally, as we will soon see.\footnote{It may be also possible to introduce the second GUT-Higgs of 
$\Phi'_{so(10), \bf 45} \equiv \Phi'_{\bf 45}$ (different from $\Phi_{\bf 45}$) 
which also contains the $\Phi_{su(5), \bf 24}$ that can break SU(5) down to $G_{\SM_6}$.\\[2mm]
Another possible choice proposed in Georgi's textbook \cite{georgi2018liebook} 
is that in addition to the first GUT-Higgs $\Phi_{so(10), \bf 45} \equiv \Phi_{\bf 45}$,
one may also introduce a scalar Higgs of a ${\bf 16}$ or a ${\bf 126}$ of Spin(10) in order to Higgs down to $G_{\SM}$.\\[2mm] 
However, these choices are \emph{not} ideal for us, due to the reason of 
quantum criticality that we pursue later.
The quantum criticality that we pursue \emph{only require} 
$\Phi_{so(10), \bf 45} \equiv \Phi_{\bf 45}$ 
and 
$\Phi_{so(10), \bf 54} \equiv \Phi_{\bf 54}$, from \eq{eq:phi45} and \eq{eq:phi54}.}
\item 
$\Spin(10) \hookleftarrow  \frac{\Spin(6) \times \Spin(4)}{\Z_2} \hookleftarrow \frac{{\SU(3)}_c \times {\SU(2)}_{\rm L} \times \U(1)_Y}{\Z_6}$ branching rules:\\
$\diamond$ 
For $\Spin(10) \hookleftarrow \frac{\Spin(6) \times \Spin(4)}{\Z_2} =  \frac{\SU(4) \times \SU(2) \times \SU(2)}{\Z_2} $, also for
$\SO(10) \hookleftarrow {\SO(6) \times \SO(4)}$ (or in terms of Lie algebra $so(10) \hookleftarrow {so(6) \times so(4)}$ or ${su(4) \times su(2)  \times su(2)}$ 
with a regular Lie subalgebra in \cite{1912.10969LieART}),
we find that:
%
%
%
%
%
\bea 
\label{eq:Spin10-SU4SU2SU2}
\left\{\begin{array}{l }
{\bf 10} \sim ({\bf 1},{\bf 2},{\bf 2}) \oplus ({\bf 6},{\bf 1},{\bf 1}).\\
{\bf 16} \sim ({\bf 4}, {\bf 2}, {\bf 1}) \oplus (\overline{\bf 4}, {\bf 1}, {\bf 2}).\\
{\bf 45} \sim ({\bf 1}, {\bf 3}, {\bf 1}) \oplus ({\bf 1}, {\bf 1}, {\bf 3}) \oplus ({\bf 6},{\bf 2},{\bf 2}) \oplus ({\bf 15},{\bf 1},{\bf 1}).\\
{\bf 54} \sim ({\bf 1}, {\bf 1}, {\bf 1}) \oplus ({\bf 1},{\bf 3}, {\bf 3}) \oplus ({\bf 6},{\bf 2},{\bf 2}) \oplus ({\bf 20}',{\bf 1},{\bf 1}).\\
{\bf 120} \sim ({\bf 1},{\bf 2},{\bf 2})  \oplus ({\bf 6}, {\bf 3}, {\bf 1}) \oplus ({\bf 6}, {\bf 1}, {\bf 3}) \oplus ({\bf 10},{\bf 1}, {\bf 1}) \oplus (\overline{\bf 10},{\bf 1}, {\bf 1}) \oplus ({\bf 15},{\bf 2},{\bf 2}) .\\
{\bf 126} \sim  ({\bf 6}, {\bf 1}, {\bf 1}) \oplus  (\overline{\bf 10},{\bf 3}, {\bf 1}) \oplus ({\bf 10},{\bf 1}, {\bf 3}) \oplus ({\bf 15},{\bf 2},{\bf 2}) .
\end{array}
\right.
\eea
$\diamond$  For $\frac{\Spin(6) \times \Spin(4)}{\Z_2} =  \frac{\SU(4) \times \SU(2) \times \SU(2)}{\Z_2} \hookleftarrow
 \frac{{\SU(3)}_c \times {\SU(2)}_{\rm L} \times \U(1)_Y}{\Z_6} $ 
 (or in terms of Lie algebra ${so(6) \times so(4)}$ or ${su(4) \times su(2)  \times su(2)} \hookleftarrow
 {su(3) \times su(2) \times u(1)}$),
we find that the
$su(4) \hookleftarrow {su(3)  \times u(1)}$
(with a regular Lie subalgebra in \cite{1912.10969LieART})
branching rule says:
\bea 
\label{eq:SU4-SU3U1}
\left\{\begin{array}{l }
{\bf 4} \sim {\bf 1}_{-3}  \oplus  {\bf 3}_{1}. \\
{\bf 6} \sim {\bf 3}_{-2}  \oplus  \overline{\bf 3}_{2}. \\
{\bf 10} \sim {\bf 1}_{-6}  \oplus  {\bf 3}_{-2} \oplus {\bf 6}_{2}.\\
{\bf 15} \sim {\bf 1}_{0}  \oplus  {\bf 3}_{4} \oplus  \overline{\bf 3}_{-4} \oplus  {\bf 8}_{0}.
\end{array}
\right.
\eea
(1) First, in order to break the Spin(10) down to 
$G_{\PS_2} \equiv \frac{\Spin(6) \times \Spin(4)}{\Z_2} =  \frac{\SU(4) \times \SU(2) \times \SU(2)}{\Z_2}$, 
we take the representation whose branching rule in \eq{eq:Spin10-SU4SU2SU2} 
contains the $({\bf 1}, {\bf 1}, {\bf 1})$ 
on the right-handed side so that $G_{\PS_2}$ is left unbroken. This means that we may take a GUT-Higgs ${\bf 54}$ that we had named it in \eq{eq:phi54}
as 
$$
\Phi_{so(10), \bf 54} \equiv
\Phi_{\bf 54}.
$$
(2) Second, in order to break $G_{\PS_2}$ further down to $G_{\SM_6} \equiv \frac{{\SU(3)}_c \times {\SU(2)}_{\rm L} \times \U(1)_Y}{\Z_6}$,
we take the representation whose branching rule in \eq{eq:SU5-SU3SU2U1}
contains the $ ({\bf 1},{\bf 1})_0$ of $G_{\SM_6}$. This means that we can take the ${\bf 15}$ of SU(4) as the second GUT-Higgs called 
$\Phi_{su(4), \bf 15}$.
But if we want to obtain this second GUT-Higgs from a higher-energy $so(10)$ GUT, it turns out that
we can find $\Phi_{su(4), \bf 15}$ from what we had named in \eq{eq:phi45} called
$$
\Phi_{so(10), \bf 45}\equiv
\Phi_{\bf 45},
$$
from \eq{eq:Spin10-SU4SU2SU2} more naturally, as we will soon see.\footnote{Another possible choice 
proposed in Georgi's textbook \cite{georgi2018liebook} 
is that in addition to the first GUT-Higgs $\Phi_{so(10), \bf 54} \equiv \Phi_{\bf 54}$,
one may also introduce a scalar Higgs of a ${\bf 16}$ or a ${\bf 126}$ of Spin(10) in order to Higgs down to $G_{\SM}$.\\[2mm] 
However, these choices are \emph{not} ideal for us, due to the reason of 
quantum criticality that we pursue later.
The quantum criticality that we pursue \emph{only require} 
$\Phi_{so(10), \bf 45} \equiv \Phi_{\bf 45}$ 
and 
$\Phi_{so(10), \bf 54} \equiv \Phi_{\bf 54}$, from \eq{eq:phi45} and \eq{eq:phi54}.}
\item $\frac{{\SU(3)}_c \times {\SU(2)}_{\rm L} \times \U(1)_Y}{\Z_6} \hookleftarrow
\frac{{\SU(3)}_c   \times \U(1)_{\rm{EM}}}{\Z_3}$ branching rules:\\
The Standard Model (SM) electroweak Higgs in the representation
\bea
\Phi_{\SM}  \text{  in }  ({\bf 1},{\bf 2})_{Y=\frac{1}{2}} = ({\bf 1},{\bf 2})_{Y_W={1}} = ({\bf 1},{\bf 2})_{\tilde{Y}={3}}  \text{ of  } su(3) \times su(2) \times u(1)
\eea
does the job to break 
$G_{\SM_6} \equiv \frac{{\SU(3)}_c \times {\SU(2)}_{\rm L} \times \U(1)_Y}{\Z_6}$ to 
$\frac{{\SU(3)}_c   \times \U(1)_{\rm{EM}}}{\Z_3}$.
Then next, we can ask how to find 
$\Phi_{\SM}$ from the representation of $su(5)$, or ${su(4) \times su(2)  \times su(2)}$, or $so(10)$.

$\bullet$ $\Phi_{\SM}$ from $su(5)$: From the branching rule in \eq{eq:SU5-SU3SU2U1},
one can try to take the $\Phi_{su(5), \bf 5}$ and $\Phi_{su(5), \bf 45}$ which contains
$({\bf 1},{\bf 2})_{-{3}}$ of ${su(3) \times su(2) \times u(1)}$ 
which is the complex conjugation of $\Phi_{\SM}$'s $({\bf 1},{\bf 2})_{\tilde{Y}={3}}$.

$\bullet$ $\Phi_{\SM}$ from ${su(4) \times su(2)  \times su(2)}$:
From the branching rule in \eq{eq:SU4-SU3U1},
one can try to take the $\Phi_{{su(4) \times su(2)  \times su(2)}, ({\bf 4},{\bf 2}, {\bf 1})}$ that contains
$({\bf 1},{\bf 2})_{-{3}}$ of  ${su(3) \times su(2) \times u(1)}$, 
which is also the complex conjugation of $\Phi_{\SM}$'s $({\bf 1},{\bf 2})_{\tilde{Y}={3}}$.
We may also need $\Phi_{{su(4) \times su(2)  \times su(2)}, (\overline{\bf 4},{\bf 1}, {\bf 2})}$
if we wish to break the $\SU(2)_{\rm R}$ completely.

$\bullet$ $\Phi_{\SM}$ from $so(10)$:\\
From the branching rule in \eq{eq:Spin10-SU5},
we can get the $\Phi_{su(5), \bf 5}$ and $\Phi_{su(5), \bf 45}$
out of ${\bf 10}$, ${\bf 120}$ or $\overline{\bf 126}$ of $so(10)$,
which we can call $\Phi_{so(10), \bf 10}$, $\Phi_{so(10), \bf 120}$,  and $\Phi_{so(10), \overline{\bf 126}}$.
These ${\bf 10}$, ${\bf 120}$ or $\overline{\bf 126}$ are particular sensible according to  \cite{georgi2018liebook},
because these Higgs can be paired up with the fermion bilinear operators $\psi_i \psi_j$
whose representations are also in the tensor product
${\bf 16} \otimes {\bf 16} = {\bf 10} \oplus {\bf 120} \oplus \overline{\bf 126}$.

From the branching rule in \eq{eq:Spin10-SU4SU2SU2},
we can get the $\Phi_{{su(4) \times su(2)  \times su(2)}, ({\bf 4},{\bf 2}, {\bf 1})}$ and $\Phi_{{su(4) \times su(2)  \times su(2)}, (\overline{\bf 4},{\bf 1}, {\bf 2})}$ 
out of ${\bf 16}$ of Spin(10),
which we can call $\Phi_{so(10), \bf 16}$.

\end{enumerate}


\section{Induce a 3d WZW term between N\'eel $so(2)$ and VBS $so(3)$ on a 4d bulk $w_2(V_{\SO(3)})w_2(V_{\SO(2)})$}
\label{sec:HomotopyCohomology-NeelVBS}

This Appendix provides a logical pedagogical account on the familiar 3d dQCP \cite{SenthildQCP0311326} proposed as a continuous quantum phase transition,  
on a 2+1d bosonic lattice model with an internal non-relativistic (iso)spin-1/2 bosons,\footnote{What condensed matter people 
call the spin-1/2 bosons on site is actually the isospin-1/2 boson which is in the representation {\bf 2} of the internal symmetry SU(2), 
as the internal SU(2) doublet, or namely the qubit. 
The spin up $|\uparrow \rangle$ and down $|\downarrow \rangle$ are mapped to $|1 \rangle$ and $|0 \rangle$ of qubit.
To emphasize again, the \emph{internal} SU(2) here is not the \emph{spacetime} SU(2) from the spacetime Spin group.} 
between two kinds of Landau-Ginzburg symmetry breaking orders 
on each lattice site: 
\begin{enumerate}[leftmargin=.mm]
\item One side has the N\'eel anti-ferromagnet order: This order breaks the $\Z^2$-spatial lattice translation to $(\Z_2)^2$ on a lattice.
It also {\bf \emph{breaks the $\SO(3)$ internal (iso)spin rotational symmetry}} (actually, breaking SO(3) faithfully, not SU(2)\footnote{There is 
an internal SU(2) spin rotational symmetry, but the center $Z(SU(2))=\Z_2$ does not act on the Hilbert space in a physical faithful or meaningful way.
What faithful representation means physically here is that 
whether we can find states as that representation, 
being acted by any physical operator such that these states can be distinguished from each other. 
The answer is that we cannot distinguish the two states charged under $Z(SU(2))=\Z_2$ physically in this bosonic system.
\label{footnote:SU2vsSO3}}).
But it respects the spatial rotational symmetry, which is $\Z_4$ spatial rotational symmetry on a square lattice,
but it {\bf \emph{preserves an enhanced $\SO(2)$ spatial rotational symmetry}} in the continuum. 

\item Another side has the Valence-Bond Solid (VBS) order, 
which  {\bf \emph{preserves a faithful $\SO(3)$ (iso)spin rotational symmetry}} (again, see footnote \ref{footnote:SU2vsSO3}),
because the VBS order pairs the two neighbor-site (iso)spin-1/2 bosons to an (iso)spin-0 state
$\frac{1}{\sqrt{2}}(|\uparrow \rangle |\downarrow \rangle - |\downarrow \rangle |\uparrow \rangle)$.
But the pattern of VBS {{breaks}} the $\Z_4$ spatial rotational symmetry on a square lattice,
so the VBS {\bf \emph{breaks an $\SO(2)$ spatial rotational symmetry}} in the continuum. 
\end{enumerate}
{If we take into account the discrete $\Z_2$ symmetry (a time-reversal or a spatial reflection symmetry), 
the above $\SO(2)$ symmetry becomes an $\O(2)= \SO(2) \rtimes \Z_2$ symmetry,
while the above $\SO(3)$ symmetry becomes an $\O(3)= \SO(3) \times \Z_2$ symmetry.}

Below we write $G$ as the original symmetry group (such as ${\SO(3)\times \SO(2)}$ valid to the UV lattice scale),
while ${G_{\text{sub}}}$ is the remained preserved unbroken symmetry in the corresponding order (N\'eel or VBS orders).
Then we have the following fibration structure:
\bea \label{eq:fibration-app}
{G_{\text{sub}}} \lhook\joinrel\xrightarrow{\quad} G \longrightarrow \frac{G}{G_{\text{sub}}},
\eea
where the quotient space $\frac{G}{G_{\text{sub}}}$ is the base manifold (i.e., the orbit) as the \emph{symmetry-breaking order parameter space}.
The $G$ is the total space obtained from the fibration of the ${G_{\text{sub}}}$ fiber (i.e., the stabilizer) over the base $\frac{G}{G_{\text{sub}}}$. 
Here is a systematic table computation 
on the homotopy group $\pi_k$ of $(\frac{G}{G_{\text{sub}}})$ for N\'eel or VBS orders,
\bea
\begin{tabular}{lccccccc}
\hline
 & $\pi_0$ & $\pi_1$ & $\pi_2$ & $\pi_3$ & $\pi_4$ & $\pi_5$ \\
\hline
N\'eel  $S^2=\frac{\O(3)\times \O(2)}{\O(2)\times \O(2)} =\frac{\O(3)}{\O(2)}$ & $0$ & $0$ & $\Z$ & $\Z$ & $\Z_2$ & $\Z_2$ \\
\;\;\,\quad\quad \quad $=\frac{\SO(3)\times \SO(2)}{\SO(2)\times \SO(2)} =\frac{\SO(3)}{\SO(2)}$ & &  &  &  &   &  \\
\hline
VBS $S^1=\frac{\O(3)\times \O(2)}{\O(3)\times \O(1)}=\frac{\O(2)}{\O(1)}$ & $0$ & $\Z$ & $0$ & $0$ & $0$ & $0$ \\
\;\;\,\quad\quad \quad $=\frac{\SO(3)\times \SO(2)}{\SO(3)\times \SO(1)} =\frac{\SO(2)}{\SO(1)}$ & &  &  &  &   &  \\
\hline
$\O(5)$ & $\Z_2$ & $\Z_2$ & $0$ & $\Z$ & $\Z_2$ & $\Z_2$  \\
\hline
$\SO(5)$ & $0$ & $\Z_2$ & $0$ & $\Z$ & $\Z_2$ & $\Z_2$  \\
\hline
\end{tabular}.
\eea

To our knowledge, the most systematic, physically intuitive, and mathematically transparent construction of the 3d dQCP and its 3d WZW term 
can be based on the following arguments:
\begin{enumerate}[leftmargin=.mm]
\item The N\'eel order {\bf \emph{breaks an $\SO(3)$ (iso)spin rotational symmetry}} down to an $\U(1)=\SO(2)$ (iso)spin rotational symmetry such as along the $z$ axis,
such that \eq{eq:fibration} in the N\'eel order becomes:
\bea \label{eq:fibration-Neel}
\Big({G_{\text{sub}}}={\SO(2)\times \SO(2)} \Big) \lhook\joinrel\xrightarrow{\quad} \Big(G={\SO(3)\times \SO(2)}\Big) \longrightarrow \Big(\frac{G}{G_{\text{sub}}}=S^2\Big). 
\eea
\begin{enumerate}[leftmargin=2.mm, label=\textcolor{blue}{(\roman*)}., ref={(\roman*)}]
\item {\bf \emph{Hedgehog core, instanton, and magnetic monopole}}:
The $\SO(3)$ symmetry breaking \emph{hedgehog core} has a 0d singularity in the spacetime.
This 0d singularity of this \emph{hedgehog core} in the 3d spacetime can be also regarded an \emph{instanton} in the 3d spacetime.
We can couple this whole configuration to SO(3) background gauge field,
this means that we can use the $w_2(V_{SO(3)})$ to measure the magnetic charge of SO(3).
We evaluate the $w_2(V_{SO(3)})$ over the N\'eel's SO(3) symmetry-breaking target space
$S^2$, it turns out that there is a $2 \pi$-flux over $S^2$.
Therefore, the \emph{hedgehog core} is not only an \emph{instanton} event but also an SO(3) \emph{magnetic monopole}, 
living on a 0d open end of some non-dynamical 1d 't Hooft line defect of SO(3) background gauge field.
\item This $\SO(3)$ symmetry-breaking hedgehog core traps 
a ``fractionalized charge-1/2 object charged under the preserved SO(2) symmetry (or $\Z_4$ symmetry on a lattice scale),'' 
namely in the projective representation of $\Z_4$,
which is in the unit integer representation $\Z_8$.
{Namely, {\bf \emph{the $\SO(3)$-symmetry-breaking topological defect, 
hedgehog core in the N\'eel phase, traps 
the $\frac{1}{2}$-fractionalization of the unbroken $\SO(2)$, or $\Z_4$, 
{charged object} of VBS order.}}}
\item The winding number of such N\'eel  hedgehog configuration can be classified by
\bea
\pi_2(\frac{\SO(3)\times \SO(2)}{\SO(2)\times \SO(2)}) =\pi_2(\frac{\SO(3)}{\SO(2)})= \pi_2(S^2)=\Z.
\eea 
This says the $S^2$ as a 2d surface in 3d spacetime 
wrapping around the target $S^2$ of the 
N\'eel's SO(3) symmetry-breaking target space (the base manifold and stabilizer in \eq{eq:fibration-Neel}).
The spatial $S^2$ circle as a homology class 
(in $\H_2 (M,\Z)$, called this 2d sphere $\varrho^2_{\text{}}$)
can be paired up with a cohomology class
$\cB \in \H^2(M,\Z)$.
To make sense the unit generator of the winding $\Z$ class, 
the $\cB$ evaluated on $\varrho^2$ (bounding a 3-disk $\Sigma^3$ by $\varrho^2$ so $\prt\Sigma^3=\varrho^2$)
must have the following:
\bea
\ointint_{\varrho^2=\prt\Sigma^3} \cB= \ointint_{\varrho^2} w_2(V_{\SO(3)})= 1 \mod 2.
\eea
\item Now imagine in a 3d spacetime picture, we can regard:\\ 
$\bullet$ the 0d hedgehog core  $\varsigma^0_{\text{N\'eel hedgehog}}$ as the \emph{charged object}, 
fractionalized charged under the preserved SO(2) (a projective representation in $\Z_4$, precisely a linear representation in $\Z_8$).\\
$\bullet$ the 2d $S^2$ called $\varrho^2_{\text{}}$ with $\cB \in \H^2(M,\Z)$ on the $\varrho^2_{\text{}}$,
as the \emph{charge operator}, or the \emph{symmetry generator} of the SO(2).\\
Then, follow the higher symmetry or generalized global symmetry language \cite{Gaiotto2014kfa1412.5148},
the measurement of the symmetry is exactly performed by evaluating 
the linking between the  $\varsigma^0_{\text{N\'eel hedgehog}}$ and $\varrho^2_{\text{}}$  in a 3d spacetime $M^3$.
Precisely, the linking number Lk, manifested as a statistical Berry phase,
is evaluated via the expectation value of path integral:
\bea \label{eq:Lk-2d-0d}
\langle \exp(\ii \pi \ointint_{\varrho^2=\prt\Sigma^3} \cB) \cdot 
\exp(\ii \pi  \varphi\big\vert_{\varsigma^0_{\text{N\'eel hedgehog}}}  )\rangle = (-1)^{{\rm Lk}(\varrho^2, \varsigma^0_{\text{N\'eel hedgehog}} )} \Big\vert_{M^3}
\eea
Here $ \varphi\big\vert_{\varsigma^0_{\text{N\'eel hedgehog}}}$ is the 0d \emph{vertex operator} evaluated around the 0d hedgehog core,
which is again the 0d magnetic monopole at the open end of the SO(3) background-gauged 1d 't Hooft line.
Related descriptions of link invariants of QFTs can be found in \cite{Putrov2016qdo1612.09298PWY,GuoJW1812.11959} and references therein.

\end{enumerate}

\item The VBS order {\bf \emph{breaks an $\SO(2)$ spatial rotational symmetry}} in the continuum (or breaks $\Z_4$ rotational symmetry on a lattice),
such that \eq{eq:fibration} in the VBS order becomes:
\bea \label{eq:fibration-VBS}
\Big({G_{\text{sub}}}={\SO(3)\times \SO(1)} \Big) \lhook\joinrel\xrightarrow{\quad} \Big(G={\SO(3)\times \SO(2)}\Big) \longrightarrow \Big(\frac{G}{G_{\text{sub}}}=S^1\Big). 
\eea
\begin{enumerate}[leftmargin=2.mm, label=\textcolor{blue}{(\roman*)}., ref={(\roman*)}]
\item
The $\SO(2)$ symmetry-breaking VBS vortex core has a 0d singularity trapping an (iso)spin-1/2 object called the (iso)spinon 
in the space (famously popularized by Levin-Senthil \cite{LevinSenthil0405702}), which indeed is a 1d vortex loop 
(called this 1d loop $\varsigma^1_{\text{VBS vortex}}$) in the spacetime.
\item The (iso)spinon with (iso)spin-1/2 trapped at the VBS order parameter vortex core is 
a ``fractionalized charge-1/2 object charged under the preserved symmetry SO(3),'' namely in the projective representation of SO(3),
which is in the fundamental representation {\bf 2} of SU(2).
{Namely, {\bf \emph{the $\SO(2)$-symmetry-breaking topological defect, the vortex in the VBS phase, traps the $\frac{1}{2}$-fractionalization of $\SO(3)$ 
{charged object} of N\'eel order.}}}
\item The winding number of such VBS vortex configuration can be classified by
\bea
\pi_1(\frac{\SO(3)\times \SO(2)}{\SO(3)\times \SO(1)}) =\pi_1(\frac{\SO(2)}{\SO(1)})= \pi_1(S^1)=\Z.
\eea 
This says the spatial $S^1$ wrapping around the target $S^1$ of the 
VBS's SO(2) symmetry-breaking target space (the base manifold and stabilizer in \eq{eq:fibration-VBS}).
The spatial $S^1$ circle as a homology class 
(in $\H_1 (M,\Z)$, called this 1d circle $\varrho^1_{\text{}}$)
can be paired up with a cohomology class
$\cA \in \H^1(M,\Z)$.
To make sense the unit generator of the winding $\Z$ class, 
the $\dd \cA$ evaluated on a 2-disk $\Sigma^2$ (bounded by $\varrho^1$ so $\prt\Sigma^2=\varrho^1$)
must have the following Stoke theorem:
\bea
\oint_{\varrho^1=\prt\Sigma^2} \cA=
\int_{\Sigma^2} \dd \cA =\int_{\Sigma^2} w_2(V_{\SO(2)})= 1 \mod 2.
\eea
\item Now imagine in a 3d spacetime picture, we can regard:\\ 
$\bullet$ the 1d vortex loop $\varsigma^1_{\text{VBS vortex}}$ as the \emph{charged object}, 
fractionalized charged under the preserved SO(3) (a projective representation in SO(3), precisely a linear representation in SU(2)).\\
$\bullet$ the 1d $S^1$ circle $\varrho^1_{\text{}}$ with $\cA \in \H^1(M,\Z)$ on the loop,
as the \emph{charge operator}, or the \emph{symmetry generator} of the SO(3).\\
Then, the measurement of the symmetry is exactly performed by evaluating 
the linking between the $\varsigma^1_{\text{VBS vortex}}$ and $\varrho^1_{\text{}}$  in 3d spacetime.
Precisely, the linking number Lk, manifested as a statistical Berry phase,
is evaluated via the expectation value of path integral:
\bea \label{eq:Lk-1d-1d}
\langle \exp(\ii \pi \oint_{\varrho^1=\prt\Sigma^2} \cA) \cdot \exp(\ii \pi \oint_{\varsigma^1_{\text{VBS vortex}}} a)\rangle = (-1)^{{\rm Lk}(\varrho^1, \varsigma^1_{\text{VBS vortex}})}
\Big\vert_{M^3}
\eea
{Here $a$ is a 1d background-gauged $\SO(2)$ connection evaluated around the 1d vortex loop.}
Related descriptions of link invariants of QFTs can be found in \cite{Putrov2016qdo1612.09298PWY,GuoJW1812.11959} and references therein.
\end{enumerate}
\item Overall, combined the above data, we have learned that the 3d dQCP construction can be induced by 
the linking number
${{\rm Lk}(\varrho^2, \varsigma^0_{\text{N\'eel hedgehog}} )}=1$
and 
${{\rm Lk}(\varrho^1, \varsigma^1_{\text{VBS vortex}})}=1$ in the 3d spacetime.
To furnish more physical intuitions, we can deduce that:
\begin{enumerate}[leftmargin=2.mm, label=\textcolor{blue}{(\roman*)}., ref={(\roman*)}]
\item
If we extend the 3d spacetime $t,x,y$ to an extra 4th dimension $z$,
the previous 0d hedgehog core $\varsigma^0_{\text{N\'eel hedgehog}}$ trajectory can be a 1d pseudo-worldline ${\varsigma'^1_{\text{N\'eel hedgehog}}}$ in the 4d spacetime $M^4$.
Similarly, the previous 1d vortex loop ${\varsigma^1_{\text{VBS vortex}}}$ trajectory can be a 2d pseudo-worldsheet ${\varsigma'^2_{\text{VBS vortex}}}$ in the 4d spacetime $M^4$.
Such two configurations can be linked in 4d, with a linking number:
\bea
{{\rm Lk}({\varsigma'^1_{\text{N\'eel hedgehog}}}, {\varsigma'^2_{\text{VBS vortex}}})} \Big\vert_{M^4}.
\eea
This describes the link in the extended 4d spacetime of two \emph{charged objects}, 
charged under SO(2) and SO(3) respectively.\\
\item 
In a parallel story, the \emph{charge operators} (of the above charged objects) are
the 1d SO(2)-background gauged $\cA$ line operator on $\varrho^1$, and 2d SO(3)-background gauged $\cB$ surface operator on $\varrho^2$.
Such two configurations can be linked in 4d, with a linking number:
\bea
{{\rm Lk}(\cA \text{ on } \varrho^1, \cB \text{ on } \varrho^2)} \Big\vert_{M^4}.
\eea
This describes the link in the extended 4d spacetime of two \emph{charge operators}, of SO(2) and SO(3) respectively.

{$\bullet$ If we open up the closed $\oint_{\varrho^1} \cA$ on $\varrho^1$ with two open ends on the 3d boundary ${M^3}$ of the bulk ${M^4}$, then one open end carries 
a $\varphi\big\vert_{\varsigma^0_{\text{N\'eel hedgehog}}}$.  Their link configuration in 3d corresponds to the earlier \eq{eq:Lk-2d-0d}:
$$
{{\rm Lk}(\varsigma^0_{\text{N\'eel hedgehog}}, \varrho^2 )} \Big\vert_{M^3}.
$$
$\bullet$ If we open up the closed $\ointint_{\varrho^2} \cB$ on $\varrho^2$ with an open end 
on the 3d boundary ${M^3}$ of the bulk ${M^4}$, then this open end carries a closed 1d vortex loop 
$\oint_{\varsigma^1_{\text{VBS vortex}}} a$.  Their link configuration in 3d corresponds to the earlier \eq{eq:Lk-1d-1d}:
$$
{{\rm Lk}( \varsigma^1_{\text{VBS vortex}}, \varrho^1)} \Big\vert_{M^3}.
$$
}
\end{enumerate}
These above facts together imply that:
\begin{enumerate}[leftmargin=2.mm, label=\textcolor{blue}{(\roman*)}., ref={(\roman*)}]
\item
The 3d dQCP construction \cite{SenthildQCP0311326} is valid if we introduce a mod 2 class 3d WZW term defined on a 3d boundary $M^3$ of a 4d manifold $M^4$.
Based on the homotopy data $\pi_1(S^1)=\Z$ and $\pi_2(S^2)=\Z$, 
schematically the WZW in a differential form or de Rham cohomology is:\footnote{Here our differential form normalization follows 
the footnote \ref{ft:normalization-WZW}. So we send $\cA/\pi \mapsto \cA$ and $\cB/\pi \mapsto \cB$. 
It can again be easily verified that this WZW has two properties: 
(1) invertible on $|{\bf Z}(M^4)|=1$ on a closed 4-manifold, (2)
this WZW term really is a 3d boundary theory on $M^3$ of the extended $M^4$.
This WZW term is meant to capture the 3d boundary anomaly of the 4d bulk invertible TQFT:
$(-1)^{\int_{M^4} w_2(V_{\SO(3)})w_2(V_{\SO(2)})}$.}
\bea \label{eq:AdBdeRham}
\exp(\ii S^\text{WZW})=\exp( \ii\pi \int_{M^4} \cA \wedge \dd \cB). \Big\vert_{{M^3=\prt {M^4}}}
\eea
More precisely, we can improve this to construct the cohomology class relying on
$\cA \in \H^1(S^1,\Z)=\Z$ and $\cB \in \H^2(S^2,\Z)=\Z$ classes,
the WZW term is written in the singular cohomology class of $\cA$ and $\cB$: 
\bea  \label{eq:AdBsingular}
\exp(\ii  S^\text{WZW})
=\exp( \ii \pi \int_{M^4} \cA \smile  \delta \cB) \Big\vert_{{M^3=\prt {M^4}}}
=\exp( \ii 2\pi \int_{M^4} \cA \smile  \Sq^1 \cB)  \Big\vert_{{M^3=\prt {M^4}}},
\eea
with the coboundary operator $\delta$, 
and the Steenrod square $\Sq^1 \equiv \frac{\delta}{2} \mod 2$ here maps the singular cohomology $\H^2(M,\Z_2) \mapsto \H^3(M,\Z_2)$,
on some triangulable manifold $M$.\footnote{The $\Z_2$ classification of the WZW 
term also comes from another quantum matter intuitive argument: 
When two copies of the WZW terms are put together, 
the system can be trivialized by an interlayer large coupling without breaking symmetry.}

\item The 3d dQCP construction \cite{SenthildQCP0311326} is supported by a 3d 't Hooft anomaly in the ${\SO(3)\times \SO(2)}$ global symmetry on a 3-manifold $M^3$, 
captured by a 4d bulk invertible TQFT \cite{WangSenthildQCP1703.02426} living on a 4-manifold $M^4$ with a boundary $\prt M^4= M^3$:
\bea \label{eq:w2w2}
\exp(\ii \pi \int_{M^4} w_2(V_{\SO(3)})w_2(V_{\SO(2)})).
\eea
This 3d 't Hooft anomaly is a mod 2 class global anomaly,
whose 4d invertible TQFT corresponds to a $\Z_2$ generator in the following cobordism group $\Omega^{d}_{G} \equiv \TP_d(G)$ (see the detailed computations in \cite{HAHSIV}):
\bea \label{eq:NeelVBS-anomaly}
&&\text{a $\Z_2$ generator $w_4(V_{\SO(5)})$
 in $\TP_4(\SO \times \SO(5))=\Z_2$,} \cr
&&\text{a $\Z_2$ generator $w_2(V_{\SO(3)})w_2(V_{\SO(2)}))$
in $\TP_4(\SO \times \SO(3) \times \SO(2))=\Z_2$.}
\eea
\end{enumerate}
\end{enumerate}
With \eq{eq:AdBsingular} and \eq{eq:w2w2}, 
these conclude our derivation of 3d WZW and 't Hooft anomaly for 3d dQCP for N\'eel-VBS transition.

\section{Perturbative Local and Nonperturbative Global Anomalies via Cobordism: Without or With $T$ or $CP$ symmetry}
\label{app:Anomalies via Cobordism}

Here we enlist the results of {perturbative local and nonperturbative global anomalies via cobordism}
mostly obtained from \cite{WanWang2018bns1812.11967, WW2019fxh1910.14668}.
Some of these results are used in \eq{eq:Spin10-anomaly}.
For some spacetime-internal symmetry group $\bar{G}$ of the SM or GUT models, we denote:
$$
\bar{G} \equiv {{G_{\text{spacetime} }} \times_{{N_{\text{shared}}}}  {{G}_{\text{internal}} } }
\equiv  ({\frac{{G_{\text{spacetime} }} \times  {{G}_{\text{internal}} } }{{N_{\text{shared}}}}}).
$$
We apply a version of cobordism group $\Omega^{d}_{\bar{G}} \equiv \TP_d(\bar{G})$ from Freed-Hopkins \cite{Freed2016}.
\Refe{WangWen2018cai1809.11171, WanWang2018bns1812.11967, WW2019fxh1910.14668, HAHSIV} had computed
some of these 5th cobordism group $\TP_{5}$ classifications of the 4d anomalies (via 
Thom-Madsen-Tillmann spectra \cite{thom1954quelques, MadsenTillmann4}, Adams spectral sequence \cite{Adams1958}, and Freed-Hopkins's theorem \cite{Freed2016}), to obtain:
\bea \label{eq:TP5cobordism}
\TP_{5}(\Spin \times_{\Z_2^F} \Z_{4,X} \times G_{\SM_q})&=&
\left\{ \begin{array}{ll} 
\Z^5\times\Z_2\times\Z_4^2\times\Z_{16}, &  q=1,3.\\
 \Z^5\times\Z_2^2\times\Z_4\times\Z_{16}, & q=2,6.
 \end{array}
\right.\cr
\TP_{5}(\Spin \times_{\Z_2^F} \Z_{4,X} \times \SU(5))&=&
\Z \times\Z_2\times \Z_{16}.\cr
\TP_{5}({\Spin\times_{\Z_2^F} G_{\PS_2}})=\TP_{5}({\Spin\times_{\Z_2^F} \frac{\Spin(6)\times \Spin(4) }{\Z_2}}) &=& \Z\times\Z_2^2. \cr
\TP_{5}({\Spin\times_{\Z_2^F} G_{\PS_1}})=\TP_{5}({\Spin \times_{\Z_2^F} {\Spin(6)\times \Spin(4) }}) &=& \Z\times\Z_2^3.  \cr
\TP_5(\Spin \times_{\Z_2^F}\Spin(10))&=&\Z_2. \cr
\TP_5(\Spin \times_{}\Spin(10))&=&0.
\eea
For details about their 5d manifold generators and 5d invertible TQFTs, see \Refe{WW2019fxh1910.14668}.
Comments on these perturbative local and nonperturbative global anomalies are in order: 
\begin{itemize}[leftmargin=.mm]
\item {\bf \emph{Perturbative local anomalies}} are classified by  integer $\Z$ classes,
detectable via the infinitesimal or small gauge or diffeomorphism transformations deformable to the identity element.
Given the chiral fermion (quarks and leptons) contents in Appendix \ref{sec:SM-GUT-table}, 
we can check that all the perturbative local anomalies (all $\Z$ classes) are cancelled in SMs and GUTs. 
These perturbative local anomaly cancellations are well-known, verified in any standard text books on SMs and GUTs.

\item {\bf \emph{Nonperturbative global anomalies}} are classified by finite torsion $\Z_n$ classes,
detectable via the large gauge or diffeomorphism transformations, not deformable to the identity element.
\begin{itemize}[leftmargin=2.mm]
\item  {\bf \emph{The $\Z_2$ and $\Z_4$ anomalies in $\TP_{5}(\Spin \times_{\Z_2^F} \Z_{4,X} \times G_{\SM_q})$ or $\TP_{5}(\Spin \times_{\Z_2^F} \Z_{4,X} \times \SU(5))$}}
include the variants or mutated versions of the Witten anomaly \cite{Witten1982fp}, by modifying the original SU(2) bundle to some principal $\SU(n)$ bundles.
Also there is a $\Z_4$ class anomaly from the hypercharge $\U(1)_Y^2$ paired with a $X$-background field with $(X)^2=(-1)^F$.
All these $\Z_2$ and $\Z_4$ anomalies are checked to be cancelled \cite{JW2006.16996, JW2008.06499, JW2012.15860}.
\item  {\bf \emph{The $\Z_{16}$ anomaly in $\TP_{5}(\Spin \times_{\Z_2^F} \Z_{4,X} \times G_{\SM_q})$ or $\TP_{5}(\Spin \times_{\Z_2^F} \Z_{4,X} \times \SU(5))$}}
can be cancelled if there are 16n Weyl fermions, each is charged under $\Z_{4,X}$ with $(X)^2=(-1)^F$. 
Since we only observe 15n Weyl fermions so far by experiments, 
\Refe{JW2006.16996, JW2008.06499, JW2012.15860} proposed alternative scenarios 
to cancel $\Z_{16}$ anomaly with 15n Weyl fermions at low energy ---
we revisit this issue separately in \Sec{sec:16nvs15nWeylfermions}
\item {\bf \emph{Several $\Z_2$ anomalies in $\TP_{5}({\Spin\times_{\Z_2^F} G_{\mathrm{PS}_{q'=1,2}}})$
or $\TP_5(\Spin \times_{\Z_2^F}\Spin(10))$}} come from 
either the variants of the Witten SU(2) anomaly \cite{Witten1982fp} (modifying the SU(2) gauge bundle to other bundles)
or the variants of the new SU(2) anomaly \cite{WangWenWitten2018qoy1810.00844}
(modifying the $w_2(TM)w_3(TM)=w_2(V_{\SO(3)})w_3(V_{\SO(3)})$ of $\SO(3)$ bundle to other $\SO(n)$ bundles).
Follow \cite{WangWen2018cai1809.11171, WangWenWitten2018qoy1810.00844}, 
we can check that the chiral fermion sectors (of quarks and leptons) of PS and $so(10)$ GUTs 
\emph{do not} suffer from any of these $\Z_2$ global anomalies. 
\end{itemize}
\end{itemize}
However, the hallmark of our 4d WZW term, and the {Fragmentary GUT-Higgs Liquid model} in \Sec{sec:FragmentaryHiggs},
relies on matching them with the $w_2w_3$ anomaly. So below, we walk through the
distinct properties of the various kinds of $w_2w_3$ anomalies listed in \eq{eq:TP5cobordism}, in more details.
\begin{enumerate}[leftmargin=2.mm, label=\textcolor{blue}{\arabic*}.]
\item $\TP_5(\Spin \times_{\Z_2^F}\Spin(10))=\Z_2$ is generated by a 5d invertible TQFT,
explained in \cite{WangWen2018cai1809.11171, WangWenWitten2018qoy1810.00844, WanWang2018bns1812.11967, WW2019fxh1910.14668}, 
$$
(-1)^{\int w_2(TM)w_3(TM)} 
=(-1)^{\int w_2(V_{\SO(10)})w_3(V_{\SO(10)})}.
$$

\item $\TP_5({\Spin\times_{\Z_2^F} G_{\PS_1}})$ includes $(\Z_2)^3$. One $\Z_2$ is closely related to the Witten SU(2) anomaly, see \cite{WW2019fxh1910.14668}.
The other $(\Z_2)^2$ are generated by 5d invertible TQFTs:
$$
\text{$(-1)^{\int w_2(V_{\SO(6)})w_3(V_{\SO(6)})}$
and
$(-1)^{\int \tilde{\eta}(\PD(w_4(V_{\SO(4)}))) }$.}
$$
The $\tilde{\eta}$ is a mod 2 index of 1d Dirac operator as a real massive 1d fermion, as a 1d cobordism invariant of $\TP_1({\Spin})=\Z_2$.
\item $\TP_5({\Spin\times_{\Z_2^F} G_{\PS_2}})$ includes $(\Z_2)^2$, which are generated by 5d invertible TQFTs:
$$
\text{$(-1)^{\int w_2(V_{\SO(6)})w_3(V_{\SO(6)})}$
and
$(-1)^{\int w_2(V_{\SO(4)})w_3(V_{\SO(4)})}$.}
$$
\item 
Now we can ask what are the relations between the $w_2w_3$ of $\SO(10)$ bundle (for the $so(10)$ GUT),
and that of $\SO(6)$ and $\SO(4)$ bundles (for the PS model)? We find that:
\bea
w_2(V_{\SO(n+m)})w_3(V_{\SO(n+m)})
=w_2(V_{\SO(n)})w_3(V_{\SO(n)})
+ w_2(V_{\SO(m)})w_3(V_{\SO(m)}) \mod 2,
\eea
where the crossing terms become
\begin{multline} \label{eq:w2w3crossing}
w_2(V_{\SO(n)})w_3(V_{\SO(m)})
+ w_2(V_{\SO(m)}) w_3(V_{\SO(n)})\\
=\Sq^1(w_2(V_{\SO(n)}) w_2(V_{\SO(m)}))=w_1(TM)(w_2(V_{\SO(n)}) w_2(V_{\SO(m)})),
\end{multline}
based on the Wu formula using the Steenrod square $\Sq^1$. 
This \eq{eq:w2w3crossing} vanishes if we restrict to the system without time-reversal $T$ symmetry (i.e., charge-conjugation-parity $CP$ symmetry) 
or on orientable manifolds so $w_1(TM)=0$ (i.e., here we only require $\Spin$ structures instead of $\Pin^{\pm}$ structures).
So if no $T$ or $CP$ symmetry, we simply relate a mod 2 anomaly of the $so(10)$, to two mod 2 anomalies of PS model: 
\bea
w_2(V_{\SO(10)})w_3(V_{\SO(10)})
=w_2(V_{\SO(6)})w_3(V_{\SO(6)})
+ w_2(V_{\SO(4)})w_3(V_{\SO(4)})  \mod 2.
\eea
\item {\bf \emph{With a time-reversal $T$ or $CP$ symmetry, or a generic $T'$ such as $CT$ symmetry}}:\\
If we hope to have the crossing term
\bea \label{eq:w2w3crossing-2}
w_2(V_{\SO(6)})w_3(V_{\SO(4)})+ w_2(V_{\SO(4)})w_3(V_{\SO(6)})
\eea
to enter the anomaly constraint in the PS models, we need to have
$\Sq^1(w_2(V_{\SO(6)}) w_2(V_{\SO(4)}))=w_1(TM)(w_2(V_{\SO(6)}) w_2(V_{\SO(4)}))\neq 0$,
this means that we need to include the time-reversal $T$ (or $CP$) symmetry,
or a generic $T'$ such as $CT$ symmetry.

In the $so(10)$ GUT, there are actually 
two kinds of time-reversal symmetry square:
\bea
T^2=(-1)^F \text{ for } \Pin^+, \quad T^2=+1 \text{ for } \Pin^-.
\eea
There are two kinds of commutation relations between time-reversal $T$ and the Spin(10) generators:
either commute (direct product ``$\times$'') or non-commute (semi-direct product ``$\ltimes$'' ).

So if we include the time-reversal $T$ into the $(\Spin \times_{\Z_2^F}\Spin(10))$-structure,
there are totally (at least) four kinds of time-reversal symmetries for the $so(10)$ GUT.
Based on the computation in \Refe{HAHSIV}, we summarize the four versions of the $so(10)$ GUT with time-reversal symmetries,
and their cobordism group $\TP_5$:
\bea
\TP_5(\Pin^+ \times_{\Z_2^F}\Spin(10))&=&\Z_2. \cr
\TP_5(\Pin^- \times_{\Z_2^F}\Spin(10))&=&\Z_2. \cr
\TP_5(\Pin^+ \ltimes_{\Z_2^F}\Spin(10))&=&\Z_2. \cr
\TP_5(\Pin^- \ltimes_{\Z_2^F}\Spin(10))&=&\Z_2. 
\eea
Interestingly, for the cases of 
$\TP_5(\Pin^+ \ltimes_{\Z_2^F}\Spin(10))=\Z_2$ 
and $\TP_5(\Pin^- \ltimes_{\Z_2^F}\Spin(10))=\Z_2$,
their 4d anomalies are generated by a subtilely distinct 5d invertible TQFT
\bea
(-1)^{\int w_2(TM)w_3(TM)} 
=(-1)^{\int w_2(V_{\O(10)})w_3(V_{\O(10)}) }.
\eea
Notice now we have $w_2(V_{\O(10)})w_3(V_{\O(10)})$ instead of
$w_2(V_{\SO(10)})w_3(V_{\SO(10)})$. The bundle constraints for 
$(\Pin^+ \ltimes_{\Z_2^F}\Spin(10))$ and 
$(\Pin^- \ltimes_{\Z_2^F}\Spin(10))$ are also different:
\bea \label{eq:Pinbundleconstraint}
&&\bullet\quad \Pin^+ \ltimes_{\Z_2^F}\Spin(10) \text{ constraint}:  \quad w_2(V_{\O(10)})=w_2(TM), \quad w_3(V_{\O(10)})=w_3(TM).\cr
&&\bullet\quad \Pin^- \ltimes_{\Z_2^F}\Spin(10) \text{ constraint}: \quad w_2(V_{\O(10)})=w_2(TM)+w_1(TM)^2, \quad \cr 
&&                           w_3(V_{\O(10)})+w_1(V_{\O(10)}) w_2(V_{\O(10)})=\Sq^1w_2(V_{\O(10)})=\Sq^1w_2(TM)=w_3(TM)+w_1(TM) w_2(TM).\cr
&&
\eea
The punchline here in \eq{eq:Pinbundleconstraint} is that because time-reversal $T$ (or $CP$) or some $T'$ is a valid global symmetry,
we can put the theory on an unorientable manifold with $w_1(TM) \neq 0$ also $w_1(V_{\O(10)}) \neq 0$.
Therefore, the crossing term in \eq{eq:w2w3crossing-2} can still contribute a potential anomaly.
This crossing term anomaly $w_2(V_{\SO(6)})w_3(V_{\SO(4)})+ w_2(V_{\SO(4)})w_3(V_{\SO(6)})$
turns out to play a possible crucial role in our construction of \Sec{sec:FragmentaryHiggs}. See more discussions in a companion work. 

Similar stories apply to a larger gauge group unification for three generations of fermions, such as the $so(18)$ GUT
with a Spin(18) gauge group. We simply replace all above discussions of $so(10)$ to $so(18)$,
and replace Spin(10) to Spin(18).

\end{enumerate}

\section{Fermionic Double Spin structure DSpin for a modified $so(10)$
GUT-Higgs liquid model}
\label{sec:DSpin}

{Here are detailed comments about our construction of spacetime-internal symmetry that involves the fermionic double spin structure DSpin 
given in \Sec{sec:QED4Parton}.
\begin{enumerate}[leftmargin=0.mm]
\item 
First, we recall that we have introduced: 
$$
\left\{
\begin{array}{l}
\text{{Weyl} fermion $\psi$ in the ${\bf 16}$ of $\Spin(10)$ for the $so(10)$ GUT,}\\
\text{{Dirac} fermion $\xi$ in the ${\bf 10}$ of $\SO(10)$ (also of $\Spin(10)$) for the fermionic parton QED$'_4$ theory.}
\end{array}
\right.
$$
\item 
The modified $so(10)$ GUT requires a $\Spin \times_{\Z_2^F}\Spin(10)$ structure in order to manifest a $w_2w_3$ anomaly.
In this structure, the fermion $\psi$ in ${\bf 16}$ is charged with $(-1)^F$ odd under the fermion parity ${\Z_2^F}$.
This meanwhile implies the constraint on the matter field spectrum under the $\Spin \times_{\Z_2^F}\Spin(10)$ structure:\\
There is a short exact sequence: $1 \to {\Z_2^F} \to Z(\Spin(10))=\Z_{4,X} \to Z(\SO(10))=\Z_2 \to 1$. 
Given
the $\Z_{4,X}$ charge state
$| X\rangle$ with $X=0,1,2,3$,
we have its representation $z^X$
such that $z \in \U(1)$ with $|z|=1$, where we embed the normal subgroup $\Z_2^F \subset  \Z_{4,X} \subset \U(1)$.\\
$\bullet$ The $\Z_{4,X}$ symmetry generator $\U_{\Z_{4,X}}$ acts on $| X\rangle$,
which becomes $\U_{\Z_{4,X}} | X\rangle = \ii^X | X\rangle$ with $z=\ii$. \\
$\bullet$ The subgroup  $\Z_{2}^F$ symmetry generator 
$\U_{\Z_{2}^F}= (\U_{\Z_{4,X}})^2$
can also act on $| X\rangle$,
which becomes $\U_{\Z_{2}^F}  | X\rangle  = (\U_{\Z_{4,X}})^2 | X\rangle = \ii^{2X} | X\rangle = (-1)^X | X\rangle $.
Thus, we read the fermion parity $(-1)^F$,
the $| 1\rangle$ and $| 3\rangle$ are fermionic with $-1$ (thus odd in $\Z_{2}^F$), while 
the $| 0\rangle$ and $| 2\rangle$ are bosonic with $+1$ (thus even in $\Z_{2}^F$).\\
$\bullet$ Any fermion charged under ${\Z_2^F}$ must have the $(-1)^F=-1$ also identified as the $\Z_2$ normal subgroup of the center $Z(\Spin(10))=\Z_{4,X}$.
Thus these fermions must have a $Z(\Spin(10))=\Z_{4,X}$ charge either 1 or 3 mod 4.\\
$\bullet$  Any boson not charged under ${\Z_2^F}$ must have a $Z(\Spin(10))=\Z_{4,X}$ charge either 0 or 2 mod 4.\\
\item 
The $\xi$ fermion in the ${\bf 10}$ of $\SO(10)$ has a charge 1 mod 2 under $Z(\SO(10))=\Z_2$. 
The $\xi$ fermion has a charge 2 mod 4 under $Z(\Spin(10))=\Z_{4,X}$, thus the $\xi$ is ``bosonic under the ${\Z_2^F}$.''
Thus the $\xi$ fermion is not compatible with the fermion parity required in $\Spin \times_{\Z_2^F}\Spin(10)$ described earlier.
Thus, we must introduce a new fermion parity $\Z_2^{F'}$ for $\xi$.
\item 
We construct the full spacetime-internal symmetry group by including the bosonic spacetime rotational symmetry 
$\SO$, the bosonic internal symmetry $\SO(10)$, and the two fermion parities $\Z_2^{F}\times \Z_2^{F'}$,
then we combine the group extensions 
\bea
1 \to \Z_2^F    \to &\Spin= \Z_2^F   \rtimes \SO& \to \SO \to 1,\cr
1 \to \Z_2^{F'}    \to &\Spin' = \Z_2^{F'}   \rtimes \SO& \to \SO \to 1,\cr
1 \to \Z_2^F \times  \Z_2^{F'} \to &\DSpin & \to \SO \to 1,\cr
1 \to \Z_2^F    \to &\Spin(10)   & \to \SO(10) \to 1,\cr
1 \to \Z_2^{F'}    \to & \Z_2^{F'} \times \SO(10)   & \to \SO(10) \to 1,
\eea
to obtain the full web \eq{eq:extension-web},
\begin{equation}
\xymatrix{
&  &1 \ar[d] & 1 \ar[d]\\
&  & {G'_{\text{int}}}_{{\supseteq \mathbb{Z}_{2}^{F'}}}   \ar[d]  & {G'_{\text{int}}}_{{\supseteq \mathbb{Z}_{2}^{F'}}}   \ar[d]\\
1\ar[r]&  \Z_2^F  \ar[r] & {{{(\DSpin  \times_{\Z_2^F}\Spin(10))} }} \times_{\Z_2^{F'}} {G'_{\text{int}}} \ar[d] \ar[r] & 
 {{({\Spin' \times\SO(10)})\times_{\Z_2^{F'}} {G'_{\text{int}}}}} \ar[r] \ar[d] &1\\
1\ar[r]& \Z_2^F \ar[r] & \Spin \times_{\Z_2^F}\Spin(10) \ar[r]^{}  \ar[d]& \SO \times \SO(10) \ar[r] \ar[d] &1\\
&  &1 & 1
}
\end{equation}
where we can choose ${G'_{\text{int}}}=\Z_2^{F'}, \U(1)'$, or $\SU(2)'$ to reproduce the required structure in  \Sec{sec:QED4Parton}.
In all cases, we have ${G'_{\text{int}}} \supseteq \Z_2^{F'}$ contains the new fermion parity as its normal subgroup.
\end{enumerate}
In addition to the DSpin structure, by including an extra discrete symmetry (such as a time-reversal symmetry),
the literature also discovers the structure known as DPin \cite{Kaidi2019tyfJulioTachikawa1911.11780} and EPin \cite{Wan2019sooWWZHAHSII1912.13504} structures. 
\begin{itemize}
\item
The DPin  \cite{Kaidi2019tyfJulioTachikawa1911.11780} is known as introducing two types of fermions (with $\Z_2^{F_+}$ and  $\Z_2^{F_-}$,
such that an extra discrete $\Z_2^T$ symmetry (e.g., called it a time-reversal symmetry) exchanges this two types of fermions.
The $\DPin(d)$ contains a  discrete dihedral group of order 8, known as $\mathbb{D}_8= (\Z_2^{F_+} \times \Z_2^{F_-})\rtimes_{\rho,0} \Z_2^T$,
where $\rho$ is a nontrivial $\Z_2^T$ action on 
${\rm Aut}(\Z_2^{F_+} \times \Z_2^{F_-})$ with two kinds of fermion parity 
$\Z_2^{F_+} \times \Z_2^{F_-}$ at the
$\mathbb{D}_8$'s center. 
Overall, the $\mathbb{D}_8$ structure
sits at the group extension $1 \to (\Z_2^{F_+} \times \Z_2^{F_-}) \to \mathbb{D}_8  \to \Z_2^T \to 1$.
\item
The EPin \cite{Wan2019sooWWZHAHSII1912.13504} is known as simultaneously imposing both $\Pin^+$ and $\Pin^-$ structure,
via introducing two types of fermions (with $\Z_2^{F_+}$ and  $\Z_2^{F_-}$) with the time-reversal symmetry
acting differently on fermions, $T^2=(-1)^{F_+}$ and $T^2=+1$ respectively 
(via the group extension $1\to\Z_2^{F_+} \to \Z_4^{TF_+} \to \Z_2^T \to 1$
and $1\to\Z_2^{F_-} \to \Z_2^{T} \times \Z_2^{F_-} \to \Z_2^T \to 1$).
\end{itemize}
See also the interpretations via the regularized quantum many-body model \cite{PrakashJW2011.13921}. 
}

\section{Bibliography}

\bibliographystyle{Yang-Mills}
\bibliography{BSM-SU3SU2U1-cobordism-GEQC.bib}

\end{document}